\documentclass[prc,11pt,nofootinbib]{revtex4}

\usepackage{amsmath}    
\usepackage{graphicx}   

\usepackage{xfrac}
\usepackage{lineno}

\newcommand{\isotope}[2]{$^{#2}{\rm #1}$}
\newcommand{\etal}{{\em et al.}~}
\newcommand{\adr}{De\,R\'{u}jula}

\usepackage{hyperref}   
\hypersetup{
    colorlinks=true,
    linkcolor=blue,
    filecolor=magenta,      
    urlcolor=cyan,
}
\begin{document}

\title{\boldmath Direct Measurements of Neutrino Mass}

\author{Joseph~A.~Formaggio}
\affiliation{Laboratory for Nuclear Science and Dept. of Physics, Massachusetts Institute of Technology, Cambridge, MA 02139, USA}
\author{Andr\'e~Luiz~C.~de~Gouv\^ea}
\affiliation{Department of Physics and Astronomy, Northwestern University, Evanston, IL 60208, USA}
\author{R.~G.~Hamish~Robertson}
\affiliation{Center for Experimental Nuclear Physics and Astrophysics, and Dept. of Physics, University of Washington, Seattle, WA 98195, USA}

\date{\today}

\begin{abstract}

The turn of the 21st century witnessed a sudden shift in our fundamental understanding of particle physics.  
While the minimal Standard Model predicts that neutrino masses are exactly zero, the discovery of neutrino oscillations proved the Standard Model wrong. 
Neutrino oscillation measurements, however, do not shed light on the scale of neutrino masses, nor the mechanism by which those are generated.  The neutrino mass scale is most directly accessed by studying the energy spectrum generated by beta decay or electron capture -- a technique dating back to Enrico Fermi's formulation of radioactive decay.  In this Article, we review the methods and techniques -- both past and present -- aimed at measuring neutrino masses kinematically. We focus on recent experimental developments that have emerged in the past decade,  overview the spectral refinements that are essential in the treatment of the most sensitive experiments, and give a simple yet effective protocol for estimating the sensitivity.  Finally, we provide an outlook of what future experiments might be able to achieve.

\end{abstract}

\maketitle

\tableofcontents

\flushbottom

\section{Introduction}
\label{sec:intro}
\setcounter{equation}{0}

The existence of neutrinos was first postulated by Pauli nine decades ago. In his famous `Dear Radioactive Ladies and Gentlemen' letter \cite{Pauli:1930:LTC,Brown:1978:IN}, Pauli also made the first non-trivial estimate of the mass of the neutrino: ``The mass of the neutrons\begin{NoHyper}\footnote{``Neutron'' was the name attributed to the hypothetical particle by Pauli. In order to avoid confusion with the modern neutron, discovered a few years later, the diminutive form `neutrino' was famously introduced by Fermi. For information on the history of the neutrino, see, for example, Ref.~\cite{Franklin:2000yt}.}\end{NoHyper} should be of the same order of magnitude as the electron mass and in any case not larger than 0.01 times the proton mass.'' His qualitative prediction -- $m_{\nu}\sim m_e$ -- based on aesthetics and minimality, turned out to be too large by many orders of magnitude. 

Laboratory searches for a nonzero neutrino mass started in the 1930s and have continued in earnest up to the present. Pauli's neutrino, now called the electron neutrino $\nu_e$, was not directly detected until the work of Reines and Cowan in the 1950s \cite{Reines:1953pu,Cowan:1992xc}. Two other neutrino flavors were discovered, $\nu_\mu$ in 1962 \cite{Danby:1962nd} and $\nu_\tau$ in 2001 \cite{Kodama:2000mp}, and the searches diversified in order to accommodate the possibility that the different neutrino species had qualitatively different masses. The most stringent upper bounds to the mass of the electron neutrino evolved from Pauli's 10~MeV qualitative upper bound -- 1\% of the proton mass -- to several electron-volts by the late 1990s.

Conclusive evidence for nonzero neutrino masses was  revealed in 1998 with the discovery of atmospheric neutrino oscillations by the Super-Kamiokande collaboration \cite{Fukuda:1998mi}, building on previous hints for atmospheric neutrino oscillations obtained by the Irvine-Michigan-Brookhaven \cite{Casper:1990ac} and Kamiokande Collaborations \cite{Hirata:1992ku}, and the different collaborations that helped define the solar neutrino puzzle: Homestake \cite{Cleveland:1998nv}, Gallex \cite{Hampel:1998xg}, SAGE \cite{Abdurashitov:1999zd}, and Kamiokande \cite{Fukuda:1996sz}. The solar neutrino puzzle was definitively resolved by the Sudbury Neutrino Observatory (SNO) collaboration \cite{Ahmad:2002jz} and is also a consequence of nonzero neutrino masses. The 2015 Nobel Prize in Physics was awarded to Takaaki Kajita -- from the Super-Kamiokande Collaboration -- and Arthur B.~McDonald -- from the SNO collaboration -- ``for the discovery of neutrino oscillations, which shows that neutrinos have mass."\begin{NoHyper}\footnote{ The Nobel Prize in Physics 2015. \url{https://www.nobelprize.org/prizes/physics/2015/summary/}.} \end{NoHyper}

Several twenty-first century oscillation experiments provide precision measurements of the neutrino oscillation phenomenon \cite{Abe:2008aa,Adamson:2008zt,Abe:2014bwa,RENO:2015ksa,An:2016ses,Abe:2017aap,Abe:2017vif,NOvA:2018gge}. These translate into rather precise measurements of neutrino mass-squared differences and reveal that at least two of the three neutrinos are massive and the heaviest neutrino mass is at least 0.05~eV. %
Oscillation experiments, however, are powerless when it comes to measuring the individual values of the neutrino masses -- they are only sensitive to mass-squared differences. Other laboratory observables are sensitive to nonzero neutrino masses. Some of these observables are only indirectly sensitive to the masses. In those cases, the connection between measurement and neutrino masses is mediated by a theoretical framework, along with other hypotheses. Such observables include the rate for neutrinoless double-beta decay and the large-scale structure of the universe. Other observables are more directly sensitive to neutrino masses -- the kinematics of the observable are directly established by the fact that neutrinos have nonzero masses in a way that is virtually independent from the nature of the physics responsible for the observable. Among these are precision measurements of  nuclear beta decay, meson decay,  charged-lepton decays, and electron and neutrino capture in nuclei. The most sensitive among these direct probes of nonzero neutrino masses are the subject of this review. 

A number of reviews of the subject are available \cite{Otten:2008zz,Drexlin:2013lha,Holzschuh:1992xy,robertson:1988aa}.  As the most recent was seven years before the present one, it is an opportunity to review the recent progress and to consider where the field will go in the future.  The KATRIN experiment is now running, tightening the upper limit on neutrino mass by a factor 2 after only a month of operation.  The new method of cyclotron radiation emission spectroscopy has passed a crucial proof-of-principle test. Prompted by this success, the possibility of a neutrino mass experiment based on atomic tritium is once again receiving consideration.  Microcalorimetry is advancing technically to enable studies of isotopes other than tritium.  

This review is organized as follows. In Section~\ref{sec:status}, we summarize the direct and indirect information on neutrino masses that is currently available, along with some near-future expectations. In Section~\ref{sec:models}, we discuss how the discovery of nonzero neutrino masses impacted our understanding of fundamental particle physics, along with the different outstanding questions that we hope will be informed by the direct observation of nonzero neutrino masses. Nature has provided only two isotopes that continue to offer prospects for gains in sensitivity, as we describe in Section~\ref{sec:spectrum}.  The decades of progress that have brought the field to the 1-eV sensitivity level are reviewed in Section~\ref{sec:progress}.  Related research on sterile neutrinos and the relic neutrino background is covered in Section~\ref{sec:other}.  In Section~\ref{sec:refinements}, we discuss spectrum refinements that are essential in the treatment of the most sensitive experiments, and, in Section~\ref{sec:sensitivity}, provide a simple yet effective protocol for estimating the sensitivity.  In Section~\ref{sec:future}, the new techniques that are emerging to advance the field are introduced, and in Section~\ref{sec:conclusion}, we conclude.

\section{Neutrino Masses: Current Status}
\label{sec:status}
\setcounter{equation}{0}

In this section we provide an overview of the current understanding of the values of the neutrino masses. For a detailed review, see, for example, the `Neutrino Masses, Mixing, and Oscillations' chapter of the Particle Data Book \cite{Zyla:2020zbs}.

\subsection{Neutrino oscillations}

Precision measurements of the flux of solar neutrinos reveal that fewer electron-type neutrinos arrive at the Earth than predicted. What came to be known as the solar neutrino problem presented itself with the first measurements of the solar neutrino flux in the 1960s -- see references in Ref.~\cite{Cleveland:1998nv} -- and persisted until it was definitively resolved by the SNO experiment in the early 2000s \cite{Ahmad:2002jz}. Ultimately, solar neutrino data imply that electron neutrinos are, in fact, linear superpositions of at least two neutrino mass-eigenstates and that at least one of these has a nonzero mass. The difference between the neutrino masses-squared is of order $\Delta m^2_{21}\equiv m_2^2-m_1^2\sim 10^{-4}$~eV$^2$. Here $m_1$ and $m_2$ are the masses of two different neutrino mass eigenstates, labeled $\nu_1$ and $\nu_2$. 

Precision measurements of the flux of atmospheric neutrinos also reveal that fewer muon-type neutrinos survive passage through the Earth than expected. The effect depends on the distance between the neutrino production and detection points and the neutrino energy. The solution to this new problem, the atmospheric neutrino problem, first revealed by data from the IMB and Kamiokande experiments and later confirmed beyond reasonable doubt by the Super-Kamiokande experiment, was the realization that muon neutrinos are linear superpositions of at least two neutrino mass eigenstates and that at least one of these has a nonzero mass. 
In this case the difference between the neutrino masses-squared is of order $\Delta m^2_{31}\equiv m_3^2-m_1^2\sim 10^{-3}$~eV$^2$. Here $m_3$ is the mass of the third distinct neutrino mass eigenstate, labeled $\nu_3$. 

\begin{figure}[ht]
\centerline{\includegraphics[width=0.55\textwidth]{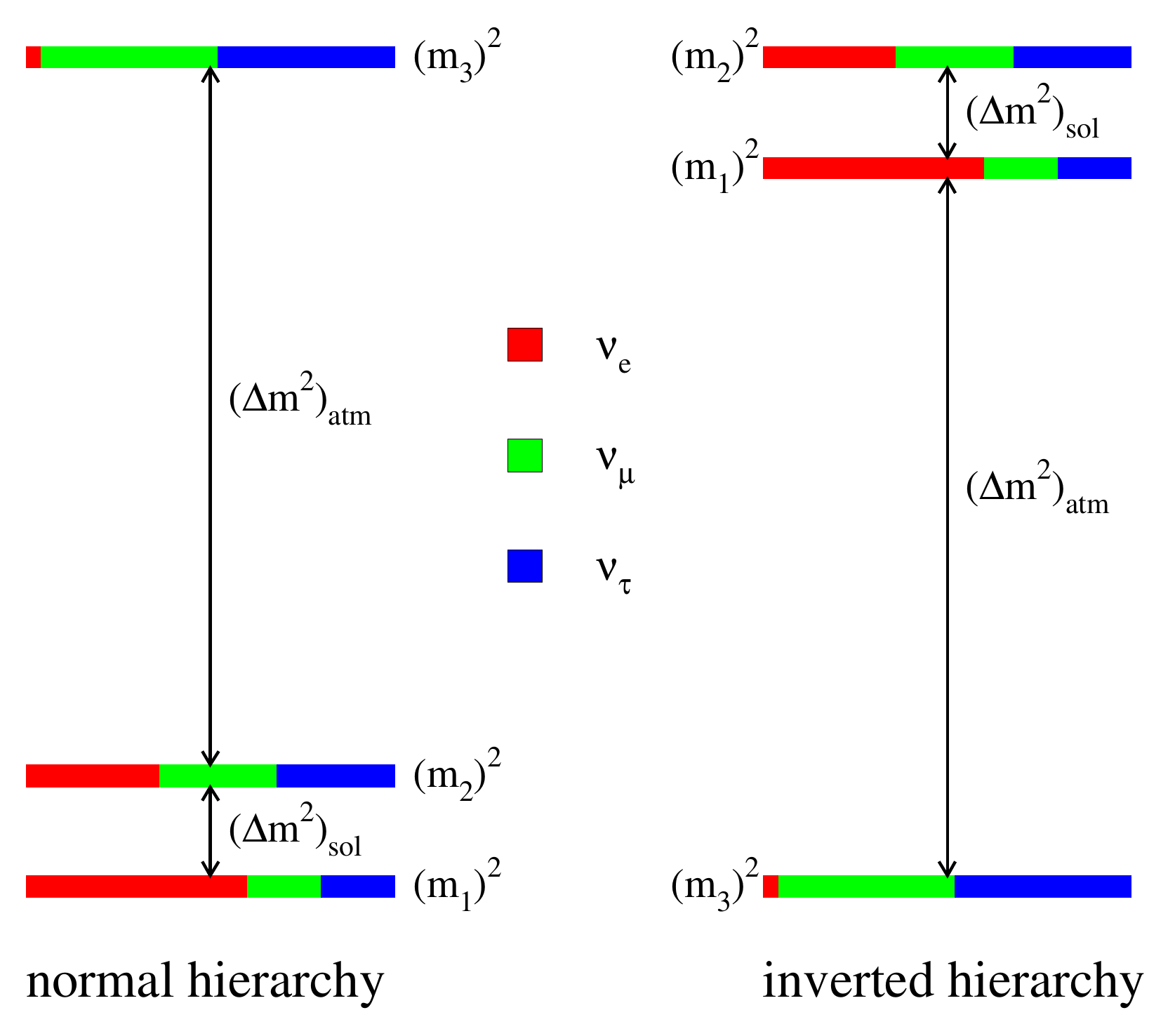}}
\caption{Illustration of the two distinct neutrino mass orderings that fit nearly all of the current neutrino data, for typical values of all mixing angles and mass-squared differences. The color coding (shading) indicates the fraction $|U_{\alpha i}|^2$ of each distinct flavor $\nu_{\alpha}$, $\alpha=e,\mu,\tau$ contained in each mass eigenstate $\nu_i$, $i=1,2,3$. For example, $|U_{e2}|^2$ is equal to the fraction of the $(m_2)^2$ ``bar'' that is painted red (shading labeled as ``$\nu_e$''). From \cite{deGouvea:2013onf}.
\label{fig:ordering}}
\end{figure}

In the last two decades, multiple experiments with multiple neutrino sources and detector technologies have confirmed the existence of neutrino oscillations and have allowed the construction of a very robust three-massive-neutrinos paradigm. It asserts that neutrinos interact as prescribed by the Standard Model of particle physics and that the neutrino charged-current-interaction eigenstates -- $\nu_{e}$, $\nu_{\mu}$ and $\nu_{\tau}$ -- are linear superpositions of the neutrino mass eigenstates, $\nu_1$, $\nu_2$, $\nu_3$, with masses, respectively, $m_1$, $m_2$, and $m_3$: 
\begin{equation}
\nu_{\alpha}=\sum_i U_{\alpha i}\nu_i,  
\end{equation}
where $i=1,2,3$, $\alpha=e,\mu,\tau$ and $U_{\alpha i}$ are the elements of the $3\times 3$ unitary leptonic mixing matrix, also referred to as the neutrino mixing matrix or the Pontecorvo-Maki-Nakagawa-Sakata (PMNS) matrix. The neutrino mass eigenstates are defined through the relative values of the neutrino masses-squared, as follows: $m_2^2>m_1^2$ and $|m_3^2-m_1^2|,|m_3^2-m_2^2|>m_2^2-m_1^2$ in such a way that $m_3^2>m_2^2>m_1^2$, termed the normal mass-ordering (NMO), or $m_3^2<m_1^2<m_2^2$, termed the inverted mass-ordering (IMO). In the NMO, $\Delta m^2_{31},\Delta m^2_{32}$ are positive while in the IMO $\Delta m^2_{31},\Delta m^2_{32}$ are negative. The two mass orderings are depicted in Figure~\ref{fig:ordering}. The mass ordering is currently unknown.  A slight preference in the world neutrino data for the NMO \cite{Esteban:2020cvm} has recently disappeared with new data \cite{Kelly_PhysRevD.103.013004}.

The PMNS matrix is parameterized with three mixing angles $\theta_{12},\theta_{13},\theta_{23}$ and one (three) CP-odd phase(s) $\delta$ ($\delta,\eta_1,\eta_2$) if the neutrinos are Dirac (Majorana) fermions. Throughout, we will use the PDG parameterization for the PMNS matrix \cite{Zyla:2020zbs} and will assume the three-massive-neutrinos paradigm is true, unless otherwise noted. 

Neutrino oscillation experiments can constrain the PMNS matrix and the neutrino mass-squared differences $\Delta m^2_{ij}=m_i^2-m_j^2$, $ij=21, 31, 32$. Only two of the three mass-squared differences are independent because $\Delta m^2_{32}=\Delta m^2_{31}-\Delta m^2_{21}$. The world neutrino data translate into robust measurements of the neutrino mass-squared differences. According to the Nufit Collaboration \cite{Esteban:2020cvm}\begin{NoHyper}\footnote{Several phenomenological collaborations regularly collect and analyze the neutrino oscillation data, estimating the values of the oscillation parameters. Here, we will use the results from \cite{Esteban:2020cvm}, unless otherwise noted. These are consistent with other global fits, including those presented in Refs.~\cite{Capozzi:2018ubv,deSalas:2020pgw}.} \end{NoHyper}
\begin{eqnarray}
& \Delta m^2_{21}=\left(7.42^{+0.21}_{-0.20}\right)\times 10^{-5}~{\rm eV}^2, \\ & \Delta m^2_{31}=\left(2.517^{+0.026}_{-0.028}\right)\times 10^{-3}~{\rm eV}^2~(\rm NMO), \\ & {\rm or} \nonumber \\
& \Delta m^2_{32}=\left(-2.498^{+0.028}_{-0.028}\right)\times 10^{-3}~{\rm eV}^2~(\rm IMO).
\end{eqnarray}
 
Neutrino oscillation experiments do not inform the values of the individual neutrino masses, only the mass-squared differences. For the individual neutrino masses, information outside of neutrino oscillations is required.

\subsection{Neutrinoless double beta decay}
\label{sec:0nubb}

With the discovery of nonzero neutrino masses, one of the most important outstanding questions in particle physics is the nature of the neutrinos: are they Majorana or Dirac fermions? If neutrinos are Majorana fermions, lepton-number $L$ conservation is not an exact law of nature and has to be violated, even if only very feebly. On the other hand, if neutrinos are Dirac fermions, lepton-number conservation is an exact law of nature or, at the very least, $\Delta L=2$ processes are strictly forbidden.   

Searches for neutrinoless double-beta decay ($0\nu\beta\beta$), $\Delta L=2$ nuclear-decay processes of the type $(Z,A)\to (Z+2,A)+e^-e^-$, where $(Z,A)$ is a nucleus with atomic number $Z$ and mass number $A$, are the most powerful probes of lepton-number conservation. Other searches include ``variations'' on $0\nu\beta\beta$, including neutrinoless double-beta-plus decay, $(Z,A)\to (Z-2,A)+e^+e^+$ and lepton-number violating electron capture, $(Z,A)+e^-\to (Z-2,A)+e^+$, along with $\mu^-\to e^+$ conversion in nuclei, forbidden meson decays \cite{Zyla:2020zbs}, including $K^+\to\mu^+\mu^+\pi^-$, and the production of same-sign di-leptons and no missing energy at hadron colliders (e.g., $pp\to e^+\mu^++X$, where $X$ is a state with zero lepton number). Here, we concentrate on constraints from $0\nu\beta\beta$.

If Majorana-neutrino exchange is the dominant contribution to $0\nu\beta\beta$, the rate for $0\nu\beta\beta$ is a function of the neutrino masses. If all neutrino masses are small relative to the typical energy scales involved in $0\nu\beta\beta$, which is of order dozens of MeV, in the absence of new physics other than nonzero Majorana neutrino masses, the amplitude for $0\nu\beta\beta$ is proportional to a linear combination of the neutrino masses:
\begin{equation}
    m_{\beta\beta}\equiv \sum_i(U_{ei})^2m_i = \cos^2\theta_{13}\left(\cos^2\theta_{12}m_1'+\sin^2\theta_{12}m_2'\right) + \sin^2\theta_{13}m_3',
\label{eq:mbb}
\end{equation}
where $m_i'=e^{i\phi_i}m_i$, where $\phi_i$ are combinations of the CP-odd phases in the PMNS matrix, see, for example, Ref.~\cite{Zyla:2020zbs} for a concrete parameterization; $m_{\beta\beta}$ is a complex parameter and experiments are sensitive to its magnitude. $\theta_{13}$ and $\theta_{12}$ are two of the mixing angles used to parameterize the PMNS matrix. Unless otherwise noted, for the mixing angles, we use the PDG parameterization~\cite{Zyla:2020zbs}.

\begin{figure}[htb]
\centerline{\includegraphics[width=1.0\textwidth]{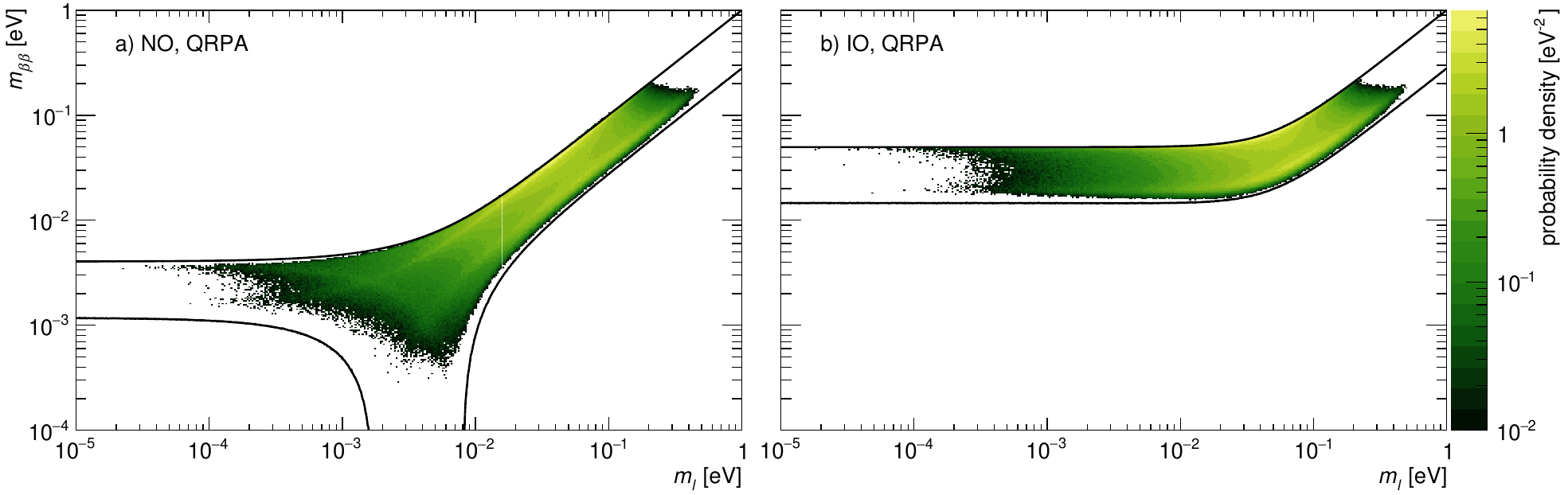}}
\caption{Marginalized posterior distributions for $|m_{\beta\beta}|$ and $m_{\rm least}$ ($m_{\rm l}$ in the figure) in the NMO (left) and IMO (right). From Agostini \etal \cite{Agostini:2017jim}.}
\label{fig:dbdbayesian}
\end{figure}

Under these conditions, a measurement of the rate for $0\nu\beta\beta$, combined with input from neutrino oscillations, provides nontrivial information on the neutrino masses. Using information from neutrino oscillation experiments, it is possible to parameterize $|m_{\beta\beta}|$ as a function of two relative phases among the $m_i'$ parameters and the value of the lightest neutrino mass $m_{\rm least}$. For the different mass orderings
\begin{equation}
m_{\rm least} = m_1~~({\rm NMO})~~~~{\rm or}~~~~m_{\rm least} = m_3~~({\rm IMO}).
\end{equation}
The fact that these so-called Majorana phases are unknown and, in practice, impossible to constrain experimentally in any other way, renders the information on $m_{\rm least}$ from $0\nu\beta\beta$ always imperfect.  Agostini, Benato and Detwiler \cite{Agostini:2017jim} have carried out a Bayesian analysis incorporating existing data to make predictions, under well-defined assumptions, of the discovery probability for true values of $|m_{\beta\beta}|$.  The distributions are shown in Fig.~\ref{fig:dbdbayesian}.
It is striking that, barring some particular physics that would drive $|m_{\beta\beta}|$ or $m_{\rm least}$ to zero, the discovery probability is not small, especially in the IMO.  It can also be seen that a direct mass measurement, essentially a measurement of $m_{\rm least}$, below 100 meV becomes highly informative in the search for neutrinoless double beta decay.  

Experimental searches for neutrinoless double beta decay measure or limit the decay rate of a particular isotope.  That rate depends on the product of $|m_{\beta\beta}|$, a phase-space factor, and a nuclear matrix element for the  $(Z,A)\to(Z+2,A)$ transition, the most commonly investigated type. Currently, the matrix elements are poorly constrained. Estimates performed using different techniques differ by a factor of a few. Qualitative and quantitative improvements are expected in the near future, but it is fair to expect that, for the foreseeable future, the theoretical uncertainty on extracting $|m_{\beta\beta}|$ from the rate for $0\nu\beta\beta$ will be sizable. For detailed, recent reviews see, for example, \cite{Agostini:2017jim,Dolinski:2019nrj}.  Notwithstanding the matrix-element problem, experimental progress in the past decade has been remarkable, limiting the effective Majorana mass to values well below the levels presently accessible to direct measurements.  Table \ref{tab:dbdexpts}, updated  from that compiled in \cite{Dolinski:2019nrj}, gives the half-life $T_{1/2}^{0\nu}$ and Majorana mass $|m_{\beta\beta}|$ limits from the most recent experiments.
\begin{table}[htb]
    \centering
    \caption{Published half-life and Majorana mass limits from recent  experiments.}
    \begin{tabular}{llllr}
    \hline \hline
    Isotope \phantom{aa} & $T_{1/2}^{0\nu}\ (\times 10^{25}$ y) \phantom{aa}& $|m_{\beta\beta}|$ (eV) \phantom{aaaa} &  Experiment &\phantom{aa} Ref. \\
    \hline
    $^{48}$Ca & $>5.8\times 10^{-3}$ & $<3.5-22$ & ELEGANT-IV & \cite{Umehara:2008ru} \\
    $^{76}$Ge & $>18$ & $<0.079-0.180$ & GERDA & \cite{Agostini:2020xta}  \\
     & $>2.7$ & $<0.200-0.433$ & {\sc Majorana Demonstrator} & \cite{Alvis:2019sil} \\
    $^{82}$Se & $>3.6\times 10^{-2}$ & $<0.89-2.43$ & NEMO-3 & \cite{Barabash:2010bd} \\
    $^{96}$Zr & $>9.2\times 10^{-4}$ & $<7.2-19.5$ & NEMO-3 & \cite{Argyriades:2009ph} \\
    $^{100}$Mo & $>1.1\times 10^{-1}$ & $<0.33-0.62$ & NEMO-3 & \cite{Arnold100Mo_PhysRevD.92.072011} \\
    $^{116}$Cd & $>1.0\times 10^{-2}$ & $<1.4-2.5$ & NEMO-3 & \cite{Arnold116Cd_PhysRevD.95.012007} \\
    $^{130}$Te & $>3.2$ & $<0.075-0.350$ & CUORE & \cite{Adams:2019jhp} \\
    $^{136}$Xe & $>10.7$ & $<0.061-0.165$ & KamLAND-Zen &  \cite{KamLANDZen_PhysRevLett.117.082503} \\
     & $>3.5$ & $<0.093-0.286$ & EXO-200 & \cite{Anton:2019wmi} \\
    $^{150}$Nd & $>2.0\times 10^{-3}$ & $<1.6-5.3$ & NEMO-3 & \cite{Arnold150Nd_PhysRevD.94.072003} \\
    \hline \hline
    \end{tabular}
    \label{tab:dbdexpts}
\end{table}

In viewing the experimental results, it must be kept in mind that the rate for $0\nu\beta\beta$ is  a function only of the neutrino masses when light-neutrino exchange is the leading contribution to $0\nu\beta\beta$. More generally, lepton-number violating physics can impact $0\nu\beta\beta$ in a way that the connection between the rate for $0\nu\beta\beta$ and the neutrino masses is either indirect or, in some cases, non-existent. For an overview, see, for example, Ref.~\cite{Pas:2015eia}.

The experimental results of the current generation have been obtained with detectors having isotopic masses in the range of tens to hundreds of kilograms.  A new generation of detectors an order of magnitude larger is now beginning.  A comprehensive summary of the plans and status may be found in the APPEC Committee Report \cite{Giuliani:2019uno} prepared for the European strategy.  The goal of the next generation is sensitivity in the range of the IMO. 

Finally, we highlight that, of course, if neutrinos are Dirac fermions, no information on neutrino masses can be extracted from searches for lepton-number violation.  Conversely, an observation of neutrinoless double beta decay is unambiguous evidence of lepton-number violation, independent of the uncertainties that affect a mass determination therefrom.

\subsection{Cosmology}
\label{sec:cosmology}

In the Standard Model of cosmology, neutrinos are predicted to be relics of the big bang. Measurements of the relic abundance of light elements \cite{Steigman:2012ve,Pitrou:2018cgg} and the large-scale structure of the universe, including precision measurements of the properties of the cosmic microwave background (CMB) \cite{Aghanim:2018eyx}, are consistent with the existence of a thermal relic-neutrino background. These neutrinos played a significant role in the expansion history of the universe even if, today, they are rather cold and make up only a tiny fraction of the universe's matter and energy budget.

The temperature of the relic neutrino background today is predicted to be of order $T^0_{\nu}\sim 2\times 10^{-4}$~eV. Hence, as the universe expanded and the relic neutrino background cooled, the behavior of neutrinos changed from that of ultrarelativistic relics -- ``radiation'' -- to that of  non-relativistic species -- ``matter'' -- as long as the neutrino masses are larger than $T^0_{\nu}$. Given information from neutrino oscillations, at least two of the three neutrino masses are known to be much larger than $T^0_{\nu}$. This transition leaves an imprint in the large-scale structure of the universe in such a way that precision measurements provide nontrivial information on the neutrino masses.   

In the absence of other light particles or new neutrino interactions, the relic neutrino background is best described as a homogeneous mixture of the neutrino mass eigenstates $\nu_1$, $\nu_2$, and $\nu_3$ and their antiparticles.\begin{NoHyper}\footnote{If neutrinos are Majorana fermions, the situation is very similar, with left-helicity neutrino states playing the role of particles and right-helicity neutrino states playing the role of antiparticles.}$^,$\footnote{There is the possibility that the asymmetry between neutrinos and antineutrinos is relatively large. A flavor-universal asymmetry is constrained to be significantly less than one percent \cite{Pitrou:2018cgg} and would not impact the discussion here. Flavor-dependent effects are subtle and could impact the picture more significantly. These are still the subject of intense exploration \cite{Dolgov:2002ab,Wong:2002fa,Abazajian:2002qx,Hansen:2020vgm}.}\end{NoHyper} Theoretically, cosmic surveys are sensitive to the values of the individual neutrino masses, $m_1, m_2, m_3$. In practice, given the expected sensitivity of next-generation experiments, data from cosmic surveys constrain the sum of the neutrino masses, assuming all neutrino masses are light (say, all masses below a few eV), labelled here for convenience as 
\begin{equation}
    \Sigma \equiv \sum_i m_i.
\end{equation}
$\Sigma$ can be expressed in terms of the known mass-squared differences and the, currently unknown, lightest neutrino mass, $m_{\rm least}$: 
\begin{eqnarray}
   & \Sigma = m_{\rm least}+\sqrt{\Delta m^2_{21}+m_{\rm least}^2}+\sqrt{\Delta m^2_{31}+m_{\rm least}^2}~~~({\rm NMO}), \\ & {\rm or} \nonumber \\
    & \Sigma = m_{\rm least}+\sqrt{-\Delta m^2_{32}+m_{\rm least}^2}+\sqrt{-\Delta m^2_{31}+m_{\rm least}^2}~~~({\rm IMO}).
\end{eqnarray}
If the neutrino mass ordering is normal (inverted), current oscillation data constrain $\Sigma>5.87\times 10^{-2}$~eV ($\Sigma>9.92\times 10^{-2}$~eV). The direct laboratory measurement by KATRIN \cite{Aker:2019uuj} constrains $\Sigma<3.3$~eV in either mass ordering.

Data from large scale structure, including the CMB, the distribution of clusters of galaxies, and the Lyman-alpha forest, most recently constrain $\Sigma<0.111$~eV (95\% CL), according to \cite{Alam:2020sor}. Other recent analyses of cosmic surveys output similar upper bounds including $\Sigma<0.12$~eV (95\% CL) \cite{Aghanim:2018eyx} and $\Sigma<0.16$~eV (95\% CL) \cite{Ivanov:2019hqk}. Bounds obtained before the Planck-2018 data became available were only slightly weaker, $\Sigma<0.19$~eV (95\% CL) \cite{Giusarma:2018jei}. These bounds do depend on the values of the individual neutrino masses and the neutrino mass ordering, but not very strongly (see recent discussion in \cite{Ivanov:2019hqk,RoyChoudhury:2019hls}). Estimates in \cite{RoyChoudhury:2019hls} output, for the NMO,  $\Sigma<0.15$~eV (95\% CL), and $\Sigma<0.17$~eV (95\% CL) for the IMO. 
These translate, roughly, into $m_{\rm least}<0.05$~eV, mostly independent from the mass ordering. 

In the next decade, it is widely anticipated that next-generation experiments, including CMB-S4 \cite{Abazajian:2019tiv}, will be sensitive to $\Sigma>2\times 10^{-2}$~eV. If expectations are realized, cosmic surveys should be able to determine that $\Sigma$ is nonzero at better than the three-sigma level \cite{Abazajian:2019tiv}, independent of the mass ordering, assuming no new degrees of freedom or interactions beyond those in the Standard Model. 

The extraction of the sum of light neutrino masses from cosmic surveys is model dependent. Cosmic surveys are sensitive, in an over-simplified way, to the expansion rate of the universe as a function of time and to the formation of large-scale structure as a function of time. While the evidence that there are relic neutrinos is very compelling (see, for example, \cite{Steigman:2012ve,Pitrou:2018cgg}), the presence of neutrinos is only indirectly inferred and so are statements about their properties. New neutrino properties can impact the sensitivity to $\Sigma$ significantly. The authors of Ref.~\cite{Escudero:2020ped}, for example, argued very recently that if neutrinos are unstable but still very long lived (lifetime between $10^{-8}$ and $10^{-1}$ times the age of the universe), bounds on $\Sigma$ can be relaxed by an order of magnitude. The same can be said for more new ingredients to the Standard Model of cosmology. For example, the nature of the dark energy -- often parameterized by the dark-energy equation-of-state parameter $w$ -- impacts the sensitivity to $\Sigma$ \cite{Hannestad:2005gj}. Allowing for different palatable ingredients in the Standard Model of cosmology loosens the upper bound on $\Sigma$ by about a factor of three \cite{Alam:2020sor,RoyChoudhury:2019hls} or more (this is, of course, not guaranteed. For a counter example, see, for example, \cite{Vagnozzi:2018jhn}). The current tension between early-universe and late-universe estimates of the the Hubble parameter \cite{Aghanim:2018eyx,Riess:2019cxk} has invited speculation concerning new neutrino properties and interactions (see, for example, \cite{Kreisch:2019yzn}); some of these may have a significant impact on extracting constraints on $\Sigma$.  See, for example, \cite{Sekiguchi:2020igz} for a very recent discussion.

Massive neutrinos are a required ingredient of cosmological models but since the masses are presently unknown they must be treated as fit parameters.  There are few cosmological parameters susceptible to laboratory measurement, and neutrino mass is one.  A measurement would alleviate the models of a degree of freedom and allow better determinations of those parameters that can only be determined from cosmology, such as the equation of state of dark energy and the Hubble constant \cite{DiValentinoCosmo_PhysRevD.93.083527,DIVALENTINO2016242}.

\subsection{Neutrinos from Astrophysical Sources -- Time of Flight}

Throughout the universe, neutrinos are produced in cataclysmic astronomical events, including Type II Supernova explosions. Since these are short-duration bursts, it is possible to obtain information on the neutrino velocity and hence -- since the neutrino energy can be measured -- the neutrino mass. The time-spread of the neutrinos observed from SN1987A allows one to constrain the neutrino mass to be less than a few eV. A very detailed analysis was performed in Ref.~\cite{Loredo:2001rx} -- which includes references to several other estimates -- and the authors constrained what they refer to as the ``electron neutrino mass'' to be less than 5.7~eV at the 95\% confidence level. Strictly speaking, given what is known about neutrino mixing, the analysis is more involved and should include the fact that there are three mass eigenstates with different probabilities for interacting via charged-current interactions with electrons. In practice, given what is known about the neutrino mass-squared differences, the 5.7~eV upper bound applies to all mass eigenvalues.

In a nutshell, the measurement works as follows. If a neutrino is produced at some $t_0=0$ with energy $E$ a distance $D$ away it will arrive at the detector at 
\begin{equation}
t(E,m) =  D\left[1+\frac{m^2}{2E^2}\right],
\end{equation}
assuming all three neutrino masses are degenerate and equal to $m$, and $m\ll E$. Relative to a massless particle, the time delay of a massive neutrino is 
\begin{equation}
\Delta t = D\left(\frac{m^2}{2E^2}\right) = 25.7~\rm ms \times \left(\frac{D}{50 ~\rm kpc}\right) \left(\frac{m}{\rm eV}\right)^2\left(\frac{\rm 10~MeV}{E}\right)^2.
\end{equation}
For supernova neutrinos, $t_0$ is not known but one can investigate whether neutrinos with different energies arrive at different times, leading to, for example, a larger-than-expected spread in the neutrino arrival times. 

The detection of neutrinos from the next galactic supernova will allow one to perform a similar measurement, perhaps with higher statistics and richer data given the existence of bigger and better detectors. The JUNO collaboration, for example, estimates that the JUNO experiment \cite{An:2015jdp}, currently under construction, is sensitive to $m>1$~eV if neutrinos from a supernova explosion at $D\simeq 20$~kpc were to be detected. The dependence on $D$ is rather mild and the sensitivity worsens as $D$ increases; the larger $\Delta t$ is compensated by the loss of statistics as $D$ increases. Similar sensitivity -- down to at most $m>0.5$~eV --  has been estimated for the future DUNE \cite{Rossi-Torres:2015rla,Abi:2020evt} and Hyper-Kamiokande experiments \cite{Abe:2018uyc}. Very recently, the authors of Ref.~\cite{Hansen:2019giq} explored in detail the sensitivity of different next-generation experiments and different sources. In particular, they discuss the possibility of comparing the arrival time of the neutrinos with the potential detection of gravitational waves from the same source and estimate sensitivity to neutrino masses of order 1~eV.

\subsection{Direct laboratory measurements}

In processes involving neutrinos where the total energy of the initial state is well known and the kinematics of the final state can be measured with precision, it is possible to constrain, using energy and momentum conservation, the neutrino mass. Such measurements are often referred to as direct measurements of the neutrino mass and are the main subject of this review. 

\begin{figure}[htb]
\begin{center}
\includegraphics[width=0.70\textwidth]{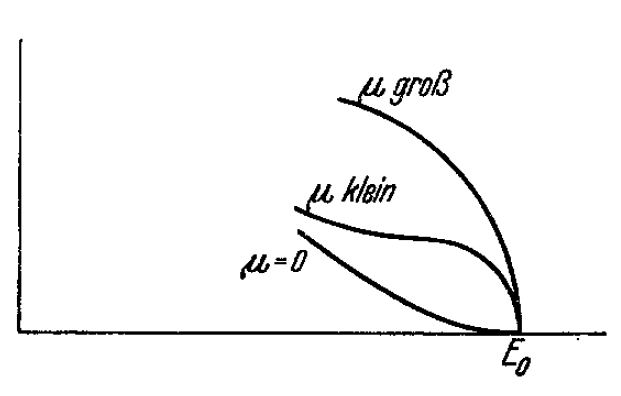}
\caption{Fermi's illustration of the impact of a nonzero neutrino mass $\mu$ on the shape of the $\beta$-ray energy spectrum, for ``large'' (gro\ss), ``small'' (klein) and $\mu=0$ \cite{fermi:1934}. Large and small are defined, loosely, relative to the mass of the electron.}
\label{fig:Fermi}
\end{center}
\end{figure}

The first direct laboratory probe of neutrino mass was suggested by Perrin in 1933 \cite{Perrin1933}: ``On peut essayer de d\'eduire de la forme des spectres continus d'\'emission une indication sur la valeur de cette masse inconnue...'' (One could attempt to deduce from the shape of the continuous emission spectra an indication of the value  of this unknown mass...).  Fermi independently reached that conclusion quantitatively in his seminal 1934 article \cite{fermi:1934},\begin{NoHyper}\footnote{See \cite{Wilson:1968pwx} for a translation of Fermi's paper to English. Fermi published preliminary work on the theory of $\beta$-decay several months before Ref.~\cite{fermi:1934} -- in  ``La Ricerca Scientifica'' \cite{Fermi:1933jpa} and ``Il Nuovo Cimento'' \cite{Fermi:1934sk} -- and submitted his theory for publication in Nature. The Nature submission was rejected, famously, because `it contained abstract speculations too remote from physical reality to be of interest to the reader' \cite{segre}. For more details on the history of Fermi's contribution to the theory of $\beta$-decay, see \cite{guerra}. We are indebted to David Kaiser for providing us most of this information.}\end{NoHyper} which introduces a so-called four-fermion interaction to describe nuclear beta-decay -- the Fermi interaction. Fermi suggested that the energy spectrum of the $\beta$-rays can be used to determine the mass of the neutrino: ``[t]he shape of the continuous $\beta$-spectrum is determined from the transition probability [computed perturbatively using the Fermi-interaction Hamiltonian]. We want to discuss first how this shape depends on the rest mass of the neutrino $\mu$, in order to determine this constant by comparison with empirical curves'' \cite{Wilson:1968pwx}. The effect of a nonzero neutrino mass is illustrated in Fig.~\ref{fig:Fermi}, from \cite{fermi:1934}. Fermi  concluded that ``[t]he greatest similarity to the empirical curves is given by the theoretical curve for $\mu=0$. \ldots Hence we conclude that the rest mass of the neutrino is either zero or, in any case, very small in comparison to the mass of the electron'' \cite{Wilson:1968pwx}.

If neutrinos are produced or absorbed via charged-current interactions associated with the charged-lepton $\ell_{\alpha}=e,\mu,\tau$ the differential rate associated with the process, which is a function of the neutrino masses-squared, can be written as 
\begin{equation}
    \sum_i|U_{\alpha i}|^2\Gamma_{\alpha}(m_i^2).
    \label{eq:general}
\end{equation}
Here $\Gamma_{\alpha}(m_i^2)$ is the differential rate of the process of interest when a neutrino with mass $m_i$ is emitted or absorbed.  
For the observables under consideration here, neutrino production is best described as incoherent, under the assumption that the neutrino masses are within the sensitivity of the experimental setup in question. As long as the final-state neutrino is not measured, however, its coherence, or lack thereof, is immaterial to our discussion. 

Eq.~(\ref{eq:general}) reveals that one is sensitive to the individual neutrino masses $m_1,m_2,m_3$ as long as all $|U_{\alpha i}|\neq 0$, which turns out to be the case. In practice, one needs to account for intrinsic and experimental uncertainties associated with the initial state and the finite resolution of the various measuring apparatuses. Taking these uncertainties into account, in the limit where the neutrino masses are small enough compared to the various energy scales of the system, one can express\begin{NoHyper}\footnote{This discussion is meant to be generic and purely for illustrative and pedagogical purposes. We return to the specific cases of the charged-lepton energy spectrum of electron-mediated charged-current processes, including nuclear beta-decay, in more detail in Sec.~\ref{sec:spectrum}.}\end{NoHyper}
\begin{equation}
    \Gamma_{\alpha}(m_i^2) =\Gamma_{\alpha}(0)+m_i^2\frac{{\rm d} \Gamma_{\alpha}}{{\rm d}m_i^2}(0)+{\cal O}(m_i^4).
\end{equation}
Here, $\Gamma_{\alpha}(m_i^2)$ stands for  the differential rate convoluted with the uncertainties associated to the measurement in question (slightly different from the object in Eq.~(\ref{eq:general})). This object is smooth around $m_i^2=0$ in such a way that the series expansion above is meaningful.

To leading order in the neutrino masses, taking uncertainties into account, 
\begin{equation}
     \sum_i|U_{\alpha i}|^2\Gamma_{\alpha}(m_i^2) \simeq \left[\Gamma_{\alpha}(0)+m_{\nu_\alpha}^2\frac{{\rm d} \Gamma_{\alpha}}{{\rm d}m_i^2}(0)\right],
\end{equation}
where 
\begin{equation}
    m_{\nu_\alpha}^2 \equiv \sum_i|U_{\alpha i}|^2m_i^2,
    \label{eq:mnue}
\end{equation}
is an effective neutrino mass-squared associated with the charged-current processes involving the charged-lepton $\ell_{\alpha}$. In the limit where all neutrino masses are very small, all such kinematical searches translate into bounds on different $m^2_{\nu_\alpha}$.\begin{NoHyper}\footnote{Different constraints can also be obtained, at least in theory, from neutral current processes. In practice, there are no low-energy, high-statistics neutral-current processes one can use to extract meaningful information. These include, for example, $\nu_{\mu}e\to\nu_{\mu} e$ scattering and the very rare $K\to \pi \nu\bar{\nu}$ decay.}\end{NoHyper}

It is often the case that experiments will quote upper bounds for the square-root of $m^2_{\nu_\alpha}$, defined to be $m_{\nu_\alpha}\equiv\sqrt{m^2_{\nu_\alpha}}$. In turn, $m_{\nu_e},m_{\nu_\mu},m_{\nu_\tau}$ are sometimes referred to in the literature as the electron-neutrino, muon-neutrino, and tau-neutrino masses. This is a practice we would like to strongly discourage since the electron neutrino, muon neutrino, and tau neutrino are not particles in the technical sense of the word and do not have well defined masses. Instead, they are interaction eigenstates and linear superpositions of the neutrino mass eigenstates. The latter are propagating particles in the strict sense of the word.  

The strongest bound on $m_{\nu_{\tau}}^2$ comes from precision measurements of $\tau$-decays into multi-pion final states. The strongest such bound was reported by the ALEPH collaboration \cite{Barate:1997zg}: $ m_{\nu_\tau}<18.2$~MeV at the 95\% confidence level. The result is obtained by combining precision measurements of $\tau^-\to\nu_{\tau}\pi^{-}\pi^{+}\pi^-$ -- around 3,000 events -- and $\tau^-\to\nu_{\tau}2\pi^{-}2\pi^{+}\pi^-(\pi^0)$ -- around 60 events. It has been estimated that an order-of-magnitude improvement is possible if one were to take advantage of the $\tau$-samples recorded by the B-factories \cite{Kobach:2014hea} (almost $10^7$ $\tau^-\to\nu_{\tau}\pi^{\pm}\pi^-$).

The strongest bound on $m_{\nu_{\mu}}^2$ comes from precision measurements of pion decay at rest, $\pi^+\to\mu^+\nu_{\mu}$. The authors of Ref.~\cite{Assamagan:1995wb}, analysing the decay at rest of positively charged pions at PSI, extracted the upper bound $ m_{\nu_\mu}<0.17$~MeV at the 90\% confidence level, along with the measurement $m_{\nu_\mu}^2=(-0.016\pm0.023)$~MeV$^2$.  From this experiment, taking the new knowledge of neutrino-flavor oscillations into account, the currently most precise value for the charged pion mass is deduced \cite{DAUM201911}.

Given what is known about neutrino masses from neutrino oscillations and constraints on $m_{\nu_e}^2$, to be discussed momentarily, the constraints on $m_{\nu_{\mu}}^2$ and $m_{\nu_{\tau}}^2$ discussed above are not especially relevant when it comes to informing the values of the light neutrino masses. Indeed, there are no processes involving muons or tau leptons -- today or in the foreseeable future -- capable of competing with current and future information from electron-mediated charged-current processes.   

Before proceeding, we highlight that a variety of alternatives to the notation $m_{\nu_e}$ are found in the literature, including $m_\beta$, $m_\nu$, and $m_0$. Henceforth, we will make use of $m_{\beta}\equiv\sqrt{\sum_i|U_{ei}|^2m_i^2}$, for a few reasons. As already discussed, $m_{\nu_e}$ should be deprecated because the electron neutrino is not a particle and does not have a mass. The term $m_{\nu}$ is better but is used often in many different contexts. The term $m_0$ is sometimes defined as the mass of the lightest eigenstate ($m_1$ or $m_3$ depending on the ordering) and we want to avoid confusing the two different objects. The choice $m_\beta$ is also the one made by the Particle Data Group \cite{Zyla:2020zbs}. The quantity $m_\beta$ is a particular combination of the masses of real (propagating) neutrinos, as distinct from the virtual or effective mass $m_{\beta\beta}$ in $0\nu\beta\beta$, introduced in Sec.\ref{sec:0nubb}, that does not correspond to propagating neutrinos. As will be shown, $m_\beta\simeq m_1$.  A kinematic measurement of $m_\beta$ simultaneously determines all 3 eigenmasses, up to a binary uncertainty in the mass ordering.

Assuming the neutrino mixing matrix is unitary, 
\begin{equation}
m_{\beta}^2= m_1^2 + |U_{e2}|^2\Delta m_{21}^2 + |U_{e3}|^2 \Delta m^2_{31},
\label{eq:mb_m1}
\end{equation}
for either mass ordering, keeping in mind that $\Delta m^2_{31}$ is positive for the NMO and negative for the IMO. Since, it turns out, $|U_{e3}|^2$ and $\Delta m^2_{21}$ are both quite small the approximation $m_{\beta}\simeq m_1$ works well unless $m_1$ is very small. Quantitatively it holds at the percent level or better for $m_1$ values down to 0.05 eV, and never differs by more than 8~meV even in the limit $m_1=0$.\begin{NoHyper}\footnote{The second and third terms in Eq.~(\ref{eq:mb_m1}) are $|U_{e2}|^2\Delta m_{21}^2\sim 3\times 10^{-5}$~eV$^2$, and $|U_{e3}|^2\Delta m_{31}^2\sim \pm 6\times 10^{-5}$~eV$^2$.}\end{NoHyper} In brief, one can say that, to a good approximation, beta decay and electron capture measure the mass $m_1$, independent from the mass ordering.

The strongest bound on $m_{\beta}$ comes from precision measurements of tritium beta decay. This is the subject of the bulk of this review. The KATRIN experiment, after collecting data for four weeks, established the strongest bound to date,  $m_{\beta}<1.1$~eV at the 90\% confidence level \cite{Aker:2019uuj}, associated with the measurement $m_{\beta}^2=-1.0^{+0.9}_{-1.1}$~eV$^2$, consistent with zero. The ultimate sensitivity of KATRIN is to $m_{\beta}>0.2$~eV (90\% confidence level). KATRIN is discussed in more detail in Sec.~\ref{sec:KATRIN}.

Direct searches for kinematic effects of nonzero neutrino masses are, for the most part, model independent and rely only on the conservation of energy and momentum to measure the neutrino mass. They do not depend, for example, significantly on whether neutrino scattering is exactly described by the Standard Model of particle physics. For a careful, recent exploration of new-interaction effects on the extraction of $m_{\beta}$, see \cite{Ludl:2016ane}. To leading order, and for all practical purposes, direct searches for kinematical effects of nonzero neutrino masses also do not depend on the nature of the neutrinos -- Majorana or Dirac fermions. The fact that there is a charged lepton with a well defined charge in the final or initial state renders the leading order amplitudes identical for Majorana and Dirac neutrinos. The same is true of all relevant QED corrections. At higher order in the weak interactions, however, there are unobservably small differences between Majorana and Dirac neutrinos. These differences are not only suppressed by the Fermi constant to some power but are also proportional to the neutrino masses. A concrete example is the electron spectrum associated with the five-body final-state neutron decay, $n\to p e^- {\bar\nu}_i\nu_j\bar{\nu}_k$ in the Dirac case, $n\to p e^-\nu_i\nu_j\nu_k$ in the Majorana case, where $i,j,k=1,2,3$, the different mass eigenstates. In the Majorana case, for example, there are interference effects between $\nu_i$ and $\nu_j$ when $i=j$, if $m_i$ is not zero. These are clearly not present in the Dirac case. For illustrations of this phenomenon, see \cite{Berryman:2018qxn} for a recent discussion of very low-energy $e\gamma$ scattering into neutrinos or \cite{Millar:2018hkv} for a discussion of the end point of the bremsstrahlung spectrum of coherent neutrino scattering on nuclei ($\nu +N \to\nu +N+\gamma$). 

The presence of new ``neutrino'' states, however, modifies the interpretation of results from these types of experiments. Indeed, the existence of new, relatively heavy, ``neutrino'' states is strongly constrained by precision measurements of $\beta$-decay, meson-decay, tau-decay, etc. For recent reviews see, for example, \cite{deGouvea:2015euy,Drewes:2015iva,Fernandez-Martinez:2016lgt,Drewes:2016jae,Bryman:2019ssi,Bryman:2019bjg,Bolton:2019pcu}. We return to this issue later in this subsection and in Sec.~\ref{sec:other}.

\begin{figure}[htb]
\begin{center}
\includegraphics[width=0.70\textwidth]{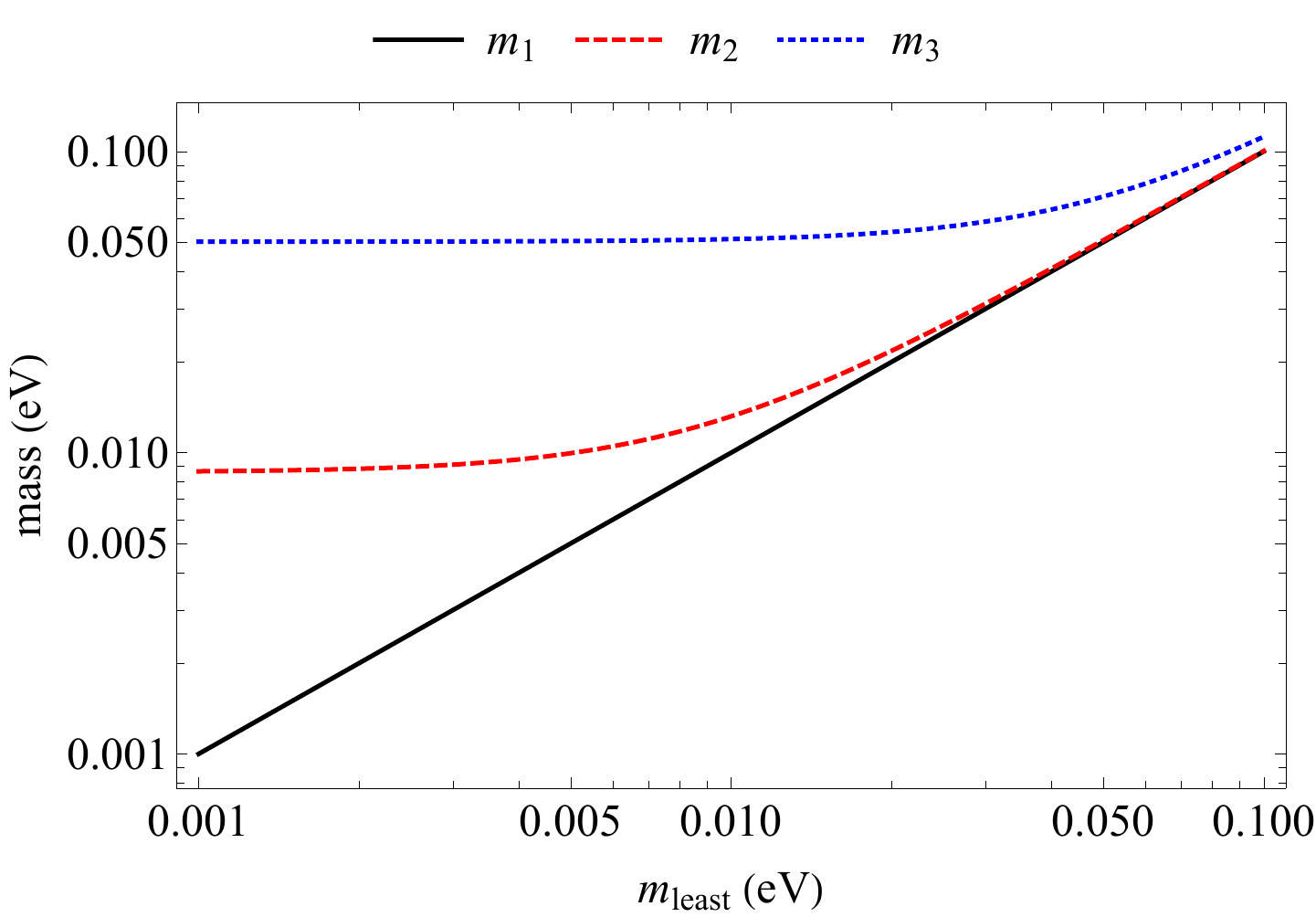}
\includegraphics[width=0.70\textwidth]{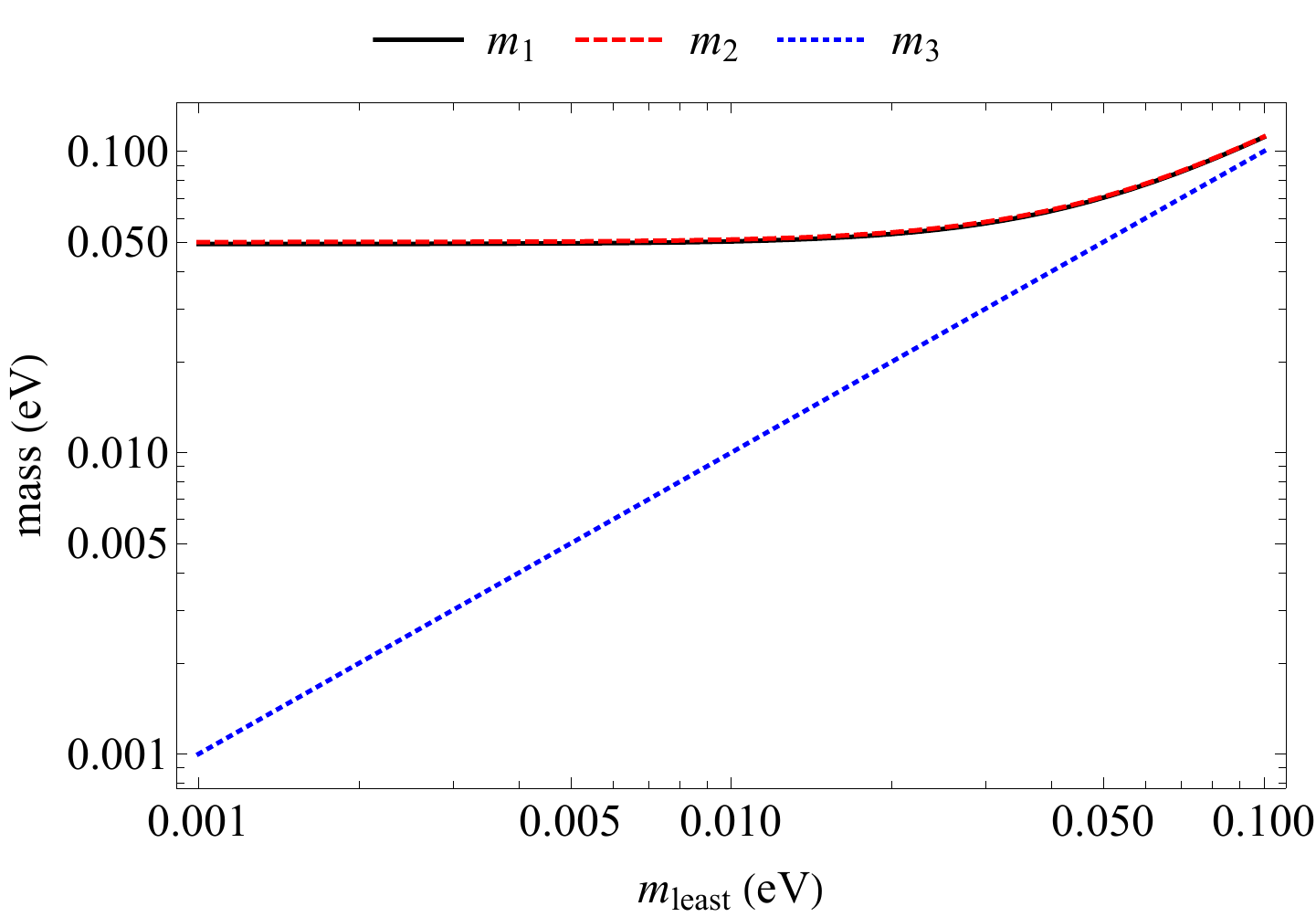}
\caption{Current best-fit values of the neutrino masses $m_1,m_2,m_3$ as a function of the lightest neutrino mass, for the normal mass-ordering (top) and the inverted mass ordering (bottom). 
}
\label{fig:masses}
\end{center}
\end{figure}

In the absence of new, light degrees of freedom, the KATRIN result can be used to set a robust upper bound on the neutrino masses. The KATRIN bound, combined with results from the current oscillation data, translates into (two significant digits)
\begin{eqnarray}
0<&~m_1^2~&<1.2~{\rm eV}^2, \\
7.4\times 10^{-5}~{\rm eV}^2<&~m_2^2~&<1.2~{\rm eV}^2,~~~~~~~(\rm NMO) \\
2.5\times 10^{-3}~{\rm eV^2}<&~m_3^2~&<1.2~{\rm eV}^2, \\
&\rm or & \nonumber \\
2.4\times 10^{-3}~{\rm eV}^2<&~m_1^2~&<1.2~{\rm eV}^2, \\
2.5\times 10^{-3}~{\rm eV}^2<&~m_2^2~&<1.2~{\rm eV}^2,~~~~~~~(\rm IMO) \\
0<&~m_3^2~&<1.2~{\rm eV}^2,
\end{eqnarray}
where the bounds are heavily correlated given the constraints on the mass-squared differences. These are, arguably, the most robust, model-independent upper bounds on all three neutrino masses. 
Figure~\ref{fig:masses} depicts the values of the three light neutrino masses as a function of $m_{\rm least}$, for both neutrino-mass orderings. 

Above, we highlighted the fact that the relation between precision measurements of the $\beta$-decay spectrum and the neutrino masses, given what we know from oscillation experiments, is very robust. The main exception to this robustness is the presence of new neutrino mass eigenstates. In a nutshell, these manifest themselves in two different ways. If the new neutrino masses -- referred to, here, as $m_4$ -- are ``large,'' the presence of the extra heavy neutrino will distort the $\beta$-decay spectrum. If, instead, the new neutrino masses are ``small,'' the presence of the extra neutrino will simply add to $m_{\beta}^2$, i.e., the sum in Eq.~(\ref{eq:mnue}) would encompass all mass eigenvalues $m_i$, $i=1,2,3,\ldots$ for all $m_i$ less than a few eV. For more details see, for example, \cite{deGouvea:2006gz,Formaggio:2011jg,Esmaili:2012vg}. The KATRIN collaboration recently made available the results of a search for new neutrino states \cite{aker2020sterilebound}; we discuss it in detail later in this review, see Sec.~\ref{sec:sterile}. As an aside, new neutrino mass states that admix with electron-flavor neutrinos only render the electron-weighted mass-squared parameter  $m^2_{\beta}$ larger. Hence upper bounds to $m^2_{\beta}$ are especially robust and cannot be bypassed by postulating the existence of new particles.

\section{Neutrino Mass Models and New Physics}
\label{sec:models}
\setcounter{equation}{0}

In this section, we briefly discuss how direct searches for neutrino masses inform our understanding of the origin of neutrino mass and can be used to discover other new physics. 

\subsection{Neutrino mass models}

Similar to all fermions in the Standard Model of particle physics, the known neutrinos can  acquire nonzero masses only after electroweak symmetry breaking. Unlike charged fermions, the dynamical mechanism behind nonzero neutrino masses is unknown. Identifying the physics responsible for neutrino masses is among the most important questions in particle physics today.

There are several qualitatively different models capable of explaining why neutrinos have mass. While all of them require the existence of new degrees of freedom, the nature of the new degrees of freedom -- one or several new states, fermions or bosons, light or heavy new states, etc -- varies dramatically. Given the high degree of uncertainty, information on the origin of neutrino masses may come from a large range of experimental efforts,  from searches for rare muon processes (e.g. $\mu\to ee^+e^-$ decays) to the Large Hadron Collider to next-generation neutrino-oscillation experiments. Direct measurements of the neutrino mass, along with pursuits of lepton-number violation, are guaranteed to provide nontrivial information.

Parallel to the origin of neutrino mass, there is the issue of the pattern of lepton mixing. Unlike quarks, the mixing angles that parameterize the PMNS matrix are all large -- the smallest lepton mixing angle is almost as large as the largest quark mixing angle -- and the potential organizing principles responsible for its observed features may be qualitatively different.  Several of the theoretical approaches to the problem of lepton flavor also make predictions for the values of the neutrino masses, which will be informed most straightforwardly by direct searches for the kinematical effects of masses. 

Direct measurements of the neutrino masses can help reveal if the lightest neutrino is massless. Knowledge of the masslessness of the lightest neutrino would impact, very significantly, our understanding of the origin of neutrino masses. For example, if neutrinos are Dirac fermions, $m_{\rm least}=0$ allows one to contemplate that there are only two right-handed neutrino fields, in stark contrast to all other fermionic degrees of freedom in the Standard Model that come in three flavors. The same is true if the neutrinos are Majorana fermions and their masses are a consequence of the so-called  Type-I seesaw mechanism \cite{Minkowski:1977sc,GellMann:1980vs,Yanagida:1979as,Glashow:1979nm,Mohapatra:1979ia,Schechter:1980gr}. In this case, $m_{\rm least}=0$ translates into the possibility that there are only two right-handed neutrinos.\begin{NoHyper}\footnote{Strictly speaking, even if there are only two right-handed neutrinos, the lightest neutrino mass is expected to be nonzero, generated at the two-loop level even in the absence of new neutrino interactions. In this case, however, $m_{\rm least}$ is expected to be many orders of magnitude lighter than the other neutrino masses \cite{Davidson:2006tg}.}\end{NoHyper} In many other scenarios, there is no natural way to ``explain'' why the lightest neutrino should be massless or much lighter than the other two. 

Experimentally, of course, it is impossible to determine that $m_{\rm least}$ is exactly zero since $m_{\rm least}=0$ and $m_{\rm least}\ll \Delta m^2_{21}, |\Delta m^2_{31}|$ are, in practice, indistinguishable. Instead, one could determine that $m_{\rm least}\ne0$ with some confidence. More quantitatively, 
\begin{eqnarray}
    m_{\beta}^2 & = & m^2_{\rm least} + |U_{e2}|^2\Delta m^2_{21} + |U_{e3}|^2\Delta m^2_{31}, \\
   & = & m^2_{\rm least} + 7.74\times 10^{-5}~{\rm eV}^2, \hspace{1cm}\rm (NMO) \label{eq:nu_e,NMO} \\
   && \rm or \nonumber \\
    m_{\beta}^2 & = & m^2_{\rm least} + |U_{e1}|^2(-\Delta m^2_{32}-\Delta m^2_{21}) + |U_{e2}|^2(-\Delta m^2_{32}), \\
    & = & m^2_{\rm least} + 2.47\times 10^{-3}~{\rm eV}^2, \hspace{1.1cm} \rm \label{eq:nu_e,IMO} (IMO)
\end{eqnarray}
using, in accordance with Ref.~\cite{Esteban:2020cvm}, $|U_{e1}|^2=0.681$, $|U_{e2}|^2=0.297$ and $|U_{e3}|^2=0.022$. 
Hence, to establish that $m^2_{\rm least}\neq 0$, one needs to constrain, in a statistically significant way, $m_{\beta}^2>7.74\times 10^{-5}$~eV$^2$ ($m_{\beta}^2>2.47\times 10^{-3}$~eV$^2$)\begin{NoHyper}\footnote{The magnitudes of the mass-squared differences and $|U_{ei}|^2$, $i=1,2,3$ entries of the PMNS matrix are currently known at the 4\% level or better. Taking uncertainties into account, the upper bounds quoted here are known at the 5\% level (one sigma).}\end{NoHyper} assuming the neutrino mass ordering is known to be normal (inverted).

Another qualitatively different hypothesis that, if confirmed, would impact our understanding of the origin of neutrino masses is the possibility that all three neutrino masses are quasi-degenerate. Even if all current bounds on neutrino masses are taken at face value, this is, experimentally, still an option. For example, if $m_{\rm least}=0.1$~eV, only in slight tension with the most stringent indirect constraints from cosmic surveys \cite{Aghanim:2018eyx}, 
\begin{equation}
m_{\rm most}-m_{\rm least}\sim 0.12{\frac{(m_{\rm most}+m_{\rm least})}{2}} ,
\end{equation}
where $m_{\rm most}$ is the heaviest neutrino mass, for both mass orderings. Current constraints allow neutrino masses that are almost degenerate -- all of the same order of magnitude -- especially if one considers that the cosmology bounds can be significantly alleviated with the introduction of new ingredients. Experiments sensitive down to $m_{\beta}^2\sim0.01$~eV$^2$ can definitively test the hypothesis that the neutrino masses are almost degenerate.

\subsection{Sensitivity to new phenomena}

As discussed in Sec.~\ref{sec:status}, cosmic surveys, searches for $0\nu\beta\beta$, and direct kinematic measurements of neutrino masses are all sensitive to the values of the neutrino masses. The first two probes are indirect. They rely on other ingredients that govern the expansion history of the universe, on the absence of new interactions involving neutrinos, on the Majorana nature of neutrinos, on the absence of more, directly accessible lepton-number violating interactions, etc. This means that by combining these different probes of the values of the neutrino masses we can verify whether the assumptions that go into relating cosmic surveys and the rate for $0\nu\beta\beta$ to the values of the neutrino masses are valid. 

\begin{figure}[htb]
\begin{center}
\includegraphics[width=0.75\textwidth]{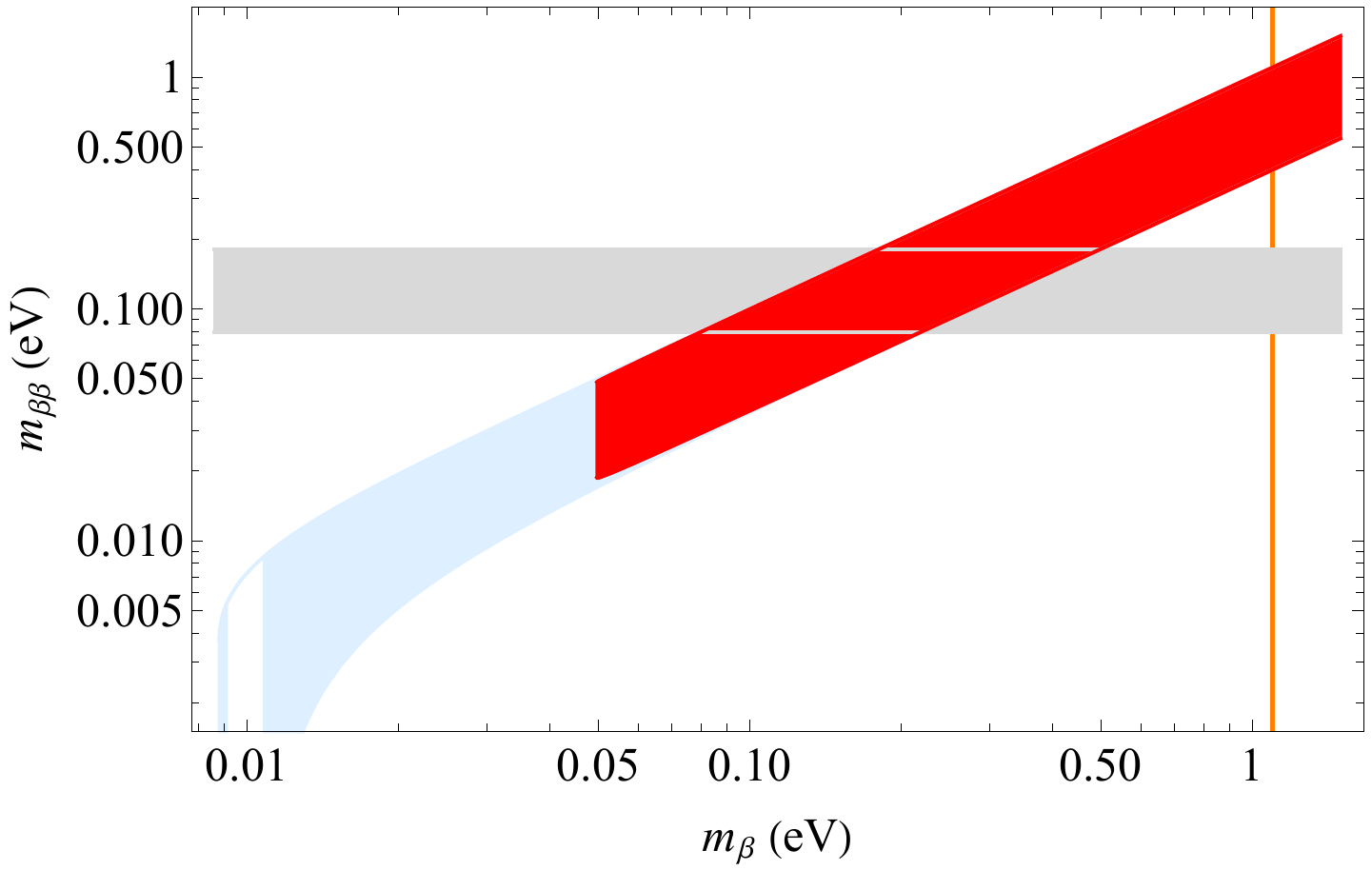}
\caption{$m_{\beta\beta}$ as a function of $m_{\beta}$, for both the normal (lighter, blue) and inverted (darker, red) mass orderings. The bands are a consequence of allowing for all possible values of the relative Majorana phases. For everything else, we use the current best-fit values of the oscillation parameters from \cite{Esteban:2020cvm}. The whited-out region inside the light-blue contour is meant to highlight the values of $m_{\beta}$ for which $m_{\beta\beta}$ can vanish exactly. We assume the neutrinos are Majorana fermions. If neutrinos are Dirac fermions, $m_{\beta\beta}=0$. The grey, horizontal band corresponds to the 95\% CL upper bound on $m_{\beta\beta}$ from GERDA \cite{Agostini:2020xta}. The width of the band is a consequence of uncertainties in the nuclear matrix element for the neutrinoless double-beta decay of $^{76}$Ge. The vertical line corresponds to the current 90\% upper bound on $m_{\beta}$ \cite{Aker:2019uuj}.
}
\label{fig:mass_corr1}
\end{center}
\end{figure}

\begin{figure}[htb]
\begin{center}
\includegraphics[width=0.75\textwidth]{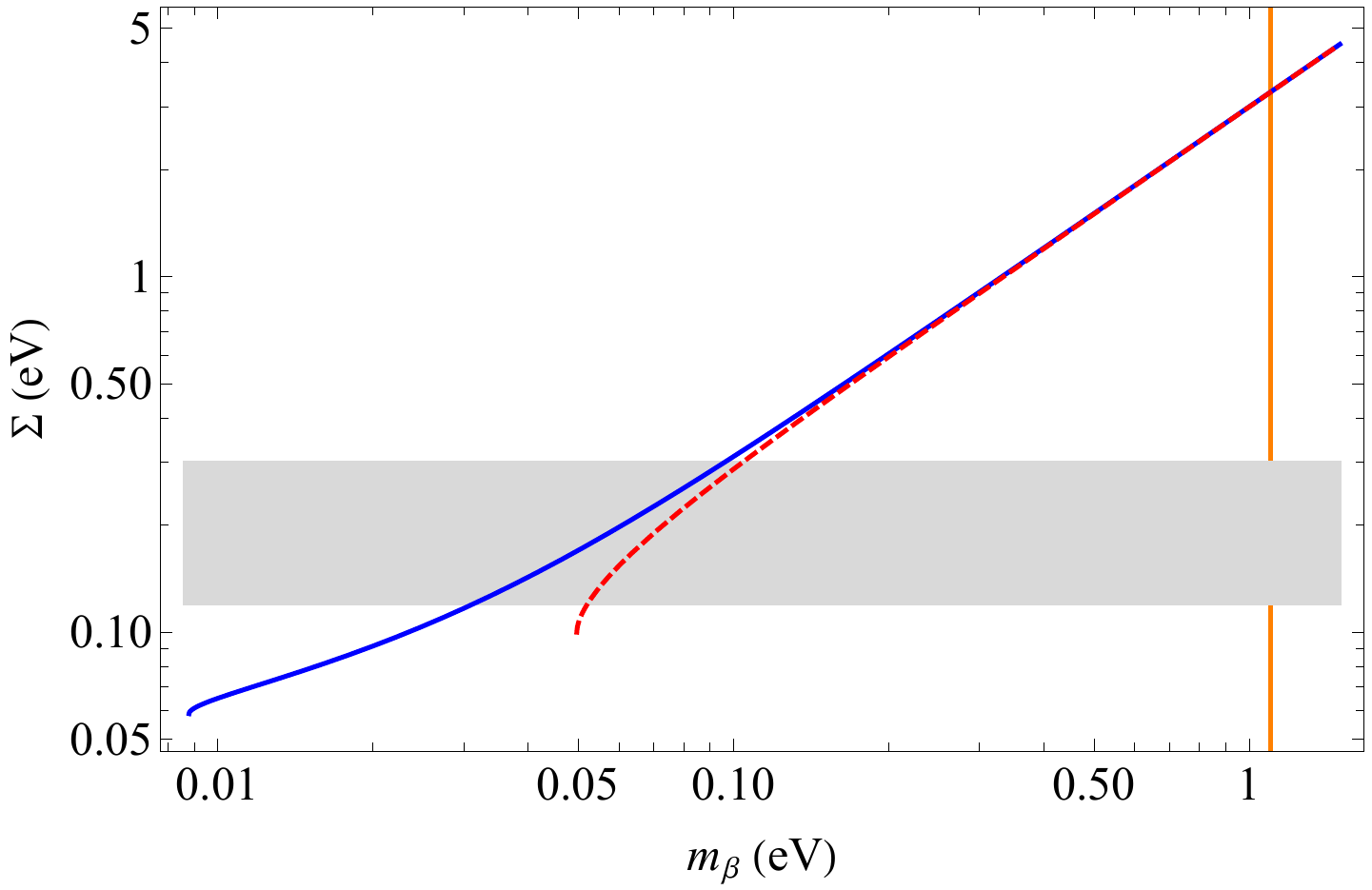}
\caption{$\Sigma$ as a function of $m_{\beta}$, for both the normal (blue,solid) and inverted (red,dashed) mass orderings. We use the current best-fit values of the oscillation parameters from \cite{Esteban:2020cvm}.
The, horizontal band corresponds to the range of
95\% CL upper bounds on $\Sigma$ discussed in
\cite{RoyChoudhury:2019hls}. Different upper bounds correspond to different ingredients added to the Standard Model of cosmology. The vertical line corresponds to the current 90\% upper bound on $m_{\beta}$ \cite{Aker:2019uuj}.
}
\label{fig:mass_corr2}
\end{center}
\end{figure}

In the absence of new particle physics and new cosmology-related ingredients, $m_{\beta}$, $m_{\beta\beta}$, and $\Sigma$ are strictly correlated. In particular, assuming the neutrino oscillation parameters are known, it is trivial to express both $m_{\beta\beta}$ and $\Sigma$ as functions of $m_{\beta}$ -- see Eqs.~(\ref{eq:nu_e,NMO}, \ref{eq:nu_e,IMO}). These are depicted in Figs.~\ref{fig:mass_corr1} and \ref{fig:mass_corr2}, for both mass orderings. Here we assume neutrinos are Majorana fermions; for Dirac fermions, the rate for $0\nu\beta\beta$ is zero. In the case of $m_{\beta\beta}$, the bands are a consequence of all possible values of the relative Majorana phases, currently completely unconstrained. For everything else, we use the current best-fit values of the oscillation parameters from \cite{Esteban:2020cvm}. The relevant oscillation parameters -- $\sin^2\theta_{13}, \sin^2\theta_{12}, \Delta m^2_{21}, \Delta m^2_{31}$ -- are all known at better than the 4\% level.

Figs.~\ref{fig:mass_corr1} and \ref{fig:mass_corr2} allow one to identify circumstances that would imply the existence of new phenomena. For example, independent from the neutrino mass ordering, if precision measurements of tritium $\beta$-decay revealed that $m_{\beta}$ is larger than 0.05~eV, cosmic surveys imply the existence of new cosmology-related ingredients. This is a very robust statement. Even if, ultimately, we find that new neutrino mass eigenstates are the dominant contribution to $m_{\beta}$, their existence, given that their mixing with active neutrinos is rather large, is ruled out by cosmic surveys in the absence of new ingredients. See, for example, the bounds on the effective number of neutrinos, $N_{\rm eff}$, in, for example, \cite{Aghanim:2018eyx}. 

If, on the other hand, $m_{\beta}$ is constrained to be smaller than 0.1~eV and one finds $m_{\beta\beta}$ to be larger than 0.1~eV -- the sensitivity of current experiments approaches $m_{\beta\beta}\sim 0.1$~eV (Table \ref{tab:dbdexpts}) -- one will be required to conclude, again independent from the mass ordering, that there are contributions to $0\nu\beta\beta$ other than neutrino exchange.

\section{Kinematic determination from beta decay}
\label{sec:spectrum}
\setcounter{equation}{0}

In beta decay the energy available from the nuclear mass difference is carried away by the electron and the neutrino.  The two particles share the energy in a statistical way, determined quantum mechanically by the available phase space for each.  Because the electron cannot abscond with all the energy if the neutrino has rest mass, that small amount of energy alters the electron spectrum near its endpoint where it would otherwise have taken all the energy.  The beta spectrum in the presence of neutrino mass has a simple analytic form that reflects the available phase space. 

The relative influence of  neutrino mass on the spectrum compared to the available energy is maximized by choosing isotopes with the smallest Q-values.  As we discuss, however, a low Q-value alone does not guarantee a good basis for an experiment.

\subsection{Beta spectrum}

As discussed in Sec.~\ref{sec:status}, the fact that the three eigenmasses $m_1$, $m_2$, and $m_3$ are linked by neutrino-oscillation data has simplified the experimental task of determining the mass scale because it is now possible to work with beta decay alone; separate determinations involving $\mu$ and $\tau$ leptons are no longer needed.  The most sensitive direct searches for $m_{\beta}$ to date are based on the investigation of the electron spectrum
of tritium $\beta$-decay. The electron energy spectrum of $\beta$-decay for a neutrino with
component masses $m_1, m_2,$ and $m_3$   is the incoherent sum of the contributions from each mass eigenstate:
\begin{eqnarray}
\frac{d\Gamma}{dE} &=&  \frac{G_F^2|V_{ud}|^2}{2\pi^3}(G_V^2+3G_A^2)F(Z,\beta) \beta (E+m_e)^2(E_0-E) \nonumber \\&& \times \sum_{i=1,3}|U_{ei}|^2\left[(E_0-E)^2-m_i^2\right]^\frac{1}{2} \Theta (E_0-E-m_i), \label{eq:mother}
\end{eqnarray}
where $G_F$ is the Fermi coupling constant, $V_{ud}$ is an element of the CKM matrix \cite{Zyla:2020zbs}, $E$ ($\beta$) denotes the electron's kinetic energy (velocity), $E_0$, the `endpoint energy,' corresponds to the maximum kinetic energy in the absence of neutrino mass, $F(Z,\beta)$
is the Fermi function, taking into account the Coulomb interaction
of the outgoing electron in the final state, and $\Theta (E_0-E-m_i)$ is the step function that ensures energy conservation. The vector and axial-vector matrix elements are $G_V=1$ and $G_A=-1.2646(35)$ for tritium, respectively \cite{Akulov:2005umb}.  
Henceforth we gather the leading constants into a single one,
\begin{eqnarray}
C&=& \frac{G_F^2|V_{ud}|^2}{2\pi^3}(G_V^2+3G_A^2).
\end{eqnarray}
The relationships between $E_0$, the Q-value, and the atomic mass difference are detailed in \cite{Bodine:2015sma} and one is given in Eq.~(\ref{eq:qvalue}). As both the matrix elements and $F(Z,\beta)$ are independent
of $m_i$, the dependence of the spectral shape on $m_i$ is given by the phase space factor only.

\begin{figure}[htb]
    \centering
    \includegraphics[width=4in]{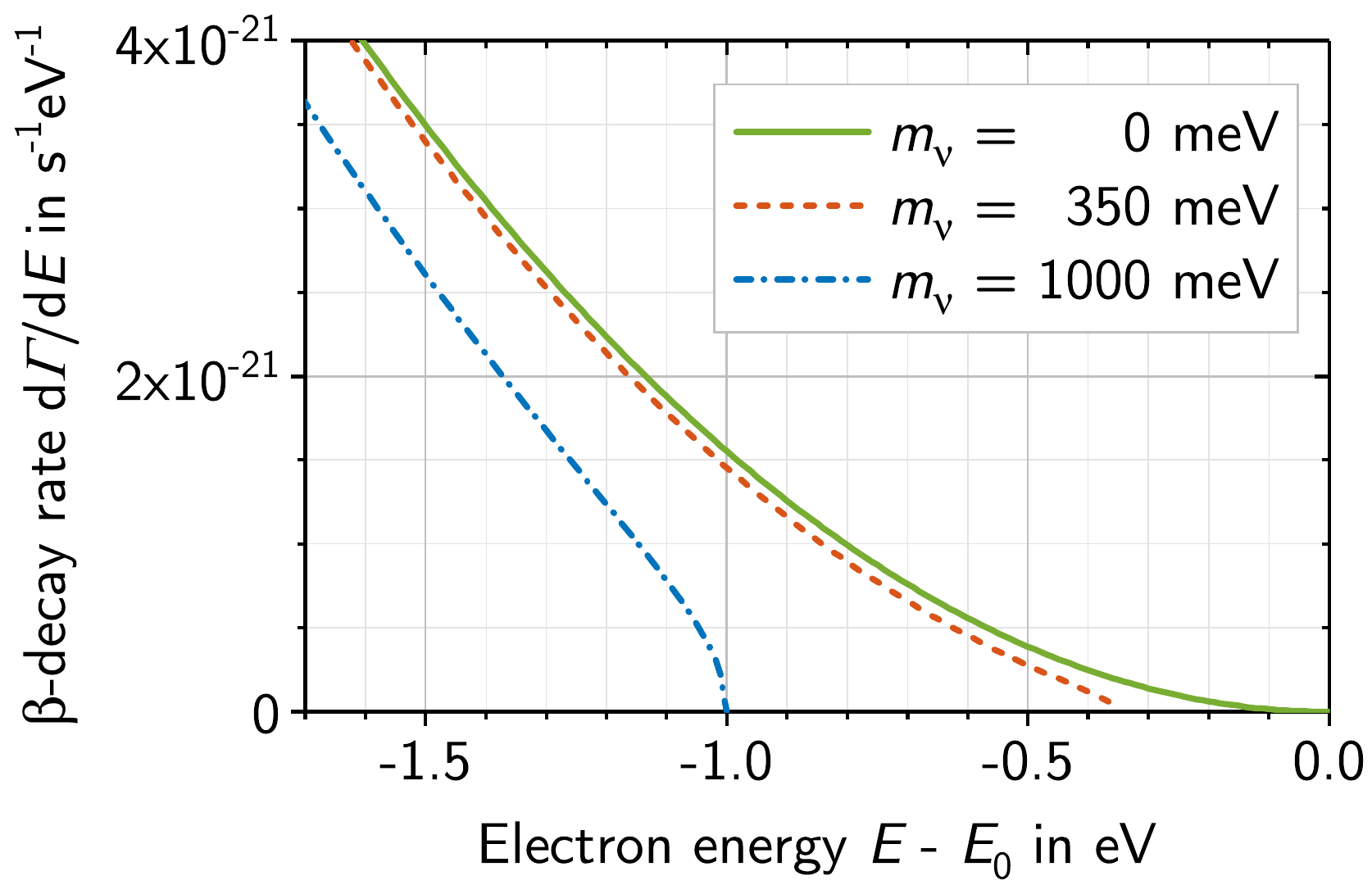}
    \caption{Beta spectrum of the decay of atomic tritium  near the endpoint, as given by Eq.~(\ref{eq:simplebetaspectrum}), from \cite{Kleesiek:2018mel}. In the figure, $m_\nu\equiv m_\beta$.}
    \label{fig:diffspecendpoint}
\end{figure}

The beta spectrum near the endpoint can be written in a simplified form for discussion,
\begin{eqnarray}
\frac{d\Gamma}{dE} &\approx& 3 r_0(E_0-E)\sum_{i=1,3}|U_{ei}|^2[(E_0-E)^2-m_i^2]^{1/2}\Theta (E_0-E-m_i), 
\label{eq:simplespectrum}
\end{eqnarray}
where $r_0$ is the detected event rate per atom in the last eV of the spectrum in the absence of mass (if $E$ and $E_0$ are also in eV).  The variables $\beta$ and $E+m_e$ and the function $F(Z,\beta)$ are evaluated at the endpoint and absorbed in the constant $r_0$. 
\begin{eqnarray}
r_0&=&\frac{\lambda}{C_tE_0^3}
\end{eqnarray}
where $\lambda=1.784\times 10^{-9}$ s$^{-1}$ is the tritium decay constant.   The constant $C_t$ is the ratio of the integral of the spectrum of the form of Eq.~(\ref{eq:mother}) to one of the form of Eq.~(\ref{eq:simplespectrum}), renormalized by any molecular or atomic branching to states populating the final eV of the spectrum. The value of $C_t\simeq1$ is more precisely 0.943 for molecular tritium, 0.767 for atomic tritium, and 0.537 for a bare tritium nucleus, when Simpson's form \cite{Simpson:PhysRevD.23.649} of the Fermi function is used:
\begin{eqnarray}
F(Z,\beta)&\simeq& 2\pi Z\alpha \frac{1.002037-0.001427\beta}{\beta(1-e^{-2\pi Z\alpha /\beta})},
\end{eqnarray}
where $\alpha$ is the fine-structure constant and $Z$ is the charge on the daughter nucleus.  For the bare nucleus, the extrapolated endpoint energy in the laboratory is 18522.44 eV \cite{Bodine:2015sma}, and the fractional intensity in the last eV of the spectrum is $2.93\times 10^{-13}$.  One can also write the constant $C$ in terms of the decay constant,
\begin{eqnarray}
C&=& \frac{3\lambda}{C_{t({\rm nuclear})}E_0^3F(Z,\beta_0)\beta_0(E_0+m_e)^2}
\end{eqnarray}
where $\beta_0$ is the value of $\beta$ at the endpoint.
If there are $N$ atoms in the source, the total rate in the last eV is $r=Nr_0$ and the total activity is $C_t rE_0^3$. Figure~\ref{fig:diffspecendpoint} shows the shape of the spectrum near the endpoint, where the effects of neutrino mass are most pronounced.
Neutrino mass experiments in beta decay are fundamentally just counting experiments.  The neutrino mass $m_\beta$ can in principle be determined or limited from a single measurement of the number of events in a suitably chosen interval $\Delta E$, the `analysis window', that ends at the extrapolated endpoint energy, as long as other parameters, namely the rate, time, endpoint energy, and background, are known well enough from other information.  This is an idealization but  not unrealistic for  experiments like Project 8 and calorimetric detectors (described below) where data both on the background above and the spectrum below the endpoint are automatically taken ``for free'' because all events are recorded as they occur.  The principle would also apply to an experiment like KATRIN \cite{Angrik:2005ep} that collects integral spectral data point--by--point, but with additional time spent to obtain the needed information. This `time expansion ratio' is discussed in Sec.~\ref{sec:estimation}.    The endpoint energy is not needed in an absolute sense; it need only be determined relative to  $\Delta E$ from the shape of the spectrum outside that window. 

The total number of signal events $N_s$ in time $t$ in this window is obtained by integrating Eq.~(\ref{eq:simplespectrum}),
\begin{eqnarray}
N_s &=& rt\sum_{i=1,3}|U_{ei}|^2[(\Delta E)^2-m_i^2]^{3/2} \\
&\simeq&rt (\Delta E)^3\left[1-\frac{3}{2}\frac{\sum_{i=1,3}|U_{ei}|^2m_i^2}{(\Delta E)^2}\right]. \label{eq:signalrate}
\end{eqnarray}
  In the last step we invoke the unitarity of the PMNS matrix and the assumption that $(\Delta E)^2 \gg \sum_{i=1,3}|U_{ei}|^2m_i^2 $ to allow a first-order expansion.  The latter assumption exploits the fact that neutrino mass experiments generally explore masses considerably smaller than the instrumental widths and backgrounds for which $\Delta E$ is optimized.  The summation in Eq.~(\ref{eq:signalrate}) was defined earlier as
\begin{eqnarray}
m_\beta^2&=&\sum_{i=1,3}|U_{ei}|^2m_i^2
\nonumber
\end{eqnarray}
that is used as a single parameter representing the result of a beta-decay measurement wherein the individual mass eigenstates are not resolved, and Eq.~(\ref{eq:signalrate}) motivates the replacement. With this replacement, the simplified beta spectrum   becomes
\begin{eqnarray}
\frac{d\Gamma}{dE} &\approx&  3 r_0(E_0-E)[(E_0-E)^2-m_\beta^2]^{1/2}\Theta (E_0-E-m_\beta). 
\label{eq:simplebetaspectrum}
\end{eqnarray}

 \subsection{Isotopes of Interest}
Neutrino mass affects the shape of the beta spectrum only near the endpoint.  Because the spectrum rises quadratically from the endpoint, the fraction of decays that produce events in a region of width $m_\beta$ at the endpoint scales as $(m_\beta/Q)^3$, which means that a low Q-value is advantageous, other things being equal.  This generalization ignores details of the spectral shape at lower energies, but is sufficient to guide attention to suitable isotopes. A low Q value is very desirable \cite{PhysRevC.81.045501}, but another important factor is the specific activity of the source. In Table~\ref{tab:isotopes} low-Q-value candidates are compared via a benchmark decay rate in the last eV of the spectrum.  
\begin{table}[htb]
    \centering
    \caption{Source mass required to produce 1 event per day in the last eV of the spectrum. $Q_A$ is the atomic mass difference.}
    \begin{tabular}{cccccccc}
    \hline \hline
    Isotope & Spin-Parity & Half-life & Specific Activity & $Q_A$ & Branching ratio & Last eV & Source Mass \\
    & & y  &    Bq/g & eV & & & g \\
    \hline
    $^3$H$_2$  & \sfrac{1}{2}$^+\rightarrow\ $\sfrac{1}{2}$^+$  & 12.3 &  $3.6\times 10^{14}$ & 18591 & 0.57 & $2.9\times 10^{-13}$ & $2.0\times 10^{-7}$ \\
    $^{115}$In &\sfrac{9}{2}$^+\rightarrow\ $\sfrac{3}{2}$^+$ & $4.4\times 10^{14}$ & 0.26 & 147 & $1.2\times 10^{-6}$ & $5.0\times 10^{-7}$ & $7.5\times 10^{7}$ \\
     $^{135}$Cs &\sfrac{7}{2}$^+\rightarrow\ $\sfrac{11}{2}$^-$ & $1.5\times 10^{6}$ & $6.8\times 10^{7}$ & 440 & $(0.04-16)\times 10^{-6}$ & $2.2\times 10^{-8}$ & 0.4 - 217 \\
     $^{187}$Re & \sfrac{5}{2}$^+\rightarrow\ $\sfrac{1}{2}$^-$ & $4.3\times 10^{10}$ & $1.6\times 10^{3}$ & 2470 & 1.0 & $1.2\times 10^{-10}$ & 57 \\
     \hline 
    $^{163}$Ho &\sfrac{7}{2}$^-\rightarrow\ $\sfrac{5}{2}$^-$ & $4750$ & $1.8\times 10^{10}$ & 2858 & &$\sim 10^{-12}$ & $\sim 1.0 \times 10^{-5}$ \\
    \hline \hline
    \end{tabular}
    \label{tab:isotopes}
\end{table}
Even though tritium has the highest Q-value of the four, its superallowed beta decay and low atomic mass have made it the isotope of choice through 70 years of direct mass searches.  The low Q-values of  $^{115}$In, $^{135}$Cs, and $^{187}$Re are outweighed by the forbidden nature of the beta decays.  Calorimetric measurements of the $^{187}$Re decay were carried out successfully down to a neutrino mass limit of 15 eV \cite{Ferri:2015dla}, but pressing on much further would require prohibitively large source masses.   Similarly, were it not for the source mass, the $^{115}$In \cite{Cattadori:2004vi,Cattadori:2005by,Mount_115IN_PhysRevLett.103.122502,UrbanPhysRevC.94.011302,Zheltonozhsky_2018} and $^{135}$Cs \cite{deRoubin:2020eol} decays would be compelling for their low Q-values.   The In transition is accompanied by a prompt gamma that could be used for background reduction. 

A different approach was suggested by \adr\   \cite{de-rujula:1982}, who noted that $^{163}$Ho has a very low Q-value.  This isotope decays by electron capture and emits neutrinos instead of antineutrinos.  The visible energy release is dominated by sharp lines corresponding to vacancies created in various atomic shells, but the Lorentzian tails of the lines extend to a kinematic endpoint that is sensitive to neutrino mass, just as in beta decay.  There is no simple prescription for the branch to the last eV, but recent work \cite{Ranitzsch:PhysRevLett.119.122501} reports a Q-value of 2858(11) eV.  A rough estimate of the source mass needed for an equivalent sensitivity of about 1 eV has been extracted from \cite{Nucciotti:2018vyc}, and this information is also included in Table~\ref{tab:isotopes}. Holmium is a viable candidate at this basic level. 
Recent experimental work from the ECHo collaboration \cite{Velte:2019jvx} has yielded a Q-value of 2838(14) eV, and a limit on the neutrino mass of 150 eV.

\section{Experimental Progress}
\label{sec:progress}
\setcounter{equation}{0}

Experiments focused specifically on determining the `mass of the  neutrino' (as it was then thought to be) began in 1948 with two contemporaneous experiments on the beta decay of tritium, which was known to have a low decay energy.  Tritium  is the simplest radioactive isotope and has the highest specific activity.  By the 1970s it was clear that neutrino mass effects were small enough that molecular and atomic effects competed, leading to the ``final-state'' problem.  If anything, this cemented the role of tritium because of its simple atomic structure.  However, a different approach that could circumvent the final-state problem completely, the microcalorimeter, emerged and has been the scene of intensive technical development.  In this section the chronology and status of the experimental research on the three viable isotopes, tritium, $^{163}$Ho, and $^{187}$Re, are presented.

The final-state problem takes a different shape for each isotope and has not been completely circumvented, as we describe in Sec.~\ref{sec:refinements}.  Henceforth for brevity we replace $^3$H with T to denote tritium symbolically.

\subsection{Tritium beta decay}

\begin{figure}[htb]
    \centering
    \includegraphics[width=5in]{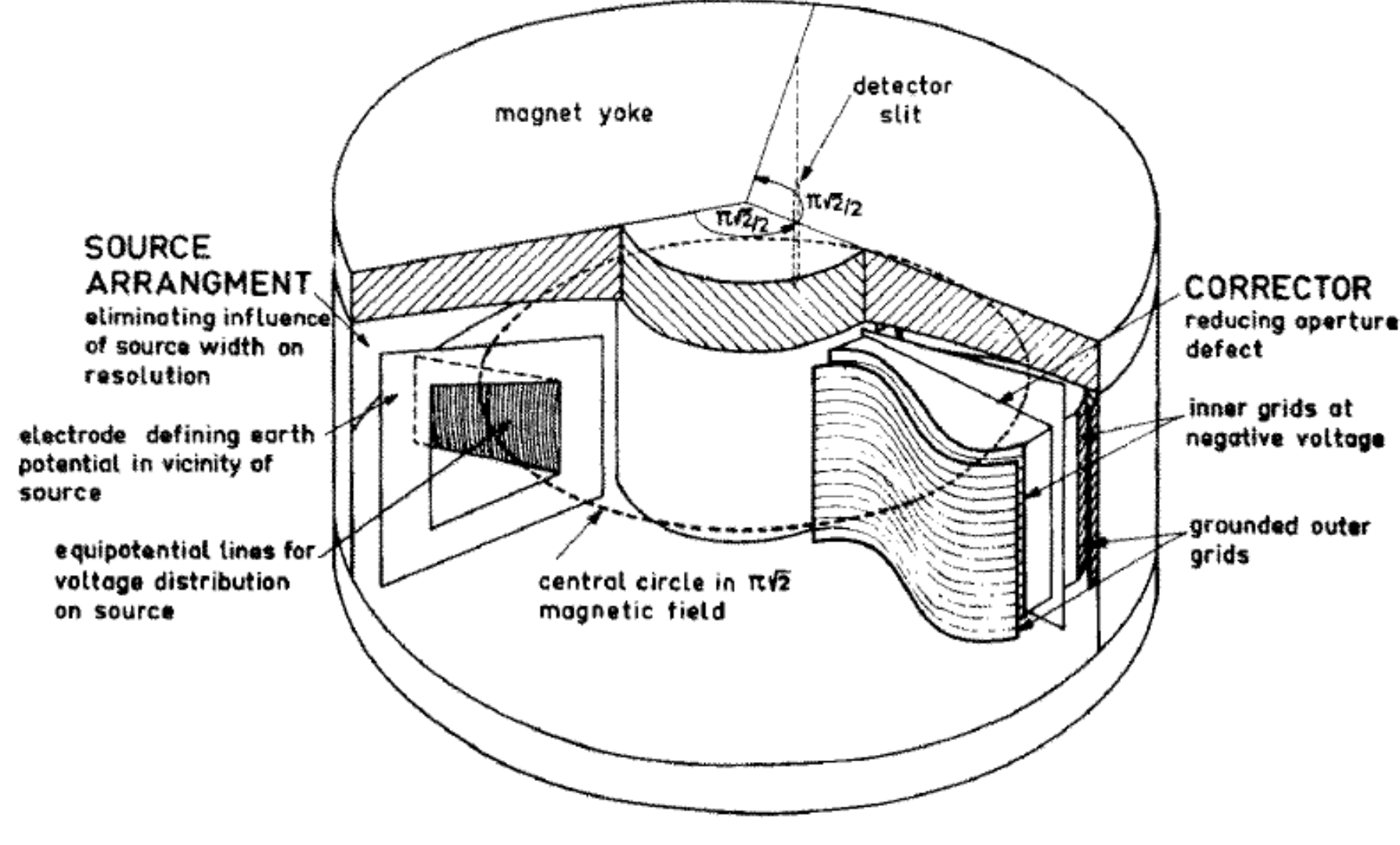}
    \caption{Illustration of the electrostatic-magnetic spectrometer used by Bergkvist in 1972 \cite{bergkvist:1971}.} 
    \label{fig:Bergkvist}
\end{figure}

\begin{figure}[htb]
    \centering
    \includegraphics[width=6.4in]{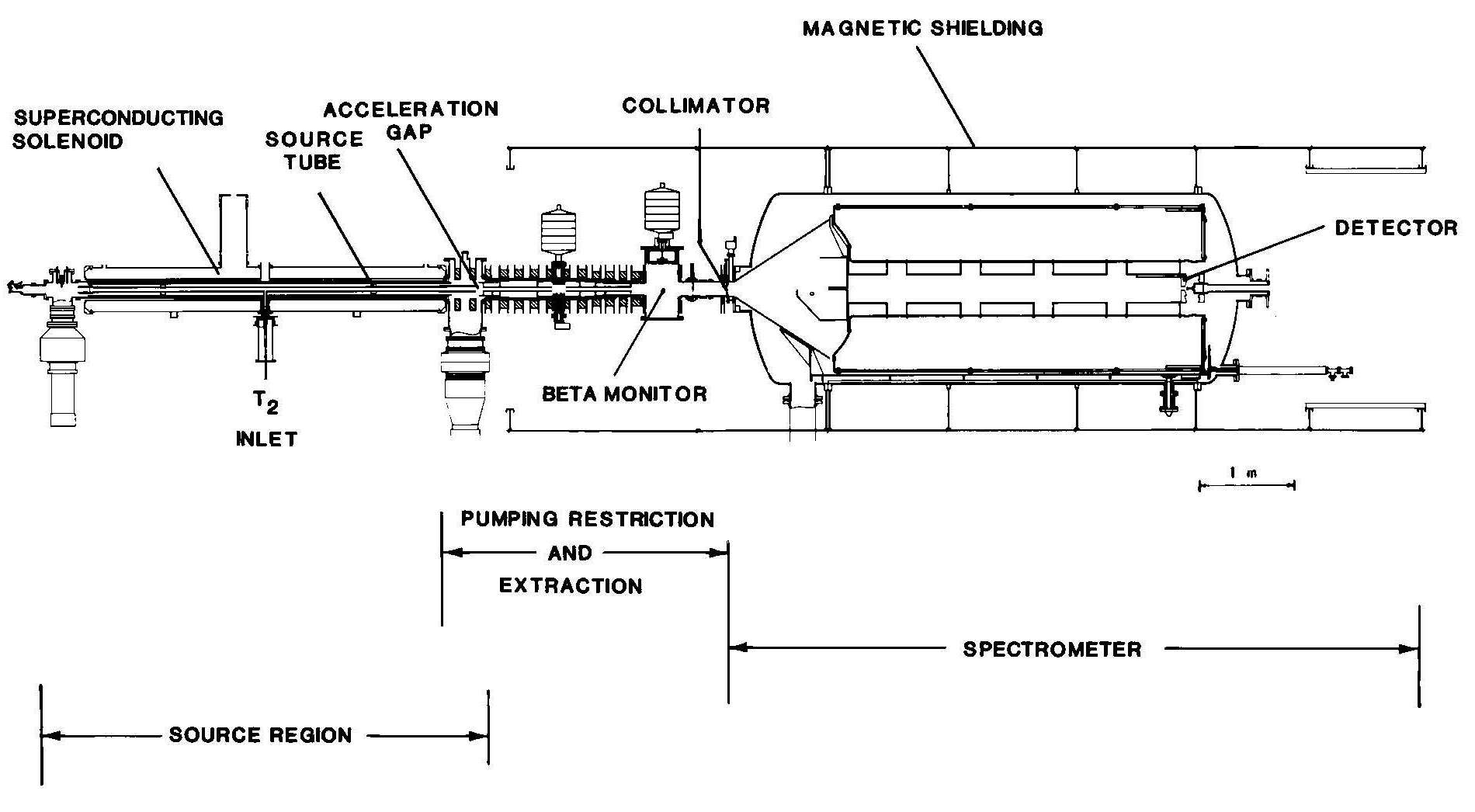}
    \caption{The LANL tritium beta decay experiment \cite{Knapp_thesis,wilkerson:1987aa,robertson:1991aa}.  A windowless gaseous source of molecular T$_2$ on the left produces beta electrons that are guided by solenoid coils to a focal-point collimator at the entrance to a magnetic spectrometer.  The spectrometer is the Tret'yakov cascaded iron-free toroidal type \cite{Tretyakov:1975} in which electrons cross the axis four times before reaching the silicon multipixel detector at the right-hand end.}
    \label{fig:lanl}
\end{figure}

The first experiments to quantitatively constrain the mass of the neutrino took place in 1948 with one in Glasgow and the other in Chalk River.  Both made use of gaseous tritium in a proportional counter.  Over the subsequent half-century, limits on the mass were pushed down thanks to experimental and conceptual improvements, with a wide variety of instruments.  The experiments are summarized in  Table~\ref{tab:tritium_experiments} (some experiments, for example \cite{SunHancheng1993Ulfe}, are not included for lack of information).
\begin{table}[phtb]
    \centering
   \caption{Neutrino mass experiments with tritium. Units: eV.}
   \begin{tabular}{llccccr}
\hline \hline
Group&Date&Source&Spectrometer&Limit or mass&&Ref.\\
\hline
Curran \etal&1948&T$_2$&Proportional counter&$<1700$&&\cite{Curran1948}\\
Hanna \& Pontecorvo&1949&T$_2$&Proportional counter&$<500$&&\cite{hanna:1949aa}\\
Curran \etal&1949&T$_2$&Proportional counter&$<1000$&&\cite{Curran:1949pvk}\\
Langer \& Moffat&1952&T:Succinic acid&Magnetic&$<250$&&\cite{Langer:1952til}\\
Hamilton \etal&1953&T:Zr&Electrostatic&$<200$&&\cite{Hamilton:1953ofv}\\
Salgo \& Staub&1969&T$_2$O&Electrostatic&$<320$&&\cite{Salgo:1969bil}\\
Daris \& St. Pierre&1969&T:Al&Magnetic&$<75$&&\cite{Daris:1969ius}\\
Bergkvist&1972&T:Al&Magnetic&$<55$&&\cite{bergkvist:1972aa,bergkvist:1972ab}\\
R\"{o}de \& Daniel&1972&T:Polystyrene&Magnetic&$<86$&&\cite{RodeDaniel:1972}\\
ITEP&1980&T:Valine&Magnetic&$=30\pm16$&&\cite{Lyubimov:1980un,lyubimov:1981aa}\\
Simpson&1981&T:Si&Si(Li)&$<65$&&\cite{Simpson:PhysRevD.23.649}\\
ITEP&1985&T:Valine&Magnetic&$=35^{+2}_{-15}$&&\cite{Boris:1985mk}\\
Zurich&1986&T:C&Magnetic&$<18$&&\cite{fritschi:1986aa}\\
ITEP&1987&T:Valine&Magnetic&$=30^{+2}_{-13}$&&\cite{Boris:1987tq}\\
LANL&1987&T$_2$&Magnetic&$<27$&&\cite{wilkerson:1987aa}\\
INS Tokyo&1988&T:CdArachidate&Magnetic&$<29$&&\cite{Kawakami:1988cz}\\
INS Tokyo&1991&T:CdArachidate&Magnetic&$<13$&&\cite{tokyo:1991}\\
LANL&1991&T$_2$&Magnetic&$<9.3$&&\cite{robertson:1991aa}\\
Zurich&1992&T:C&Magnetic&$<11$&&\cite{zurich:1992}\\
Mainz&1993&T$_2$&MAC-E&$<7.2$&&\cite{Weinheimer:1993pd}\\
Troitsk&1994&T$_2$&MAC-E&$<4.35$&&\cite{Belesev:1995sb}\\
Mainz&1998&T$_2$&MAC-E&$<2.8$&&\cite{weinheimer_high_1999}\\
Mainz&2005&T$_2$&MAC-E&$<2.3$&&\cite{Kraus:2004zw}\\
Troitsk&2011&T$_2$&MAC-E&$<2.05$&&\cite{Aseev:2011dq}\\
KATRIN&2020&T$_2$&MAC-E&$<1.1$&&\cite{Aker:2019uuj}\\
\hline \hline
    \end{tabular}
    \label{tab:tritium_experiments}
\end{table}
The experiment of Bergkvist \cite{bergkvist:1971,bergkvist:1972aa,bergkvist:1972ab} ushered in the modern era with an advanced spectrometer shown in Fig.~\ref{fig:Bergkvist}. With this device, he reached a sensitivity that was limited by atomic and molecular effects that modify the shape of the spectrum near the endpoint.
There was a flurry of excitement in the early 1980s when a group in the Soviet Union reported a non-zero mass of 30 eV,  the right size to close the universe gravitationally.  The result was erroneous, probably owing to some combination of limited understanding of the final state spectrum of the complicated tritiated molecule used, the amino acid valine, and the energy loss in the source. Two experiments,  at Los Alamos and Livermore National Laboratories, made use of gaseous molecular T$_2$, for which the final-state spectrum could be well calculated. The Los Alamos apparatus is shown in Fig.~\ref{fig:lanl}. 
While those experiments could rule out the Soviet result with much greater sensitivity, they ultimately produced mass-squared values that were apparently negative.  That happens when there are more counts in the endpoint region than expected, and it was eventually shown in 2015 \cite{Bodine:2015sma} that the problem was, once again, the final-state distribution.  More recent calculations of this distribution \cite{saenz00} resolve this problem, and with the new calculations the Los Alamos and Livermore data are consistent with zero mass.  More information on this is given in the section on final-state distributions, Sec.~\ref{sec:fsd}.  Other experiments also reported negative mass-squared values that were eventually traced to different systematic errors.  The early results from the Mainz experiment, which used as a source a frozen film of tritium \cite{Weinheimer:1993pd}, were affected by dewetting of the film at temperatures near 4K.  The microcrystalline `frost' that resulted caused the line broadening from energy loss to be anomalously large. Lowering the temperature to 2K stabilized the films and eliminated the negative mass squared effect \cite{Kraus:2004zw}. The Troitsk experiments exhibited evidence for a step in the spectrum a few eV below the endpoint. While a specific explanation for the spectral shape has not been found, the effect was found to be associated with runs during which the source pressure was not monitored \cite{Aseev:2011dq}.  Excluding those runs from analysis eliminated the step.

\begin{figure}[htb]
    \centering
    \includegraphics[width=5in]{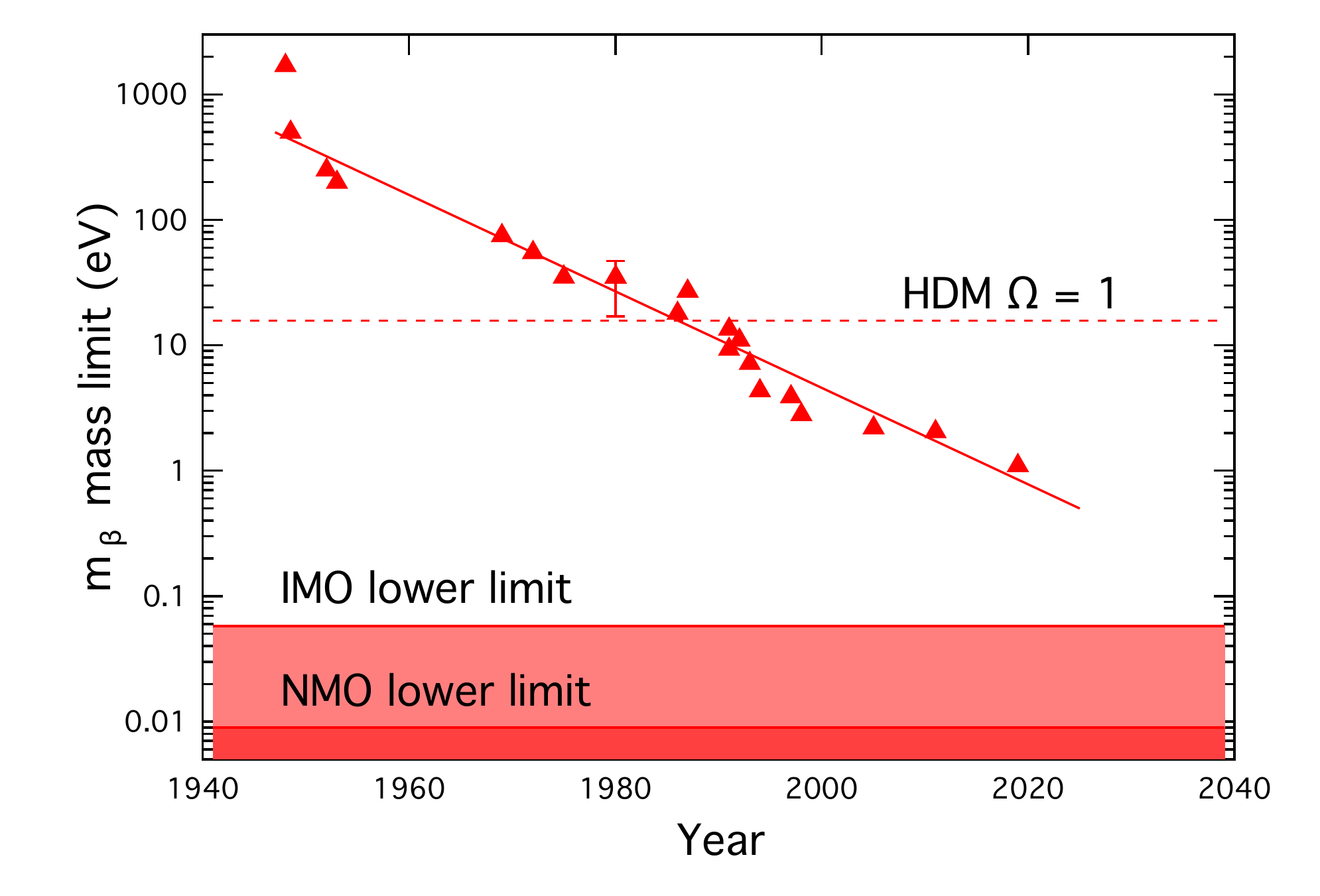}
    \caption{Upper limits on the neutrino mass obtained from tritium beta decay.  The point with error bars is a non-zero result (see text). HDM: Hot Dark Matter.}
    \label{fig:mooreplot}
\end{figure}

The limits on neutrino mass from tritium beta decay as a function of time are shown in Fig.~\ref{fig:mooreplot} (not all results in Table~\ref{tab:tritium_experiments} are included). 
The plot reveals a striking Moore's-Law character over 70 years.  Also indicated on the plot is the mass that would close the universe with neutrinos (hot dark matter) alone (HDM $\Omega=1$), the electron-weighted mass corresponding to the smallest possible value in the inverted mass ordering, and similarly for the normal mass ordering. 

A closely related experimental quantity is the atomic mass difference between T and $^3$He.  While neutrino mass can only be deduced from beta decay, the mass difference can be determined both from beta decay and from independent mass spectrometry methods.  The comparison serves as a uniquely valuable check on possible systematic effects that might influence the beta decay experiments, with no other symptoms.  Indeed, it was mass spectroscopy that initially supported and then finally contradicted the ITEP claim of a non-zero neutrino mass \cite{Staggs:1989zz}.  Agreement of the two kinds of determination within uncertainties is a necessary, although not sufficient, condition for a valid neutrino mass result \cite{Staggs:1989zz}.  As the state of the art in both fields of measurement advances, however, the power of this comparison is beginning to diminish because of the presence of work-function differences that can shift the beta endpoint by fractions of an eV.  This also has the consequence that highly precise atomic mass determinations cannot be used to improve the sensitivity of neutrino mass measurements by fixing a fit parameter. Table~\ref{tab:atomicmass} gives recent determinations of the atomic mass difference.
\begin{table}[phtb]
    \centering
   \caption{Atomic mass difference between T and $^3$He. Units: eV.}
   \begin{tabular}{lccc}
\hline \hline
Group& Year & Mass Difference & Ref.\\
\hline
Univ. Washington & 1993 & 18590.1(17) &  \cite{VanDyck:1993zz}\\
SMILETRAP & 2006 & 18589.8(12) & \cite{Nagy_Smiletrap:2006zz} \\
Florida State Univ. & 2015 & 18592.01(7) &\cite{Myers:2015lca}\\
\hline
KATRIN & 2019 & 18591.5(5) & \cite{Aker:2019uuj}\\
\hline \hline
    \end{tabular}
    \label{tab:atomicmass}
\end{table}
The measured quantity in the beta decay of molecular tritium is the ground-state to ground-state extrapolated endpoint energy $E_0$, which is related to the atomic mass difference $Q_A$ by 
\begin{eqnarray}
E_0 &=& Q_A - b_0 +b_{(f)0} -E_{\rm rec},
\label{eq:qvalue}
\end{eqnarray}
where $b_0 = 4.59$ eV is the binding energy of the initial molecular state, $b_{(f)0}=-11.71$ eV is the binding energy of the final molecular state and $E_{\rm rec}=1.705$ eV is the recoil energy \cite{Bodine:2015sma}.  The specific values are for T$_2$ beta decay.  With these corrections, a measurement of the endpoint from the KATRIN experiment translates to the value shown in the table, which is in good agreement with the mass spectroscopic values.  The 0.5-eV uncertainty is dominated by work functions.

\subsection{The MAC-E Filter and KATRIN}
\label{sec:KATRIN}

Until the 1990s the Tret'yakov cascaded toroidal magnetic spectrometer \cite{Tretyakov:1975}, such as that depicted in Fig.~\ref{fig:lanl}, was the premier instrument  for tritium beta decay experiments.  It had good acceptance and resolution by the standards of the day but, as experimental groups contemplated the next steps, a scale-up in size was clearly necessary for increased statistical precision.  The Troitsk and Mainz groups turned to a concept that had been developed for photoelectron spectroscopy, the retarding-field analyzer.  For neutrino mass experiments with tritium, this type of instrument has a unique advantage: energy conservation eliminates the possibility of a high-energy tail in the response function.  Instrumental tails extending beyond the endpoint are particularly deadly because they shift the mass-squared value negatively if not recognized, and even when recognized severely degrade the sensitivity to neutrino mass.  Moreover, the retarding-field analyzer combined with a magnetic field for collimation, the `MAC-E' filter (Magnetic Adiabatic Collimation -- Electrostatic), had another major advantage in the way it scaled in size, as we show next.  

In order to detect the effect of $m_\beta$ at the
endpoint of a beta spectrum of total kinetic energy $E_0$, instrumental
resolution of  order $\frac{m_\beta}{E_0}$ is needed.  The spectral fraction per
decay that falls in the last $m_\beta$ of the beta spectrum is approximately
$(\frac{m_\beta}{E_0})^3$, to within a constant of order unity.

For spectrometric experiments in which the source and the detector are
physically separated, a limit on source thickness is set by the cross
section for inelastic interactions of outgoing electrons, such that one must
have $\sigma n_s \leq 1$, where $\sigma$ is the inelastic cross section and $n_s$ the superficial number density.  More intense sources to reach smaller neutrino masses must therefore have larger areas.  On general grounds, the dimensions of the source (radius
$R_{\rm src}$) and the dimensions of the spectrometer (length or radius $R_{\rm ana}$) are related
through the resolution needed, specifically,
\begin{eqnarray}
\frac{\Delta E}{E} \simeq \left(\frac{R_{\rm src}}{R_{\rm ana}}\right)^{\alpha}.
\end{eqnarray}
For magnetic
spectrometers such as the Tret'yakov type, $\alpha = 1$.  The MAC-E filter, however, has a different scaling relationship.  The source is immersed in a high magnetic field $B_{\rm src}$, and the spectrometer in a relatively low field $B_{\rm ana}$.  Electrons move from source to analyzer adiabatically along magnetic field lines, and the energy resolution is determined by the field ratio \cite{Angrik:2005ep}: 
\begin{eqnarray}
\frac{\Delta E}{E} &=& \frac{B_{\rm ana}}{B_{\rm src}} \\
&=& \left(\frac{R_{\rm src}}{R_{\rm ana}}\right)^{2}.
\end{eqnarray}
Thus, for magnetic-electrostatic retarding-field
analyzers (MAC-E filters) $\alpha$ is a more favorable $2$.  This scaling property was decisive in shifting the  focus of the experimental community toward the MAC-E filter, leading ultimately to the KATRIN project.

\begin{figure}[htb]
    \centering
    \includegraphics[width=5in]{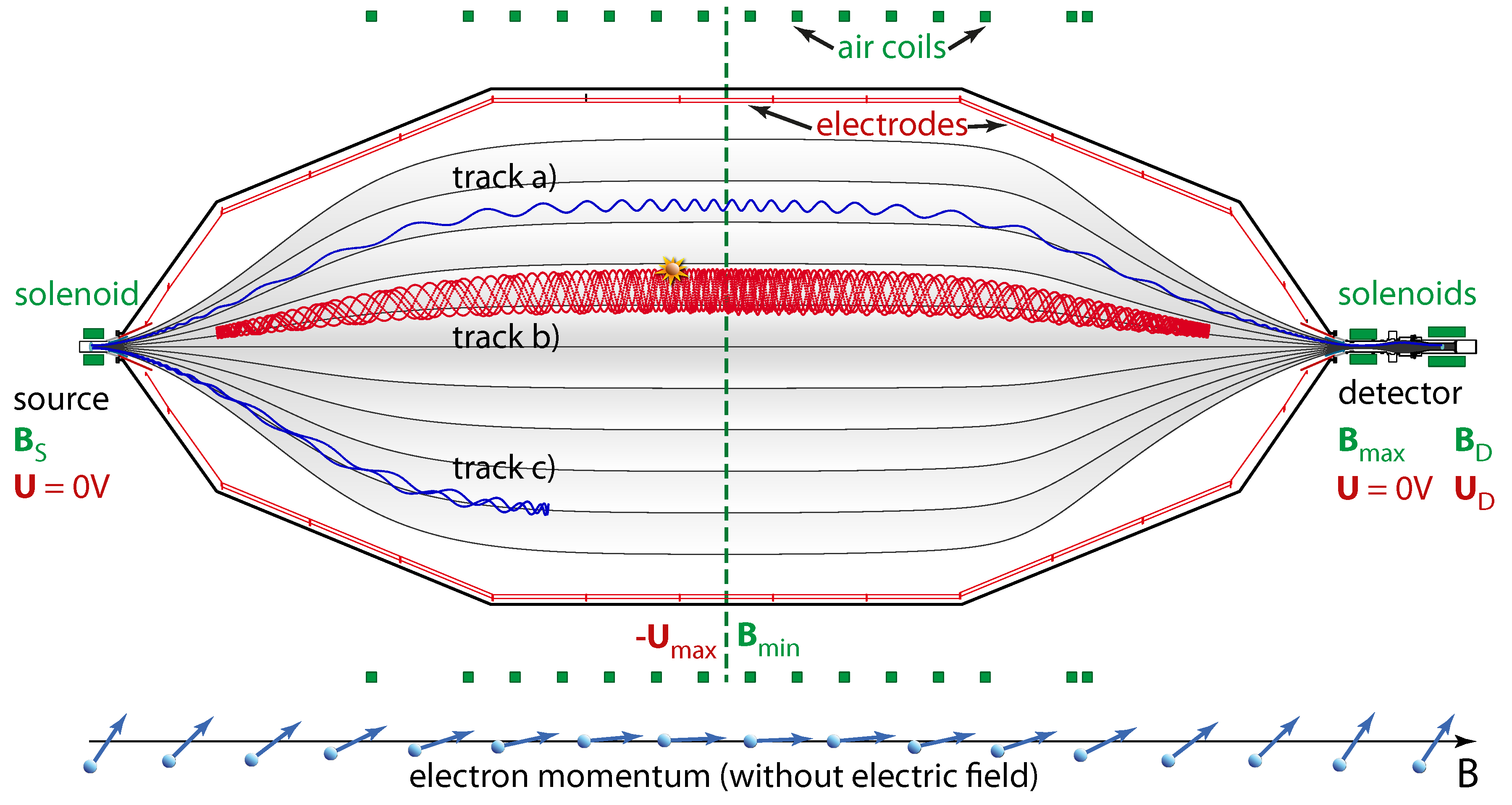}
    \caption{Principle of the MAC-E filter (from \cite{Arenz:2016mrh}).  Electrons produced in a high magnetic field region enter the large, empty high-vacuum space of the main spectrometer.  The magnetic field is lower there by a factor $\sim 10000$, and the momentum vector swings forward along the direction of the field.  At the same time a potential gradient is applied between the entry point and the analyzing plane in the center by floating the entire spectrometer shell at the (negative) analyzing potential.  Electrons with sufficient energy to surmount the barrier are re-accelerated, re-enter a high magnetic field, and are detected. }
    \label{fig:mace}
\end{figure}

 The Magnetic-Adiabatic-Collimation-Electrostatic filter concept was first described in 1976 by Hsu and Hirschfield \cite{Hsu_doi:10.1063/1.1134594} and further developed by Beamson {\em et al.} \cite{Beamson_1980} and Kruit and Read in 1983 \cite{Kruit_1983}, and it soon found adoption in many areas of electron spectroscopy.  It combines good source acceptance with high resolution.  The Mainz and Troitsk experiments were the first to adopt this new technology for tritium beta decay, with Troitsk mating it to a gaseous T$_2$ source on the lines of the Los Alamos design, and Mainz mating it to a source of tritium frozen on a substrate of highly oriented pyrolytic graphite.  The basic principle of the MAC-E filter is shown in Fig.~\ref{fig:mace}.
The MAC-E filter has very high acceptance.  All electrons from a cross-section in the high-field source that are emitted with a pitch angle smaller than a selected value are transmitted to the analyzing plane.  The pitch angle (the angle between the momentum and the field direction) is established by means of a solenoidal ``pinch'' magnet located somewhere between the source and detector. The energy resolution is determined by the ratio of the magnetic fields in source and spectrometer, and the choice of accepted pitch-angle range \cite{Kleesiek:2018mel}.  It has a simple, analytic form.  

\begin{figure}[htb]
    \begin{tabular}{c}
    \includegraphics[width=6.0in]{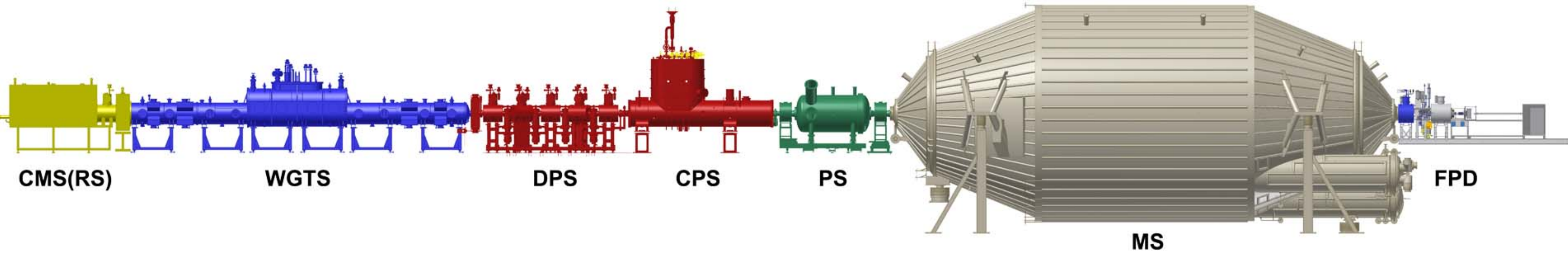} \\
    \includegraphics[width=6.0in]{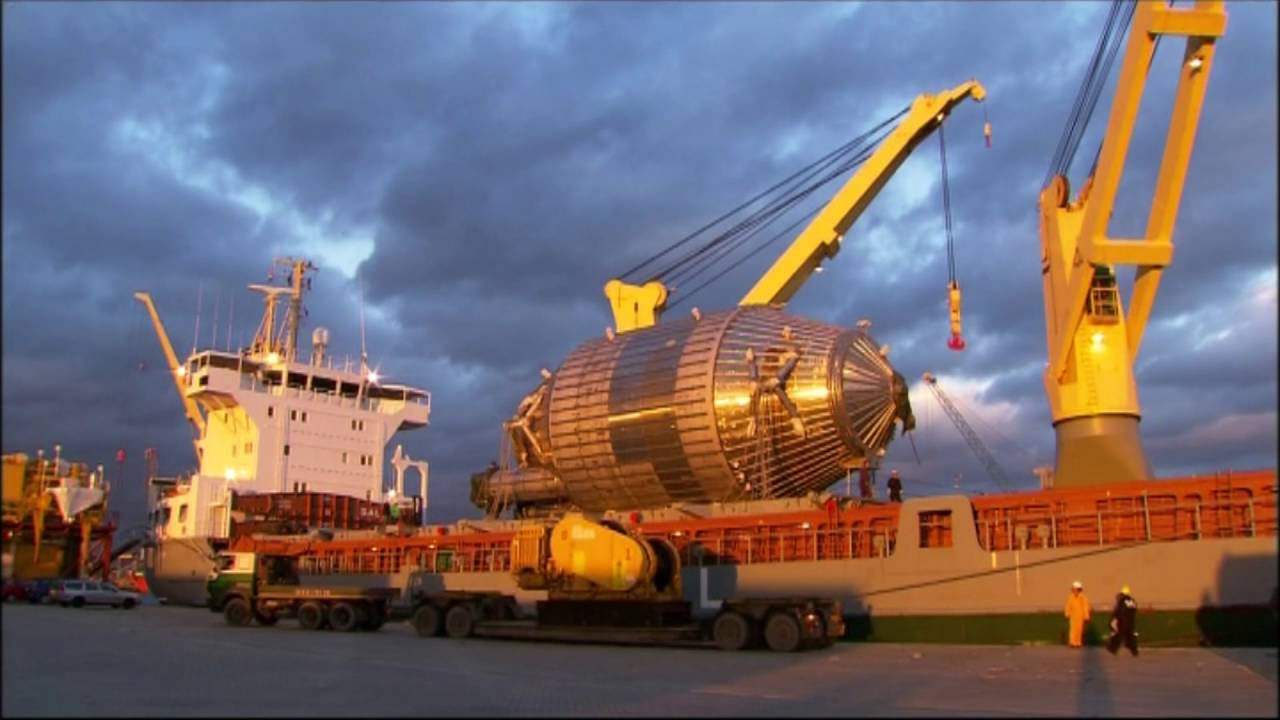} \\  
    \end{tabular}
    \caption{Top: View of the KATRIN experiment (from \cite{Arenz:2016mrh}).  Tritium gas recirculates continuously through the windlowless gaseous tritium source (WGTS). A solenoidal magnetic field guides beta electrons to the main spectrometer (MS) through a differential pumping restriction (DPS) and a cryogenic pumping restriction (CPS) that prevent tritium from entering the spectrometers.  A prespectrometer (PS) blocks lower-energy electrons from reaching the MS. Electrons that surpass the MS potential are detected in the focal-plane detector (FPD).  Calibration devices are located in the rear system (RS). Bottom:  Photograph of KATRIN's main spectrometer.}
    \label{fig:katrinoverview}
\end{figure}

The KATRIN experiment represents what is likely to be the ultimate realization of the MAC-E technology.  Its main spectrometer, 9.8 m in diameter and 23.3 m in length, is the largest ultra-high-vacuum vessel in the world and  operates at a base pressure of $10^{-11}$ mbar.  The design magnetic fields of 0.3 mT at the analyzing plane of the spectrometer and 4 T in the source lead to an integral energy resolution-function step of 0.93 eV at the tritium endpoint.  An elevation view of the KATRIN experiment is shown in Fig.~\ref{fig:katrinoverview}.
Detailed descriptions of the KATRIN approach and apparatus can be found in \cite{Otten:2008zz,Drexlin:2013lha,Arenz:2016mrh}.  

\begin{figure}[htb]
    \centering
    \includegraphics[width=5in]{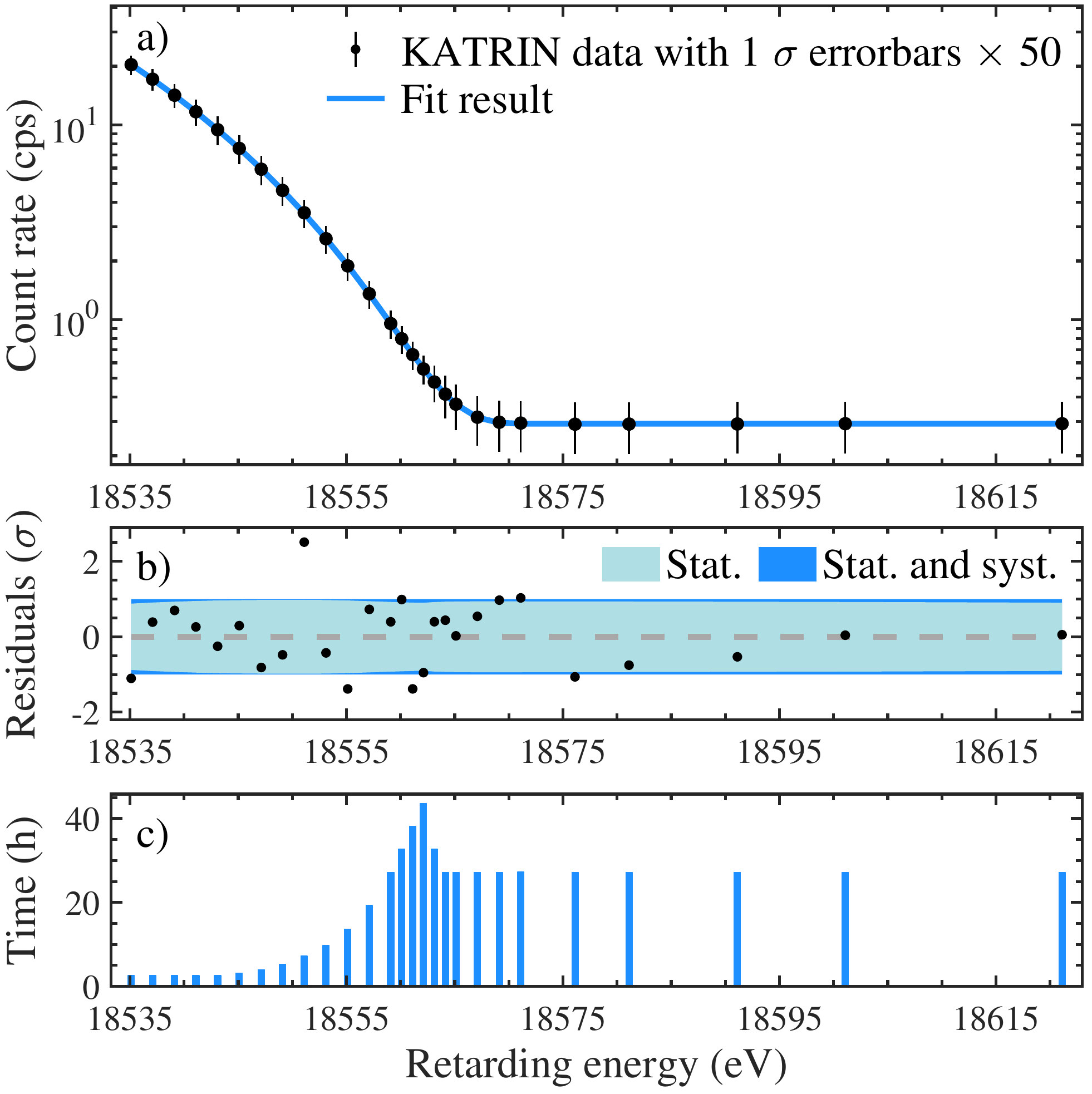}
    \caption{Integral beta spectrum of molecular T$_2$ from  the KATRIN experiment  \cite{Aker:2019uuj}.  The spectrum contains $2.03\times 10^6$ events and was taken over a period of 22 live days. The column density of tritium in the WGTS source tube was 20\% of the nominal design value. From these data, an upper limit of 1.1 eV (90\% CL) on the neutrino mass was obtained, a factor 2 below the previous world limit.  The lowest panel shows the distribution of measurement times chosen at each retarding potential.}
    \label{fig:katrinspectrum}
\end{figure}

KATRIN began commissioning with tritium at low concentrations in 2018, and in 2019 gathered 22 live days of data for its first neutrino-mass measurement.  The spectrum is shown in Fig.~\ref{fig:katrinspectrum}.

The MAC-E filter is intrinsically both a magnetic trap and a Penning trap, and both modes can be sources of background.  In KATRIN, the presence of the prespectrometer forms a second Penning trap, of the opposite sign.  The latter trap was a worry during the design phase because, as an electron trap, the repeated passage of electrons through the residual gas can produce a plasma and a catastrophic discharge.  This mode was a principal motivator for achieving ultra-high vacuum in the spectrometers, and indeed it was found that the trap could ignite at pressures in the $10^{-10}$ mbar range, but not at the operating pressure $1.2\times 10^{-11}$ mbar \cite{Aker:2019uuj}.  The magnetic mode in the main spectrometer traps relatively high-energy electrons in the keV to MeV range, and, somewhat surprisingly, $^{219}$Rn emanating from the getters was found to be the main contributor.  In this mode, the electrons circulate in the magnetic trap for times as long as an hour, slowly losing energy by ionization to the residual gas \cite{Frankle:2011xy,Mertens:2012vs}.  The resulting background of low-energy electrons is troublesome because it is non-Poissonian.  Installation of large liquid-nitrogen cooled chevron baffles in front of the getter chambers largely solved the problem \cite{Goerhardt:2018wky}, but the baffle efficiency for Rn decreases as water vapor accumulates.  Installation of a subcooler to reduce the temperature another 10 K is expected to reduce this background greatly. 

Although particles cannot easily get out of a magnetic trap, neither can they easily get in.  This `magnetic shielding' has proven to be highly effective in rejecting radioactive and cosmic ray backgrounds from the walls \cite{Altenmuller:2019xbg,Arenz:2018aly}.  In addition, layers of grids totalling 20000 wires spaced from the walls and biased to negative voltages in the 100 -- 500 V range further reject backgrounds of soft electrons, and also permit a more precise shaping of the electric field inside the spectrometer \cite{Valerius:2010zz,Prall:2008zz}.  A few of these grids became shorted to each other during a bakeout, but with no major implications for KATRIN's performance.

In KATRIN a new and unexpected kind of background was discovered, the production of Rydberg atoms and their photoionization  in the volume of the main spectrometer.  Atoms of hydrogen and heavier species are dislodged from the walls of the spectrometer by the decay of radon daughters embedded therein.  These atoms are often neutral and in a distribution of excited states, some of which are so close to the ionization edge that they can be ionized by thermal black-body radiation as they cross the spectrometer volume.  KATRIN is developing strategies to mitigate this significant background.  It exemplifies a vulnerability of the MAC-E filter method, that signal electrons are slowed almost to rest before being reaccelerated into the detector.  Therefore, in addition to the Rydberg atoms, any process that makes slow electrons (ionization of the residual gas, for example, or the decay of errant tritium) becomes a potential background.

These backgrounds, together with additional smaller contributions from the FPD, amounted at the beginning of operations to about 0.5 counts per second (cps), a factor of 50 larger than the target value in the Design Report \cite{Angrik:2005ep}.  Increasing the magnetic field in the main spectrometer reduced the background to 0.29 cps at some cost in resolution.  This configuration was used for the first KATRIN neutrino mass measurement \cite{Aker:2019uuj}. For this background rate, the optimal measurement-time distribution (bottom panel of Fig.~\ref{fig:katrinspectrum}) peaks 12 eV below the extrapolated endpoint.  The statistical contribution to neutrino-mass sensitivity, however, depends approximately on the sixth root of the background (see Eq.~\ref{eq:macesensitivity}), and if no further steps were to be taken, the final sensitivity would decrease from 200 meV to about 300 meV. Of course, KATRIN has moved aggressively to deal with backgrounds.   Shifting the analysis plane toward the detector serves to further reduce the effective spectrometer volume and therefore the main backgrounds by a factor of 3.    Additional measures are under investigation.  

Another unexpected complication to KATRIN's neutrino mass extraction has emerged from plasma potentials.  Although the presence of electric and magnetic potentials has no influence on the molecular dynamics of KATRIN's tritium source, it certainly affects the diffusion properties of the emitted electron and tritium ions created in the decay.  Electron-ion recombination and charge drift eventually neutralize the source; however, neutralization occurs over long time scales, leading to both spatial and temporal variations of the charge density of the source. These electromagnetic potential fluctuations impact the energy of the decay electron.  A campaign is currently focused on constraining the uncertainties arising from plasma potentials.  Plasma potentials can influence any technique that makes use of magnetic confinement, including cyclotron radiation emission spectrometers.

\subsection{\texorpdfstring{$^{163}$Ho}{163Ho} electron-capture decay}

Stimulated by the report of non-zero neutrino mass observed in tritium beta decay by Lyubimov {\em et al.} \cite{Lyubimov:1980un}, a search began for alternative methods as a verification.  The low Q-value for the electron-capture decay of  $^{163}$Ho to $^{163}$Dy, about 3 keV \cite{Naumann:PhysRev.171.1290}, attracted interest in the possibility of a neutrino mass measurement with this isotope.  The subshell ratios, i.e. the relative intensity of electron capture in each atomic shell, depend on the neutrino mass because the available phase space is limited for the deeper shells by the neutrino rest mass.  This formed the basis of the initial attack on the problem by Bennett {\em et al.} \cite{bennett:198119}, but they found that the theory of subshell ratios in heavy nuclei was not adequate to extract a neutrino mass. In the same year, \adr\ proposed \cite{DeRujula:1981ti} a different approach, internal bremsstrahlung in electron capture (IBEC), a radiative process producing a continuous spectrum of photons with an endpoint shape that is modified by neutrino mass, quite analogous to beta decay.  He suggested the calorimetric  technique that is the basis of experimental work today.  The term IBEC as used by \adr\  in his unified treatment covered processes that are usually considered separately, the radiative process when a real photon attaches to the electron or the $W$-boson ({\em `innere bremsstrahlung'}), and the  Lorentzian tails of X-ray transitions as determined by the vacancy lifetime.  The X-ray widths are complicated by atomic structure effects, which were treated schematically.  In a paper a year later \cite{de-rujula:1982}, however, the reference to IBEC is dropped and the X-ray widths are used to derive the transition rate near the endpoint.  Springer \etal \cite{Springer:PhysRevA.35.679} carried out the first $^{163}$Ho neutrino mass experiment in 1987 and included a  remarkably detailed theoretical study of IBEC complete with the interference effects that were found to be substantial.  They set an upper limit on the mass of 225 eV, although their use of a Q-value that is now discounted may have unduly influenced the derived limit.  In 1994, Yasumi \etal \cite{yasumi:1994229} set a less stringent limit on the neutrino mass in a study more in the spirit of the original Bennett \etal one \cite{bennett:198119} in which the subshell ratios were measured, except they circumvented the theoretical issues via a direct photoionization measurement of the X-ray yield at a synchrotron light source.  

Thereafter, experimental work converged on the microcalorimetric approach in which the de-excitation of the $^{163}$Dy, whether in the form of photons or electrons, is entirely captured and converted to heat.  The endpoint of such a spectrum is just the Q-value, apart from a small binding-energy correction in the lattice.  Neutrino phase space modifies the endpoint region of the spectrum as in beta decay.  The experimental efforts to pin down the Q-value and extract a neutrino mass or a limit, are summarized in Table \ref{tab:Ho_experiments}.
\begin{table}[htb]
    \centering
   \caption{Neutrino mass and Q-value experiments with $^{163}$Ho. Units: eV}
   \begin{tabular}{llccccr}
\hline \hline
Group&Date&Source&Spectrometer&Limit or mass&Q-value&Ref.\\
\hline
Hopke \etal&1968&Ho:Al&Proportional counter&$-$&$<9100$&\cite{Naumann:PhysRev.171.1290}\\
Bennett \etal&1981&Ho:Al&Si(Li)&$-$&$>2047$&\cite{bennett:198119}\\
Baisden \etal&1983&Ho+&Mass spectrometer&$-$&2650(250)&\cite{Baisden:PhysRevC.28.337}\\
Hartmann \& Naumann&1985&Ho(fod)$_3$&Proportional counter&$-$&2600(30)&\cite{Hartmann:PhysRevC.31.1594,Hartmann:1992jg}\\
Springer \etal&1987&HoF$_3$&Si(Li)&$<225$&2561(20)&\cite{Springer:PhysRevA.35.679}\\
Yasumi \etal&1994&Ho metal&Si(Li)&$<460$&2710(100)&\cite{yasumi:1994229}\\
Gatti \etal&1997&Ho salt&Microcalorimeter&$-$&2800(50)&\cite{gatti:1997415}\\
SHIPTRAP&2015&&ICR&$-$&2833(34)&\cite{Shiptrap_PhysRevLett.115.062501}\\
Ranitzsch \etal (ECHo)&2017&Ho:Au&Microcalorimeter&$-$&2858(11)&\cite{Ranitzsch:PhysRevLett.119.122501}\\
Velte \etal (ECHo)&2019&Ho:Au&Microcalorimeter&$<150$&2838(14)&\cite{Velte:2019jvx}\\
\hline \hline
    \end{tabular}
    \label{tab:Ho_experiments}
\end{table}
It is a testament to the difficulty of this problem that in the intervening 40 years, the mass limit has come down from 225 eV to 150 eV.  Even the Q-value has been resistant to accurate measurement, with current numbers some ten standard deviations from earlier measurements.  The development of ion cyclotron resonance (ICR) and Penning-trap mass spectrometers has made possible dramatic improvements in the accuracy of isotope mass measurements.  A review of the methods and of the progress made can be found in \cite{Dilling:2018mqa}.

Three $^{163}$Ho microcalorimeter programs have been pursued,  HOLMES,  ECHo, and NuMECS.  The HOLMES apparatus is based on superconducting transition-edge sensors \cite{Becker:2019oyu} of molybdenum-copper on a silicon nitride substrate.  The source pad is gold, which will be implanted with $^{163}$Ho \cite{HOLMES_Gallucci:2020sxk}.  A total of 1024 sources is planned, which will be multiplexed for readout by RF SQUIDs (radiofrequency superconducting quantum interference devices).  The NuMECS sensors are similarly molybdenum-copper transition-edge sensors, but  supported on nanofabricated silicon beams \cite{Croce:2015kwa}.  The $^{163}$Ho has been incorporated from aqueous solution into nanoporous gold pads at the beam ends and spectra have been obtained.  While the resolution at 35 eV FWHM is not yet sufficient, the purity and specific activity of the Ho material are good.    The ECHo collaboration \cite{echo:2014,echo:2017,gastaldo:2013,Ranitzsch:PhysRevLett.119.122501} makes use of metallic magnetic calorimeters (MMC) to read out the thermal signals via SQUIDs.  These devices have delivered the fastest risetimes,  $\lesssim 100$ ns, and also hold the record for the best energy resolution, 1.6 eV FWHM with an X-ray source of $^{55}$Fe.  The MMC devices themselves are not intrinsically faster than TES's, but risetime in TES's is usually kept longer to match the dc-SQUID readout better. The speed is determined by heat diffusion in gold in both devices.  Use of MMC's with multiplexing will unavoidably require limitations in detector speed.  Ultrapure mass-separated $^{163}$Ho ions were implanted into gold source pads to produce spectra (see Fig.~\ref{fig:haverkort}) with an instrumental resolution of 9.2 eV FWHM \cite{Velte:2019jvx}.  Only 2 events above the endpoint were observed, evidence that the background is very low.   When the $^{163}$Ho isotope is reactor-produced by irradiation of $^{162}$Er, the isotopic contaminant $^{166m}$Ho is also produced, necessitating mass separation. Both ECHo and HOLMES use this method, while the NUMECS material was produced by proton bombardment of Dy, which avoids the $^{166m}$Ho byproduct but has a lower cross section.

\subsection{\texorpdfstring{$^{187}$Re}{187Re} beta decay}

In a similar vein as for $^{163}$Ho, $^{187}$Re also offers a low Q-value, 2.5 keV, in comparison to that of tritium  and thus emerged as an attractive isotope for neutrino mass investigation. Despite its low Q-value, there are inherent difficulties in using $^{187}$Re as a neutrino mass target.  In particular, its decay process:
\begin{equation}
~^{187}{\rm Re} ({\rm J}={5/2}^+) \rightarrow ~^{187}{\rm Os} ({\rm J}={1/2}^-) ~ + e^{-} ~ + \bar{\nu}_e
\end{equation}
is a unique first-order forbidden transition ($\Delta J^\pi = 2^-$), which significantly alters the phase space of the decay electron near the endpoint and gives rise to a formidably long lifetime ($\tau = 4.12 \times 10^{10}$ y).  This places severe requirements on the amount of target material necessary to observe a significant number of decay events with electron energies near the endpoint of rhenium (Table~\ref{tab:isotopes}).

The beta decay of $^{187}$Re was originally observed using proportional counters, and a definite determination of the beta decay process was made by Brodzinski and Conway~\cite{Brodzinky_PhysRev.138.B1368} and by Huster and Verbeek~\cite{huster1967ss} using gas proportional counters.  From those early measurements, they determined a  Q value of 2.6 keV.  Inspired by the results by Lubimov indicating a positive value for neutrino mass, McCammon~\cite{Mccammon:1984cv} and later  Vitale \etal~\cite{Vitale1985} raised the possibility of using microcalorimeters to better measure the neutrino mass signal.  Vitale \etal proposed  a rhenium-based calorimeter to measure the total visible energy produced from the beta decay process, from which a neutrino mass measurement could be extracted. 

The first measurements of $^{187}$Re decay using microcalorimeters were done in Genoa by the MANU project~\cite{COSULICH1992143}.   They used metallic rhenium as a self-contained absorber, taking advantage of the fact that rhenium at temperatures below 1.7 K becomes superconducting.  Under these conditions, the electronic contributions of the heat capacity are exponentially suppressed, leaving only phonon/lattice contributions to the heat capacity.  When cooled well below the transition temperature (usually $< 100$ mK), the heat capacity should be significantly reduced, allowing for high energy resolution, as needed for an accurate endpoint measurement.  Unfortunately, the energy released in the decay by the electron at such low temperatures tends to be trapped in the form of quasi-particles, for which the recombination time to phonons (and thus a detectable signal) can be as long as several seconds. As such, metallic rhenium proved difficult to realize as a microcalorimeter for the purposes of beta decay detection.  

\begin{figure}[htb]
    \centering
    \includegraphics[width=3in]{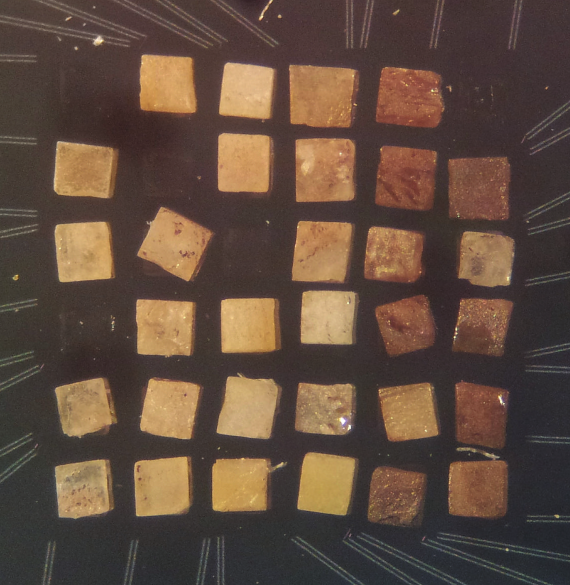}
    \caption{A view of several microcalorimeter pixels using 500 $\mu$g AgReO$_4$ absorbers for the MARE experiment~\cite{bib:MARECrystal}.}
    \label{fig:AgReO4_array}
\end{figure}

\begin{figure}[htb]
    \centering
    \includegraphics[width=5in]{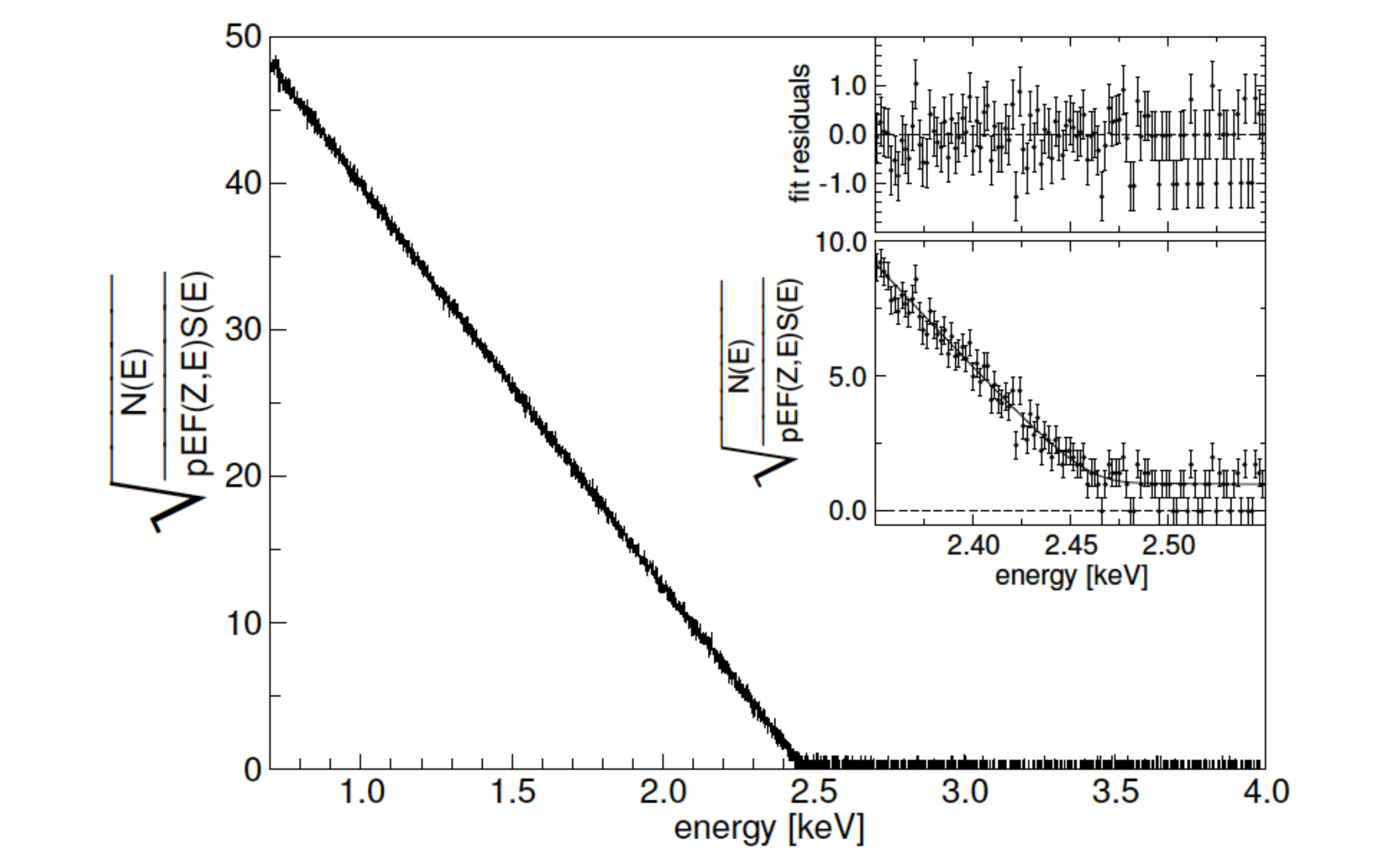}
    \caption{Kurie plot showing the measured beta spectrum from $^{187}$Re decay from the Milano group's  experiment~\cite{SISTI2004125}.}
    \label{fig:ReKurie}
\end{figure}

A parallel effort, led by the Milan group MiBETA, used a dielectric compound of rhenium to achieve a similar suppression of electron-based noise while avoiding issues associated with quasi-particle trapping  in superconductors (see Fig.~\ref{fig:AgReO4_array}).  Several compounds were tried and the MiBETA group eventually settled on silver perrhenate (AgReO$_4$) as yielding the best performance in terms of efficiency and energy resolution ($\sigma \simeq 18$ eV FWHM at 6 keV).  Both the MANU and MIBETA groups extracted measurements of the  neutrino mass, both limits below 20 eV.  The extracted Kurie plot from the decay of $^{187}$Re is shown in Fig.~\ref{fig:ReKurie}.  The MANU experiment also measured for the first time oscillations in the beta energy spectrum due to environmental fine structure from the target crystal, although also revealing a potential systematic uncertainty in their neutrino mass extraction. The effect was also measured in the AgReO$_4$ detectors by the MiBETA group~\cite{PhysRevLett.96.042503}.

\begin{table}[htb]
    \centering
   \caption{Neutrino mass and Q-value experiments with $^{187}$Re. Units: eV}
   \begin{tabular}{llccccr}
\hline \hline
Group&Date&Source&Spectrometer&Limit (90 \% C.L.) or mass&Q-value&Ref.\\
\hline
Brodzinsky \& Conway & 1965 & (C$_5$H$_5)_2$ReH & Proportional counter & N/A & 2620(90) & \cite{Brodzinky_PhysRev.138.B1368} \\
Huster \& Verbeek & 1967 & (C$_5$H$_5)_2$ReH & Proportional counter & N/A & 2650(40) & \cite{huster1967ss} \\
MANU (Genoa) & 1999& Metallic Re & Microcalorimeter  & $\leq 19$ & 2470(4) & \cite{gatti:1999} \\
MIBETA (Milano) & 2004 & AgReO$_4$ & Microcalorimeter  & $\leq 15$ & 2465.3(17) & \cite{SISTI2004125} \\
\hline \hline
    \end{tabular}
    \label{tab:Re_experiments}
\end{table}

The MANU and MIBETA groups eventually combined to propose a new calorimetric experiment using rhenium called Microcalorimeter Arrays for a Rhenium Experiment (MARE), with up to 10,000 of such microcalorimeters to reach sub-eV sensitivity to the neutrino mass \cite{nucciotti:2008}.  The MARE effort eventually gave way to the holmium efforts of ECHo and HOLMES, as discussed earlier in this section.

\section{Other Research}
\label{sec:other}
\setcounter{equation}{0}

The kinematic direct mass search method can be applied not only to the mass of the three known active neutrinos, but also to other significant physics questions.  Are there sterile neutrinos that mix slightly with the active species and thereby become observable?  They may disclose their existence in the form of kinks in the otherwise smooth beta spectrum.  Can the primordial sea of relic neutrinos created in the big bang be detected? They would produce a peak in the beta spectrum just beyond the endpoint energy.  

\subsection{Sterile Neutrinos}
\label{sec:sterile}

The width of the $Z$-boson fixes the number of active light neutrinos at 3 \cite{Zyla:2020zbs}, but the possibility remains that there might be neutral fermions with no standard-model couplings. If they admix slightly with the active neutrino states they become observable both in oscillation and direct-mass experiments.  The discovery of neutrino mass in particular renewed interest in this possibility because light right-handed singlets would be introduced into the Standard Model if neutrinos are Dirac fermions, and heavier partners might exist if neutrinos are Majorana fermions.  In direct mass searches, these `sterile neutrinos' would appear as a kink in the beta spectrum if the mass is in the kinematically allowed range.

Perhaps the most famous example is the ``17-keV neutrino'' that was first reported by Simpson in 1985, from the observation of a kink in the beta spectrum of tritium implanted into a silicon detector \cite{Simpson:1985xc}.  Much excitement ensued when the same signal seemed to show up in many experiments on tritium and other isotopes, but a convincing demonstration against it was carried out by Mortara \etal \cite{Mortara:1993iv}.  Summaries of the saga can be found in Ref.~\cite{Wietfeldt:1995ja,Robertson:1992cv,Franklin:1995pk}.

\begin{figure}[htb]
    \centering
    \includegraphics[width=3.25in]{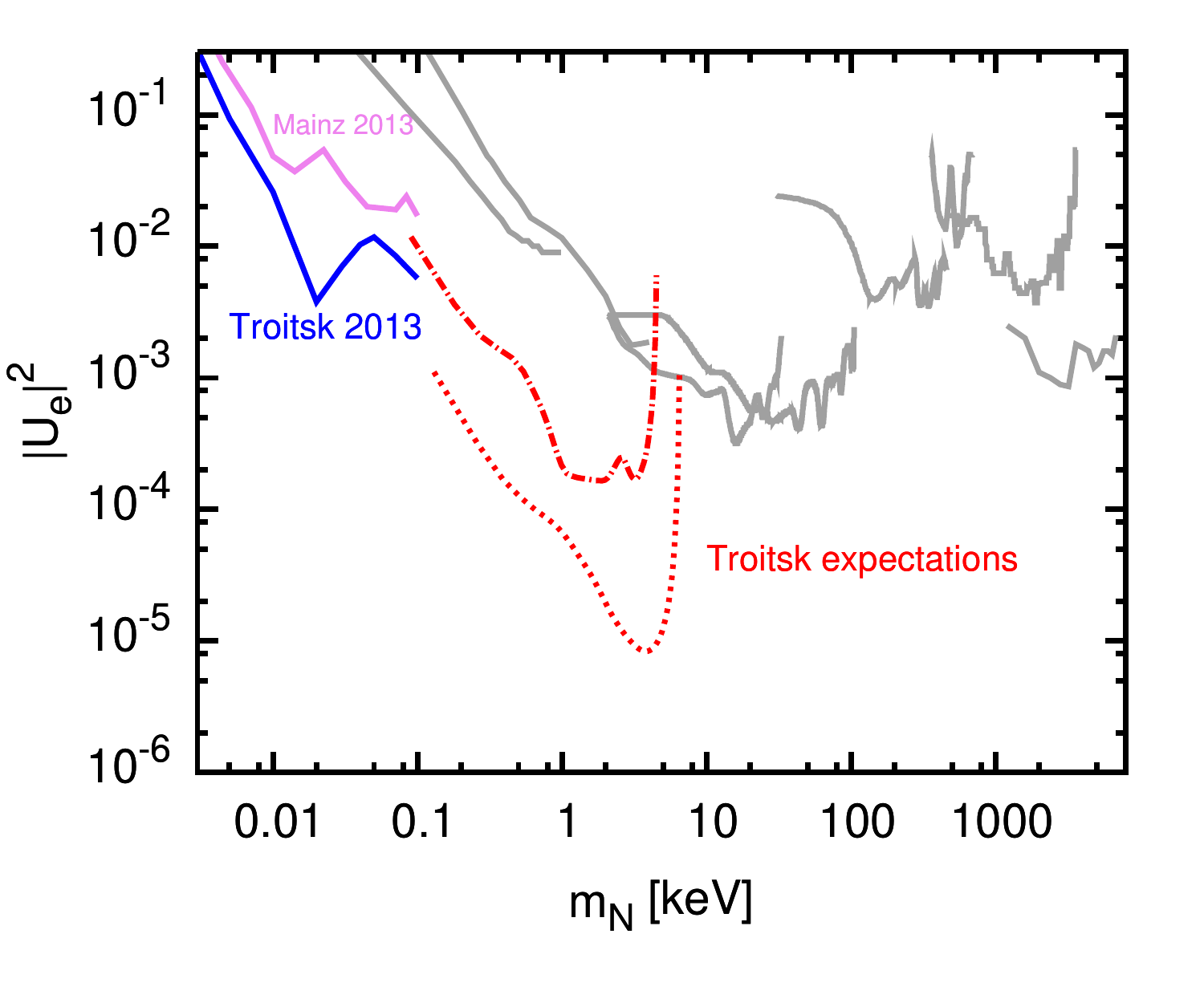}
    \includegraphics[width=3in]{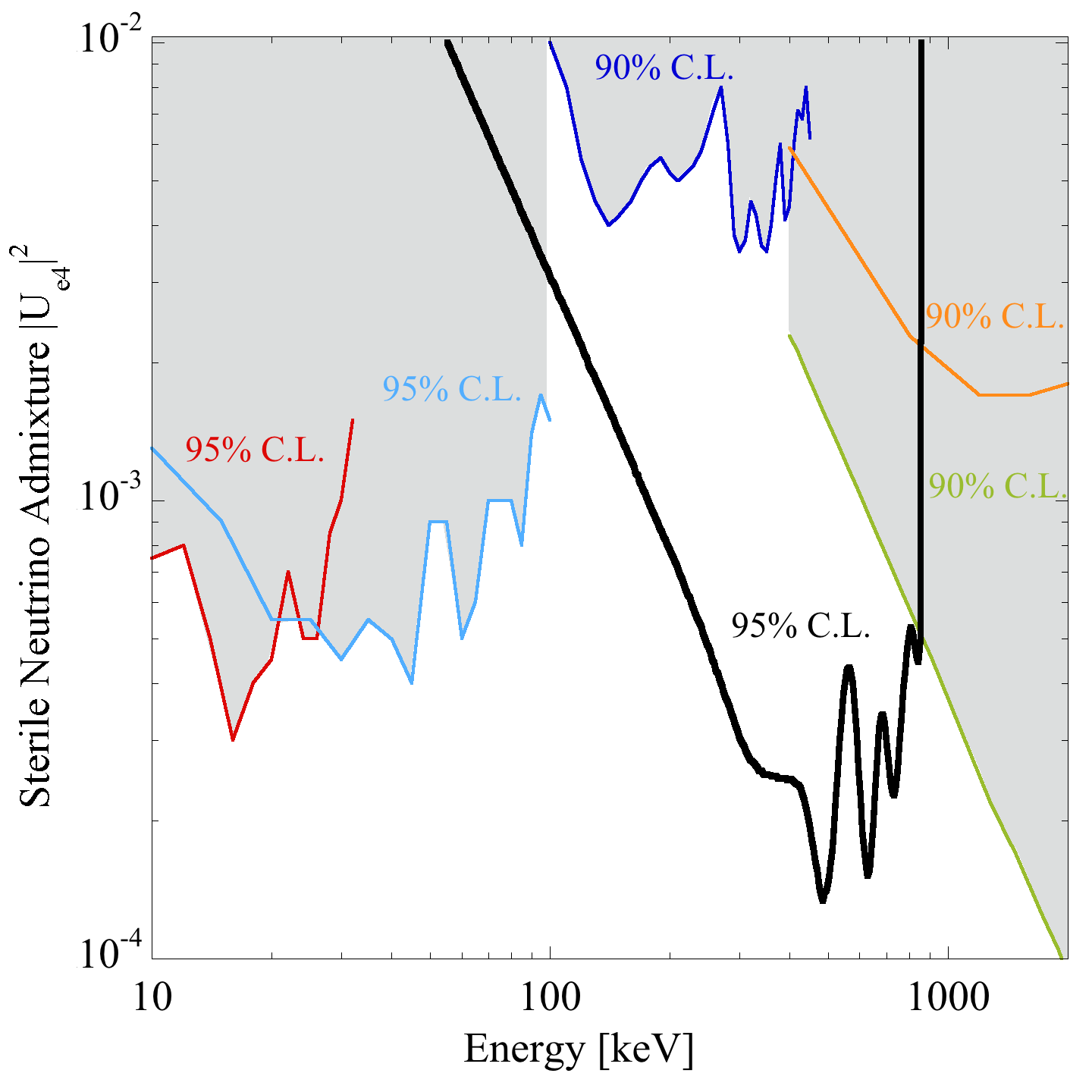}
    \caption{Limits on admixtures $|U_{e4}|^2$ of a single sterile neutrino as a function of the neutrino's mass. Admixtures above the curves are excluded.  The left panel is from \cite{Abdurashitov:2015jha}, and the right from \cite{friedrich2020limits}.  Sources of the data are given in the original publications.}
    \label{fig:sterilelimits}
\end{figure}

\begin{figure}[htb]
    \centering
    \includegraphics[width=4.25in]{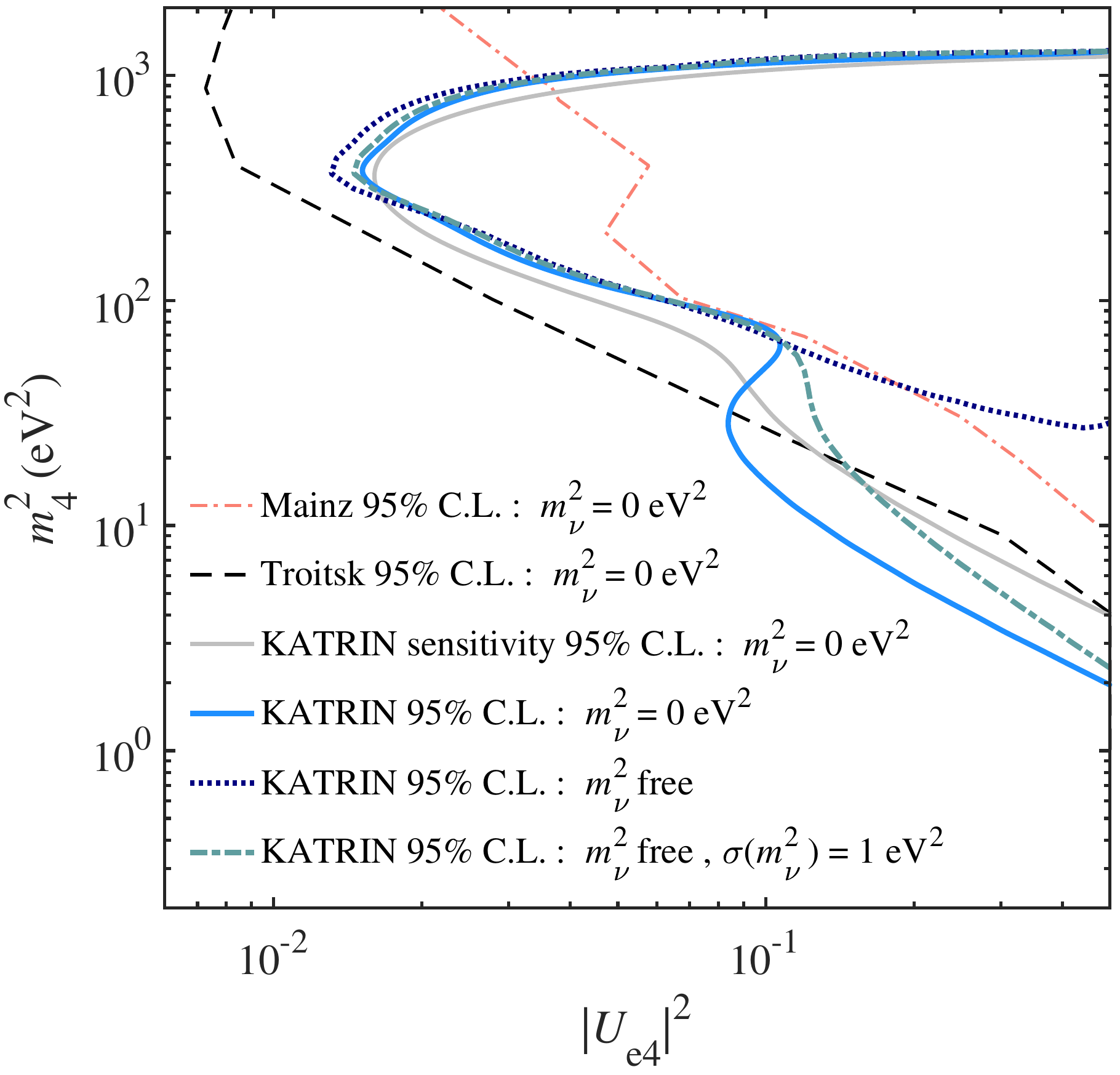}
    \caption{Exclusion curves at the 95\%~confidence level in the
($|U_{e4}|^2$, $m^2_{4}$) plane obtained from the analysis of KATRIN data including statistical and systematic uncertainties \cite{aker2020sterilebound}. The two solid lines show the expected sensitivity (light grey) and the associated exclusion (blue online) for fixed $m_\beta^2=0$~eV$^2$). The dotted line  (dark blue online) illustrates the exclusion curve obtained with a free $m_\beta^2$). The dot-dashed line (turquoise online) displays the intermediate exclusion curve with $m_\beta^2$ introduced as a Gaussian pull term  with a central value of 0 and an uncertainty $\sigma (m_\beta^2)$~=~1~eV$^2$.
Also shown are the Mainz~\cite{Kraus:2012he} and  Troitsk bounds~\cite{Belesev_2013}.  In the figure, $m_\nu^2\equiv m_\beta^2$.}
    \label{fig:katrinsterile}
\end{figure}

More recent interest in sterile neutrinos has largely been motivated by oscillation searches that have produced results that are either inconsistent with the 3-neutrino picture, or inconsistent with another observable  such as the reactor antineutrino flux.  Recent reviews may be found in Refs.~\cite{hagstotz2020bounds,Boser:2019rta,Giunti_2019}.  Direct measurements  give no indication of sterile neutrino admixtures.  The most stringent published limit contour  from tritium beta decay  comes from the Troitsk experiment \cite{Abdurashitov:2015jha,Belesev_2013}, and \cite{Abdurashitov:2015jha} also includes a summary of limits from other isotopes.  Figure~\ref{fig:sterilelimits} displays the existing limits from direct searches in beta decay.
The published KATRIN data on active neutrinos \cite{Aker:2019uuj} have been used to derive limits on sterile admixtures, but such analyses are not rigorous, leaving out correlation effects. The KATRIN collaboration has released their own analysis \cite{aker2020sterilebound}, depicted in Fig.~\ref{fig:katrinsterile}, that takes such effects into account.
The background in KATRIN precludes disentangling the effect of a sterile neutrino and an unconstrained active neutrino for masses below 10 eV.  On the other hand, the statistical power of KATRIN opens large new regions of parameter space in which to search for sterile neutrino admixtures in the mass range $>10$ eV.  In order to be able to take advantage of the available tritium source strength, a new, highly segmented, high-resolution focal-plane detector for KATRIN is being developed \cite{Brunst:2019aod,Altenmuller:2018nuw,Mertens:2018vuu} by the TRISTAN collaboration.  If the technical initiatives are successful, KATRIN would be able to reach unprecedented $|U_{e4}|^2$ levels, potentially as low as $10^{-8}$ \cite{Mertens:2014nha}.  Tiny admixtures of neutrinos in the keV mass range are particularly interesting because they are not excluded by astrophysical bounds, could be dark matter, and could help explain why supernovae explode in nature but rarely in computers \cite{Hidaka:2007se}.

It is also possible to search for sterile neutrino admixtures in electron-capture decays.  In that case, since  capture in atomic subshells produces nominally monoenergetic neutrinos,  a precision measurement of the recoil energy by calorimetric means can reveal the admixture of massive neutrinos.  Friedrich \etal \cite{friedrich2020limits} have carried out such a measurement for $^7$Be embedded in a superconducting tunnel junction, obtaining the most sensitive limits by the direct method, as low as $|U_{e4}|^2<2\times 10^{-4}$,  on the admixture of neutrinos in the mass range 100 - 1000 keV.  Their results are shown as the unshaded curve in the right-hand panel of Fig.~\ref{fig:sterilelimits}.

\subsection{Relic Neutrinos}

\begin{figure}[hbt]
    \centering
    \includegraphics[width=4in]{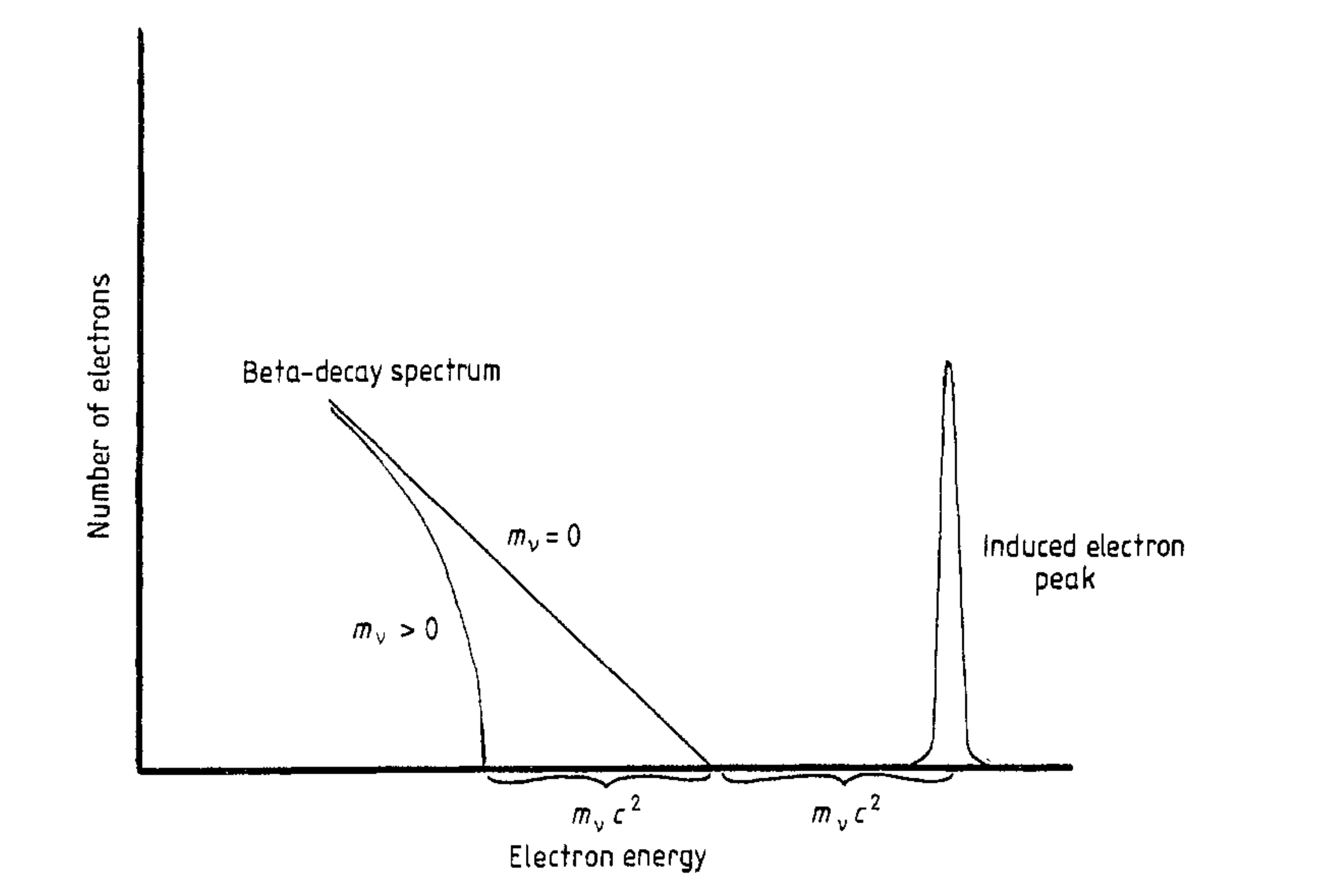}
    \caption{Spectral features from tritium beta decay and neutrino capture on tritium.  From \cite{Irvine_1983}.}
    \label{fig:relic}
\end{figure}

As Fisher has commented, ``Every neutrino physicist has wondered at some time if there might be a way to detect the relic neutrino background" \cite{fisherp}.  While not strictly a topic within our purview, since Weinberg's seminal paper in 1962 \cite{Weinberg_PhysRev.128.1457} the method of choice for such considerations has always been neutrino capture observed as a peak just beyond the endpoint of a tritium beta decay spectrum, separated from the extrapolated endpoint by a an amount of order $m_\beta$ (see Figure~\ref{fig:relic}).  It is a potential byproduct of every direct mass measurement via beta decay. 

The relic neutrino background is cold today, with a temperature of 1.95 K \cite{Hannestad_2004}, and therefore the kinetic energy is negligible in comparison to $m_\beta$.  Cosmology gives the mean relic neutrino density in the universe, 56 cm$^{-3}$ per neutrino flavor and chirality \cite{Ringwald:2004np}.  The contribution $\Omega_\nu$ of neutrinos to the closure density of the universe is \cite{Hannestad_2004}:
\begin{eqnarray}
\Omega_\nu h^2&=&  \left(\frac{\Sigma}{92.5 \rm\ eV}\right)
\end{eqnarray}
where $h$ is the Hubble constant in units of 100 km s$^{-1}$ MPc$^{-1}$.  The  rate of neutrino capture on a nuclear target depends on the neutrino density and its flavor content, the nuclear  matrix element and Q-value, and the target mass.  Betts \etal \cite{betts2013development} find a capture rate of 951(3) events per year per kg of tritium for the mean universal density modeled as a Fermi-Dirac distribution throughout space.

The local neutrino density at earth will be larger than the mean density in the universe because, having mass, neutrinos will be drawn gravitationally into galaxies.  Simulations of the overdensity to be expected (see, for example, \cite{Ringwald:2004np,zhang:naturecomm2018,de_Salas_2020})
 show that the neutrino contrast scales approximately as the square of the mass but is relatively small \cite{Ringwald:2004np,zhang:naturecomm2018}:
\begin{eqnarray}
\frac{n_\nu}{\overline{n}_\nu}-1&=& 76.5\left(\frac{m_\nu}{\rm eV}\right)^{2.21},
\end{eqnarray}
where $n_\nu$ and $\overline{n}_\nu$ are the local and universal neutrino densities, respectively.

Experimentally the difficulty is that, tritium being a shortlived radioactive isotope, the target mass is limited. In KATRIN, the largest tritium neutrino mass experiment, the mass under observation is about 100 $\mu$g, and the expected capture rate is roughly 1 every 10,000 y.  The PTOLEMY project, whose primary goal is to measure big bang relic neutrinos, is designed to put 0.1 kg, 1 MCi,  under observation \cite{betts2013development,Betti_2019}.  To achieve sensitivity for such a large target mass, PTOLEMY plans to use large surface area targets with atomic tritium adsorbed on graphene, so as to reduce charge accumulation effects.  The feasibility of using tritium-loaded graphene is currently under investigation.  An experiment with such a tritium source under observation would potentially also be capable of a sensitive measurement of neutrino mass.

\section{Spectrum Refinements}
\label{sec:refinements}
\setcounter{equation}{0}

We are accustomed to the idealized appearance of the beta spectrum near the endpoint, represented by Eq.~(\ref{eq:simplebetaspectrum}) and the corresponding figure (Fig.~\ref{fig:diffspecendpoint}, Fig.~\ref{fig:relic}).  In practice, however, one never deals with the decay of an isolated  nucleus of tritium or any other isotope, and the surrounding electrons and atoms can be excited in the decay.  These excitations, referred to loosely as the ``final-state distribution'' (FSD) reduce the energy available to the leptons. The excitations are not a result of electron energy loss in a surrounding medium following the decay, an effect that must also be considered, but occur simultaneously with the decay because the initial and final states are both complex quantum systems with many internal degrees of freedom. Thermal excitations in the initial state may even increase the lepton energy.  Nor can one escape these considerations by turning to electron capture, as will be discussed below.  Bergkvist, in his pioneering experiments on tritium (see \cite{bergkvist:1971,bergkvist:1972aa,bergkvist:1972ab} and Fig.~\ref{fig:Bergkvist}), was the first to point out that progress beyond the 55-eV limit he obtained would demand an understanding of the FSD.

There are also spectrum corrections that arise in the nuclear decay itself. We discuss those below, finding that most are negligible for the purposes of neutrino mass measurement, with the exception of radiative corrections.

\subsection{Final State Distributions}
\label{sec:fsd}

To include the final (and initial) states, a spectrum in the form of Eq.~(\ref{eq:mother}) for each initial-state $j$ to each final-state $k$   is represented in the total spectrum, 
weighting each transition by a matrix element $W_{kj}$.  Omitting small recoil-order corrections, the spectral density for each transition becomes \cite{Bodine:2015sma}
\begin{eqnarray}
\left(\frac{d\Gamma}{dE_e}\right)_{kj} &=& C F(Z,\beta)\left|W_{kj}\right|^2\beta E_e^2(\Delta_{kj}-E_e)^2\times \nonumber \\
& & \times \sum_{i}|U_{ei}|^2\left[(1-\frac{m_{i}^2}{(\Delta_{kj}-E_e)^2}\right]^{1/2}\Theta(\Delta_{kj}-E_e). \label{eq:betaspectrum}
\end{eqnarray}
This expression is written in terms of the total electron energy, $E_e=E+m$.
  The energy $\Delta_{kj}$ is the maximum energy available for each transition.
 An expression for the matrix element $W_{kj}$ is given by Jonsell, Saenz, and Froelich~\cite{jon99}. The transition matrix element for a final-state molecular ion excitation $k\equiv (v_{(f)}, J_{(f)}, M_{(f)}, n_{(f)})$ from an initial molecule state $j\equiv (v, J, M, n)$ may be written,  
\begin{eqnarray}
\label{eqn:overlap}
\left|W_{kj}(K)\right|^2 &=&\left|\int\left[\chi_{v_{(f)} J_{(f)}M_{(f)}}^{n_{(f)}}(\bf R)\right]^*S_n(R)e^{i{\bf K\cdot R}}\xi_{vJM}^{n}  ({\bf R})d^3R\right|^2.
\end{eqnarray}
In this expression, $\chi$ and $\xi$ are the rotational-vibrational wave functions of the final-state molecular ions and initial-state  molecules, respectively, and $S_n(R)$ is an electronic overlap integral.   The exponential of the dot product of the recoil momentum ${\bf K}$ and the nuclear separation ${\bf R}$ is a consequence of the recoil motion of the daughter nucleus, $^3$He in the case of tritium decay.  It may be seen from the form of Eq.~(\ref{eq:betaspectrum}) that because each transition has a slightly different endpoint energy, the shape of the spectrum in the region where one seeks evidence of neutrino mass is strongly modified by the final-state distribution.

\subsubsection{Tritium}
The ITEP result reporting a 30-eV neutrino mass with a source of tritiated valine (an amino acid) precipitated a substantial theoretical effort to calculate the FSD but in the end it was clear that such molecules are too complex.  The 30-eV result either arose from shortcomings in the FSD theory for valine, or from inexact energy-loss corrections.  Evidence  emerges from the fact that these  affect both the neutrino mass and the extrapolated endpoint energy, and the latter can be checked experimentally via mass spectroscopy \cite{Staggs:1989zz}.  An early measurement of the T -- $^3$He mass difference indeed seemed to support the FSD and energy-loss theory, but was afflicted with systematic errors that were controlled in further experiments.  The valine experiment did not give the correct endpoint energy.

The Los Alamos group turned to atomic and molecular tritium as nearly ideal sources.  Molecular hydrogen had been a topic of theoretical interest since the studies by Cantwell in 1956 \cite{cantwell:1956aa}.  In principle, since the electromagnetic potential is known exactly, the calculations can be carried out to any desired accuracy, but in practice even this simple molecule is very challenging theoretically.  In parallel with the development of gaseous tritium experiments at Los Alamos \cite{robertson:1991aa} and Livermore \cite{stoeffl:1995aa}, a number of new calculations of the molecular FSD were undertaken, and the work of the Quantum Theory Project \cite{fackler:1985} was used to interpret the results.  Both experiments  yielded negative fit values for $m_\beta^2$, and it was not until 2015 that this was traced \cite{Bodine:2015sma} to inadequacies in the FSD. 
\begin{figure}[hbt]
    \centering
    \includegraphics[width=5in]{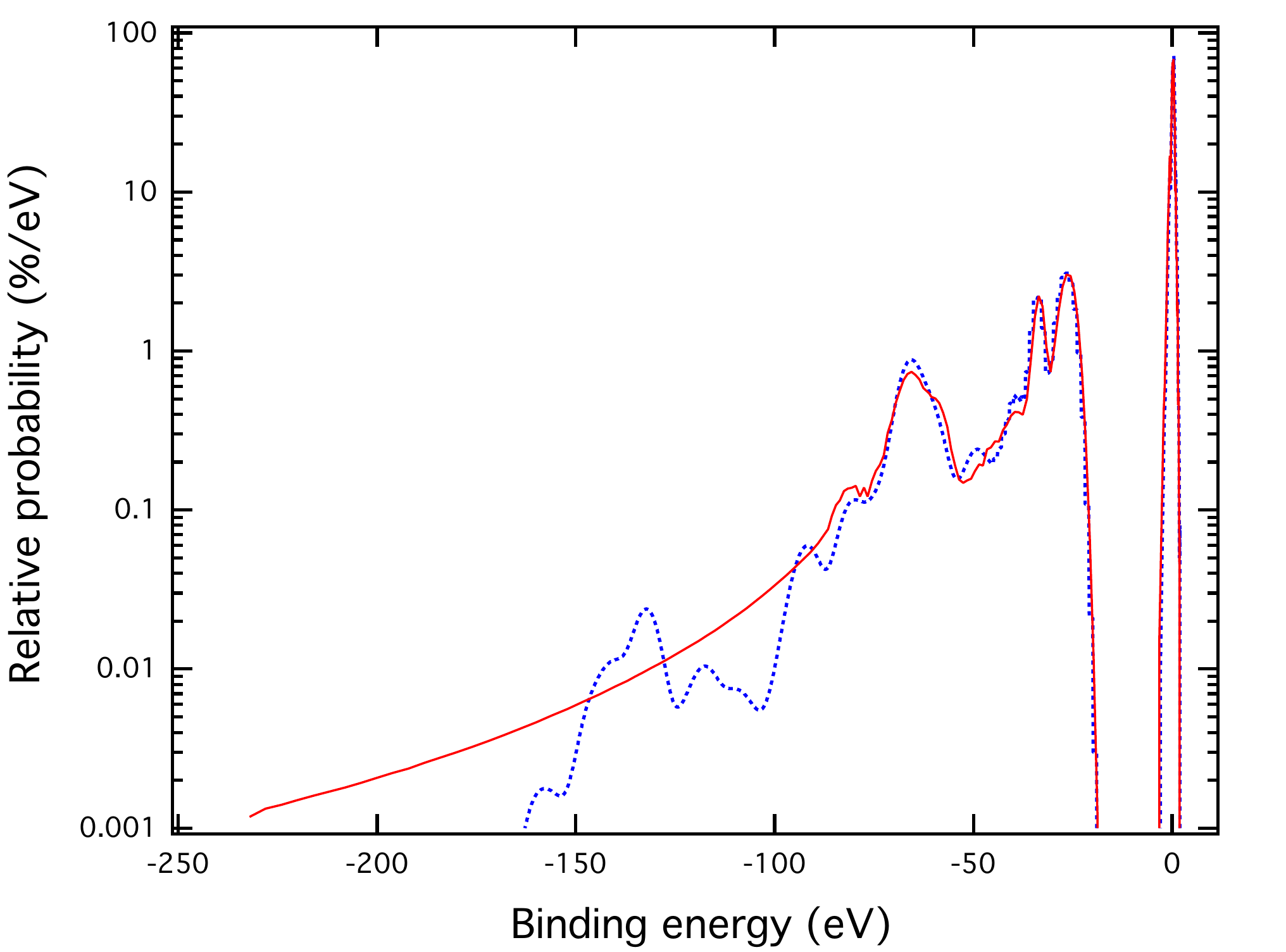}
    \caption{Comparison of the FSD spectra for the T$_2$ molecule calculated by Fackler {\em et al.} \cite{fackler:1985} (dashed blue line) and Saenz \cite{saenz00} (solid red line).  From \cite{Bodine:2015sma}.}
    \label{fig:facklersaenz}
\end{figure}

Theoretical work continued apace in the 1990 -- 2005 interval with significant progress.  Detailed assessments of the status can be found in \cite{Bodine:2015sma,Kleesiek:2018mel}. One of the most difficult aspects of the problem is the continuum, wherein one or more of the molecular electrons are ejected during the decay.   In the model of Fackler {\em et al.}~\cite{fackler:1985} continuum states have discrete energies like bound states.  An important advance by Froelich \etal \cite{froelich:1993,saenz00} was the development of `complex scaling' wherein the radial variable of continuum wave functions is made complex.  Another major advance was the inclusion of nuclear motion by Saenz and Froelich in 1997 \cite{saenz:1997ii}.    The two FSD spectra are shown for comparison in Fig.~\ref{fig:facklersaenz} and in the figure the discrete states of the Fackler \etal spectrum in the continuum have been arbitrarily assigned a 3-eV standard deviation for display and comparison with the  calculation of Saenz \etal \cite{saenz00}.  The complex scaling approach directly gives realistic distributions in the continuum, joining smoothly (with a modest order-unity normalization) to the Levinger shakeoff distributions \cite{PhysRev.90.11} that are analytic for hydrogenic atoms. 
 Table \ref{tab:moments} lists the first three moments of the binding-energy distributions for the two theories. 
\begin{table}[tb]
\centering
\caption{Comparison of zeroth, first, and second moments (columns 3, 4, and 5, respectively) of theoretical final-state distributions (see \cite{robertson:1988aa,Bodine:2015sma}).}
\label{tab:moments}
\medskip
\begin{tabular}{ccccc}
\hline
Reference & \phantom{aa} Energy range  \phantom{aa}  &  \phantom{aa} $\sum_k\left|W_{k0}\right|^2$  \phantom{aa} &  \phantom{aaa} $\langle b_{(f)k}\rangle$  \phantom{aaa} &  \phantom{aaa} $\sigma_b^2$  \phantom{aaa} \\
&    eV  &  &    eV &  eV$^2$ \\
\hline
Fackler {\em et al.} \cite{fackler:1985} & 0 to 165 & 0.9949 & -17.71 & 611.04 \\
Saenz {\em et al.}  \cite{saenz00} & 0 to 240 & 0.9988 & -18.41 & 694.50 \\
\hline
\end{tabular}
\end{table}
  
\begin{table}[htb]
\centering
\caption{Neutrino mass squared extracted from two experiments, in one case with the original 1985 theoretical calculations of the FSD and in the second case with a more modern calculation.}
\label{tab:lanlllnl}
\medskip
\begin{tabular}{lccl}
\hline
\phantom{aaaa} &  \phantom{aaa} LANL \cite{robertson:1991aa} & LLNL \cite{stoeffl:1995aa}  & \\
\hline
\multicolumn{3}{l}{{\bf As published.} Theory: Fackler {\em et al.} \cite{fackler:1985}} &  \\
 \phantom{aaa} $m_\beta^2$ & \phantom{aaa} -147(79) & -130(25) &   eV$^2$ \\
\hline
\multicolumn{3}{l}{{\bf Re-evaluated.} Theory: Saenz {\em et al.} \cite{saenz00}} &  \\
 \phantom{aaa} $m_\beta^2$ &  \phantom{aaa} 20(79) & 37(25) &   eV$^2$ \\
\hline
\end{tabular}
\end{table}
The data for the Los Alamos and Livermore experiments are no longer available, but it is possible to estimate the changes that would result had the theory of Saenz \etal  been used instead of the Fackler \etal theory, with the aid of the following relationship between an error in the variance of the FSD or other resolution contribution and the consequent error in the neutrino mass squared \cite{robertson:1988aa}:
\begin{eqnarray}
\Delta m_\beta^2 & \simeq & -2 \Delta \sigma_{\rm FSD}^2. \label{eq:errorest} 
\end{eqnarray}
  The results are shown in Table  \ref{tab:lanlllnl} with the results of the LANL and LLNL experiments as originally reported, both having been analyzed with the theory of Fackler {\em et al.} \cite{fackler:1985} and re-evaluated via Eq.~(\ref{eq:errorest}) with the theory of Saenz \etal \cite{saenz00}.   The large negative value of $m_\beta^2$ is eliminated in both experiments, subject to the limitations of Eq.~(\ref{eq:errorest}).   These results provide a striking, and essentially `blind', measure of experimental confirmation of the calculations of Saenz {\em et al.}, especially in the difficult regime of electronic excited states.

Doss \etal  in 2006 \cite{doss06} carried out a calculation in the same geminal basis used by Froelich \etal\  Their results are the same as  those obtained by Saenz \etal \cite{saenz00} up to 40 eV excitation, but should not be used above that point owing to inclusion of only the 6 lowest electronic bound states.  The effects of nuclear motion above 45 eV in a 2008 publication (Doss \etal \cite{doss:2008}) also appear anomalously broad in their treatment, and it is therefore recommended to use the calculations of Saenz \etal whenever the full spectral range is needed \cite{sibille:2020}. 

Theoretical advances notwithstanding, the most powerful weapon to reduce reliance on the calculations is actually the statistical power of state-of-the-art experiments such as KATRIN.  If an experiment can gather sufficient data in the last 25 eV of the spectrum, then most of the uncertain aspects of the FSD, particularly the continuum, drop away.  Only the rotational-vibrational manifold of the electronic ground state remains (Fig.~\ref{fig:facklersaenz}, the peak at zero binding energy).  Quantitatively, the variance of the full FSD is 695 eV$^2$, but the variance of the ground-state group is only 0.16 eV$^2$.  Even that small variance must be known to a precision of 2\% for KATRIN to meet its goal of 200-meV neutrino mass sensitivity. 

There is no known way to measure the FSD directly (other than in beta decay itself), but several kinds of experimental verification of the FSD are possible \cite{Bodine:2015sma}.  Interestingly, the Los Alamos and Livermore experiments that covered the full range of the FSD spectrum provide direct experimental confirmation of the correctness of the variance of the Saenz FSD at approximately the 2\% level of accuracy needed by KATRIN.  Since this includes the difficult continuum region, it is likely that the rotation-vibration manifold of the electronic ground state is even more reliable. 

\begin{figure}[hbt]
    \centering
    \includegraphics[width=5in]{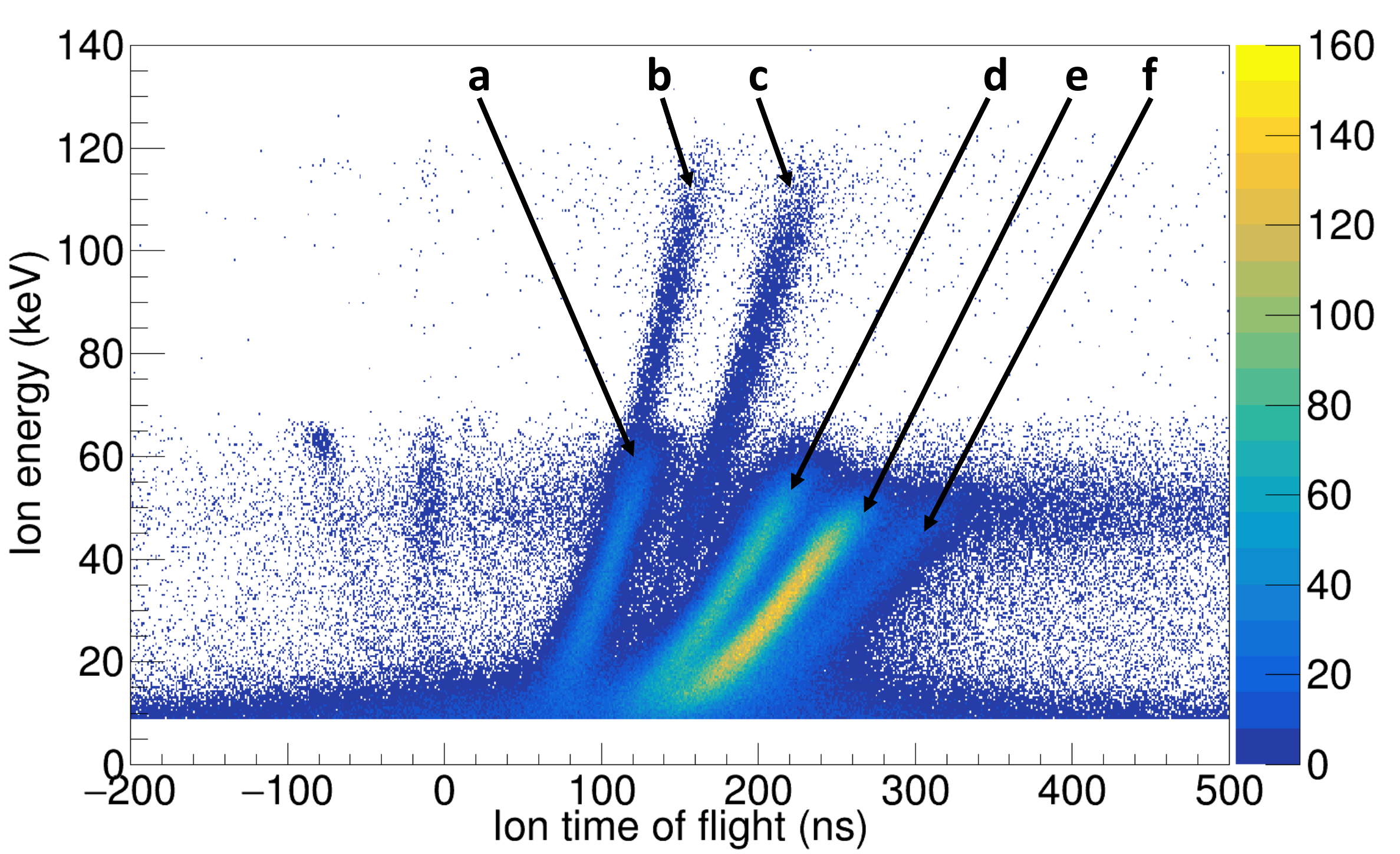}
    \caption{Experimental data from the TRIMS time-of-flight mass spectrometer experiment showing the rich complexity of ionic final states populated in the beta decay of HT. The bands originate from a) protons, b) He$^{++}$, c) H$^+ +$He$^{+}$, d) He$^+$ and T$^+$, e) HHe$^{+}$, and THe$^+$. (From \cite{Lin:2020coe})}
    \label{fig:trims}
\end{figure}

With the negative neutrino mass-squared problem resolved, one other conflict between theory and experiment remained.  When a tritium atom in the T$_2$ or HT molecule decays, the daughter $^3$He can remain bound in an ion, or can be liberated. The electronic ground state of the molecular ion THe$^+$ is bound by 1.9 eV, and corresponds to the peak near zero binding energy shown in Fig.~\ref{fig:facklersaenz}.  Rotation and vibration spread the distribution partly into the energetically unbound regime.  Two experiments carried out in the 1950s \cite{snell57,wex58} reported that the intensity of this ground-state manifold, as measured by the bound-ion fraction, was 90 -- 95\%.  The theoretical prediction, however, is only 57\%, a serious disagreement. This has now been resolved in a new experiment with a novel instrument, TRIMS (Tritium Recoil-Ion Mass Spectrometer) \cite{Lin:2020coe}. TRIMS is a time-of-flight mass spectrometer, and Fig.~\ref{fig:trims} shows data from runs with gas that is predominantly HT, with a small admixture of T$_2$.  

As shown in Table~\ref{tab:trims}, TRIMS finds close agreement with the theoretical prediction, and gives much detail about the ionic final states that can be used in further tests of the theory. 
\begin{table}[hbt]
\centering
\caption{ Branching ratio to the bound molecular ion for HT and TT.}
\label{tab:trims}
\begin{tabular}{ccccccc}
\hline \hline
   	 &  Snell {\em et al.} & Wexler & \multicolumn{3}{c}{\phantom{a}Theory \phantom{a}} & TRIMS \\   
 & \cite{snell57} & \cite{wex58}& \multicolumn{3}{c}{\cite{jon99,saenz00}} & \cite{Lin:2020coe} \\
Molecule & & & Quasibound & Bound & Total & \\
\hline 
HT  & 0.932(19)& 0.895(11) & 0.02 & 0.55 & 0.57 & 0.565(6) \\ 
TT  & -- & 0.945(6) & 0.18 & 0.39 & 0.57 & 0.503(15)\\
\hline \hline
\end{tabular}
\end{table}
In the Table, the column `Quasibound' is the part of the `Total' branching ratio to the ground-state manifold that lies above the dissociation threshold.  Many of these states have high angular momentum and therefore long lifetimes, and may travel through the TRIMS apparatus without breaking up.  The lifetimes of these states are not known.  In the case of the HT molecule, the quasibound fraction is small, and TRIMS provides the most decisive experimental test there.  Since the theory of the FSD is the same for both HT and TT, the close agreement is confirmation of the FSD for both isotopologues.

\begin{figure}[hbt]
    \centering
    \includegraphics[width=4in]{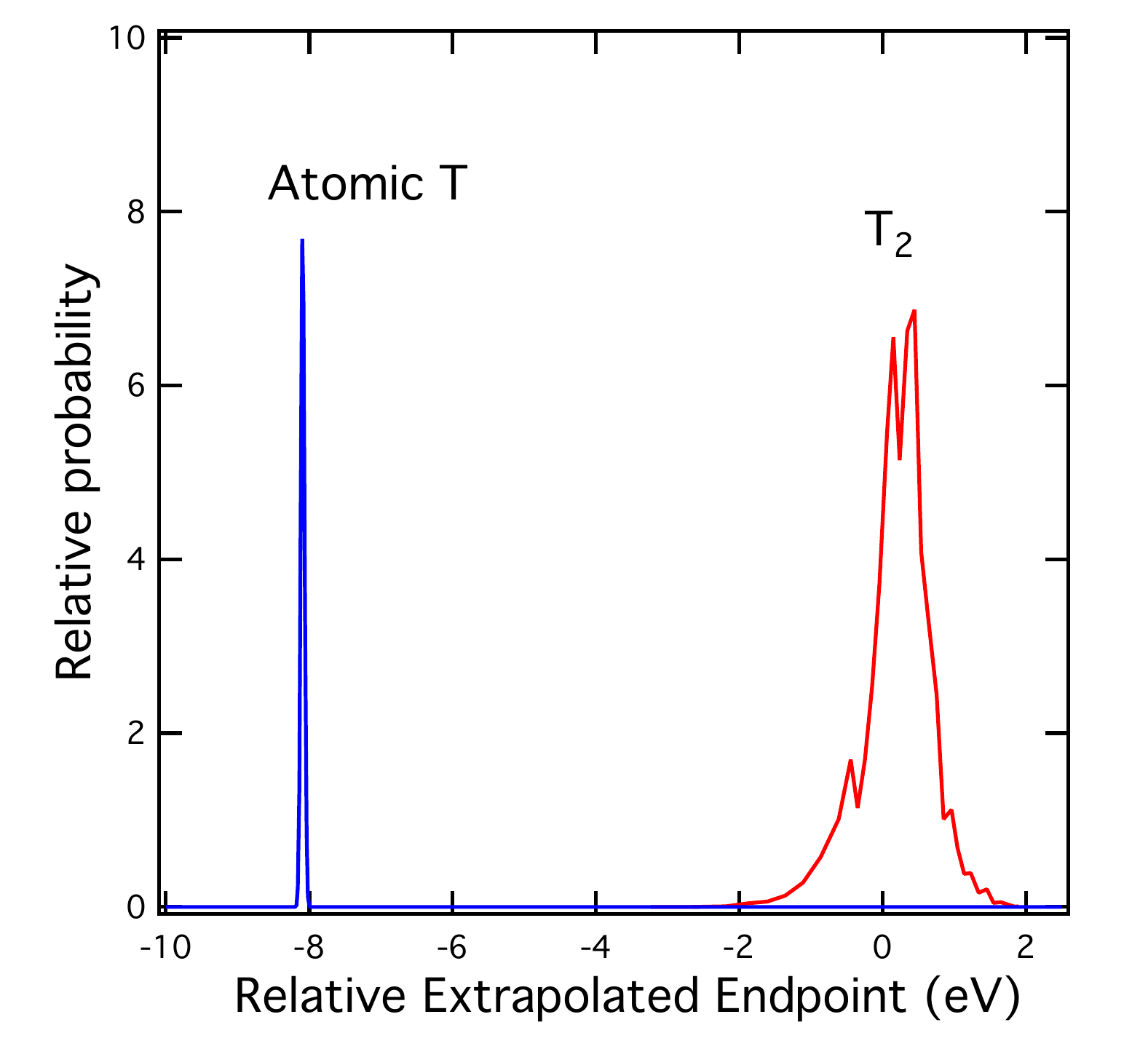}
    \caption{The ground-state manifold of molecular tritium compared to the ground state of atomic tritium.  The atomic line is here placed on the same scale as the molecular binding energy of Fig.~\ref{fig:facklersaenz}.  The ground-state Q-values differ by 8.29 eV. Because of recoil effects, the ground-state extrapolated endpoint values ($\Delta_{00}-m_e$ in \cite{Bodine:2015sma}) differ by 9.99 eV, but the molecular rotational and vibrational excitations that broaden the molecular peak make the endpoint energy for the atomic decay effectively about 8 eV smaller than the molecule.  Translational Doppler broadening corresponding to a temperature of 1 K is included in the atomic line.  The molecular line is from \cite{saenz00}.}
    \label{fig:atmol}
\end{figure}

The FSD for molecular tritium is now very robust and no longer contributes significantly to the systematic uncertainty budget even in the KATRIN experiment.  Nevertheless the broadening of the response function caused by the FSD exacts a statistical penalty and makes it more difficult to reach small neutrino masses.  The appeal of a source of atomic tritium is illustrated in Fig.~\ref{fig:atmol}.
The atomic line is very narrow, broadened only by thermal motion, and the first excited state is at -49 eV on this scale.  The FSD for atomic tritium in the sudden approximation is simply the squared overlap of the radial wavefunction $R_{nl}(r,Z)$ of the hydrogen ground state and the radial wavefunctions of the $ns$-states of He$^+$, which are analytic functions.  For example, the ground-state transition probability is $(8/9)^3=0.7023$.  More generally the transition amplitude is,
\begin{eqnarray}
T_{fi}^{(0)} &=& \int_0^\infty R_{n0}(r,Z=2)R_{10}(r,Z=1)r^2 dr \\
&=&8n\sqrt{2(n-1)!\,n!}\ \sum_{k=0}^{n-1}(-1)^k\frac{4^k(k+2)}{(n-1-k)!\,k!\,(n+2)^{k+3}}.
\label{eq:atomicstates}
\end{eqnarray}
Table~\ref{tab:atomicstates} lists the transition probabilities as calculated by Williams and Koonin \cite{Williams:1983zz}.
\begin{table}[htb]
\centering
\caption{Population of excited final states in the beta decay of atomic tritium \cite{Williams:1983zz}.}
\label{tab:atomicstates}
\medskip
\begin{tabular}{lccc}
\hline
He$^+$ state & Excitation energy (eV) & $|T_{fi}^{(0)}|^2$ & $|T_{fi}^{(0)}+T_{fi}^{(1)}|^2$  \\
\hline
1s & 0 & 70.23\% & 70.06\% \\
2s & 40.817079 & 25.00\% & 25.17\% \\
3s & 48.375798 & 1.27\% & 1.28\% \\
4s & 51.021349 & 0.38\% & 0.39\% \\
5s & 52.245862 & 0.17\% & 0.17\% \\
Continuum & 54.422773 & 2.63\% & 2.62\% \\
\hline
\end{tabular}
\end{table}
The third column is the zeroth-order sudden approximation value as in Eq.~(\ref{eq:atomicstates}), and the fourth column includes the interaction of the outgoing beta with the atomic electron.  The sudden approximation is well supported. 

The spectrum in the continuum when the atomic electron is ejected can be calculated following the method of Levinger \cite{PhysRev.90.11}, but without making the large-$Z$ approximations he used, where $Z$ is the charge on the daughter nucleus ($^3$He in this case).
The transition probability per energy interval $dW$ is:
\begin{eqnarray}
P(1s,\kappa)dW&=&64(Z-1)^3Z^{-7}(1-e^{-2\pi \kappa})^{-1}\kappa^8\exp{\left[-4\kappa\arctan{\frac{1}{x}}\right]}(x^2+1)^{-4}dW.
\end{eqnarray}
 In this expression:
\begin{eqnarray}
\kappa&=& \sqrt{\frac{E_{K(Z)}}{W}} \\
x&=&\kappa\left(1-\frac{1}{Z}\right),
\end{eqnarray}
and $E_{K(Z)}$ is the binding energy of an electron in the ground state of the daughter ion.

We return to consideration of an atomic source and experimental aspects below.

\subsubsection{Rhenium}

In the beta decay of the isotopes $^{187}$Re,  $^{115}$In, and $^{135}$Cs, the low-energy parent-progeny transitions are not allowed or super-allowed: for $^{187}$Re, it involves a first-order forbidden transition ($\Delta J^\pi = 2^-$).  Unlike its tritium counterpart, the matrix element inherits an angular momentum dependence that alters the spectral shape, including the region around the endpoint of the spectrum. The decay spectrum will incorporate contributions from the $s_{1/2}$ and $p_{3/2}$ electrons.  The decay spectrum from these two contributions (ignoring smaller recoil-order effects) can be written as~\cite{bib:forbidden, PhysRevC.83.045502}:

\begin{eqnarray}
\frac{d\Gamma}{dE} &=&  \frac{G_F^2|V_{ud}|^2}{2\pi^3} C_\beta(E) F(Z,\beta) \beta (E+m_e)^2(E_0-E) \nonumber \\&& \times \sum_{i=1,3}|U_{ei}|^2\left[(E_0-E)^2-m_i^2\right]^\frac{1}{2} \Theta (E_0-E-m_i),
\end{eqnarray}

where $C_\beta(E)$ is the shape correction factor,
\begin{eqnarray}
    C_\beta(E) =\frac{B R^2}{3} (\lambda_p(Z,\beta) p_e^2 + p_\nu^2), \\
    B = \frac{g_A^2}{6 R^2} \langle 1/2^-|| \sum_n {\bf \tau_n^+} \{{\bf \sigma_n} \otimes r_n\} ||5/2^+\rangle.
\end{eqnarray}
The shape factor itself depends on the axial-vector coupling constant $g_A$,  the coordinates of the $n$th nucleon, $r_n$ and the nuclear radius $R$.  The Fermi function is also altered, and $\lambda_{p}$ represents the ratio of a generalized Coulomb Fermi function normalized against that used in Eq.~\ref{eq:mother}.  Although the lifetime and spectral shape well below the endpoint can help constrain the uncertainties that arise from the theoretical uncertainties, they certainly complicate evaluation of the spectrum at the precision typically required for neutrino mass evaluation. Drexlin \etal \cite{Drexlin:2013lha} find that the electron term proportional to $p_e^2$ is 4 orders of magnitude larger than the neutrino one proportional to $p_\nu^2$ and also note that in calorimeters the signal is the sum of the beta energy and any electromagnetic energy from excited states.  In principle this can modify the shape of the spectrum, but the effects do not appear to be at the level of experimental significance.

\begin{figure}[hbt]
    \centering
    \includegraphics[width=5in]{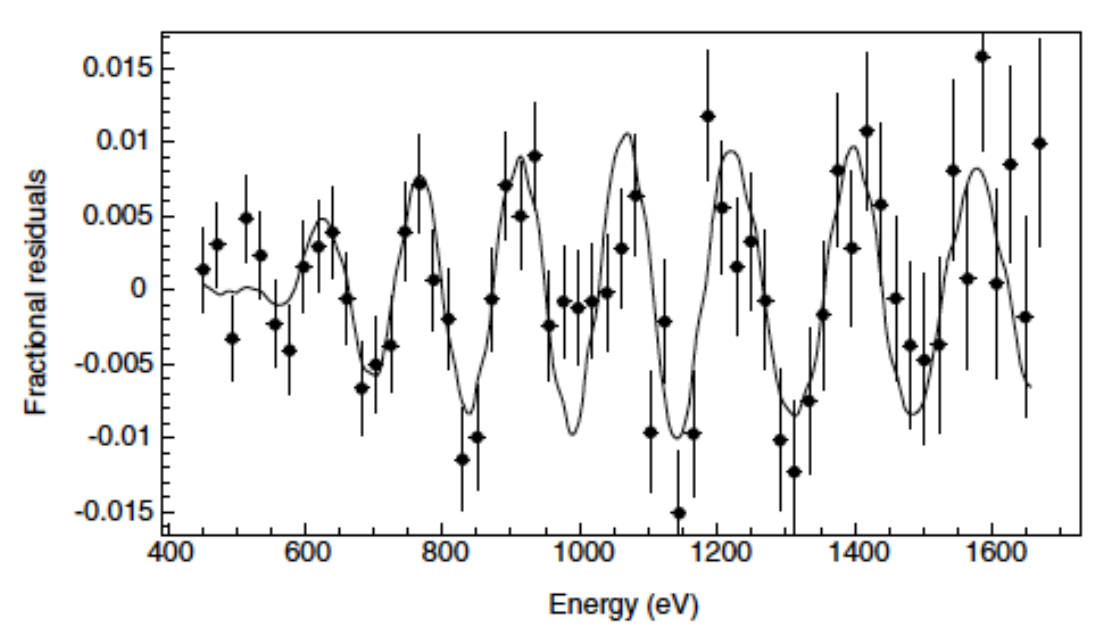}
    \caption{Experimental fraction residuals from the beta decay of $^{187}$Re, fit to the theoretical spectrum distortion from beta environmental fine structure~\cite{gatti:1999}}
    \label{fig:BEFS}
\end{figure}

The calorimetric evaluation of the beta decay spectrum from $^{187}$Re suffers from an additional complication, that of {\em environmental} alteration of the spectrum due to interactions of the emerging beta electron with the crystal structure of the absorber.  The presence of the lattice itself modifies the shape of the beta spectrum by reshaping the final-state phase space for beta electrons, producing ``beta electron fine structure'' (BEFS) in the spectrum of tritium or  $^{187}$Re  \cite{koonin:1991zz,galeazzi-PhysRevLett.86.1978,gatti:1999} embedded in a solid.   The oscillation scale is set by the momentum of the electron and the atomic separation of the crystal.  Figure~\ref{fig:BEFS} shows the oscillation spectrum as measured in rhenium crystals~\cite{gatti:1999}.  Although in good agreement with theoretical predictions, its presence introduces another complexity in a detailed understanding of the endpoint spectrum.  The spectral difficulties associated with rhenium beta decay and the total source masses necessary to achieve the  statistical accuracy for a neutrino mass measurement are among the reasons why the calorimetry groups have shifted their focus to $^{163}$Ho, as discussed in the next section.

\subsubsection{Holmium}

In electron capture, an atomic electron is captured by the nucleus and a neutrino is ejected.  Because the electronic binding energies are quantized, the neutrino spectrum nominally consists of lines of energy $Q-E_b$, the difference between the Q-value and the binding energy.  Several factors modify this simple picture.

The atomic vacancies created by electron capture have short lifetimes and refill by X-ray, Auger, and Coster-Kronig transitions.  As a result they have widths that impart to the neutrino lines a Lorentzian shape modified by phase-space restrictions in the wings.  The wings of the lines extend to the Q-value and provide a means for measuring the neutrino mass because a non-zero mass modifies the phase-space distribution near the endpoint just as it does in beta decay \cite{de-rujula:1982}.

 \adr\ and Lusignoli calculate the spectrum to be expected \cite{DeRujula:1981ti,DeRujula1982429,DeRujula:2013jba}.  Expanding the neutrino flavor eigenstate in the mass basis  the spectrum takes the following form \cite{DeRujula:1981ti,Nucciotti:2014raa,Robertson_PhysRevC.91.035504}:
\begin{eqnarray}
\frac{d\lambda_{\rm EC}}{dE_c}&=&\frac{G_F^2|V_{ud}|^2}{2\pi^3}(Q-E_c)\times \nonumber \\
&& \sum_i\left|U_{ei}\right|^2\left[(Q-E_c)^2-m_{i}^2\right]^{1/2}\times \nonumber\\
&& \sum_j\beta_j^2C_j\left|M_{j0}\right|^2\frac{\Gamma_j}{4(E_c-E_j)^2+{\Gamma_j}^2}, \label{eq:spectrum}
\end{eqnarray}
where  $\beta_j$ is the amplitude of the electron wave function at the origin, $C_j$ is the nuclear shape factor, and $E_j$ and $\Gamma_j$ are the excitation energy and natural width of atomic configuration $j$.  The `visible energy' $E_c = Q-E_\nu$, where $E_\nu$ is the total neutrino energy.  The quantity $M_{j0}$ is an overlap (monopole) electronic matrix element between the ground state of the decaying atom and state $j$ of the daughter atom.  Exchange effects \cite{Faessler:2014xpa} and orbital occupancies can be absorbed into $M_{j0}$.  

\begin{figure}[hbt]
    \centering
    \includegraphics[width=4.5in]{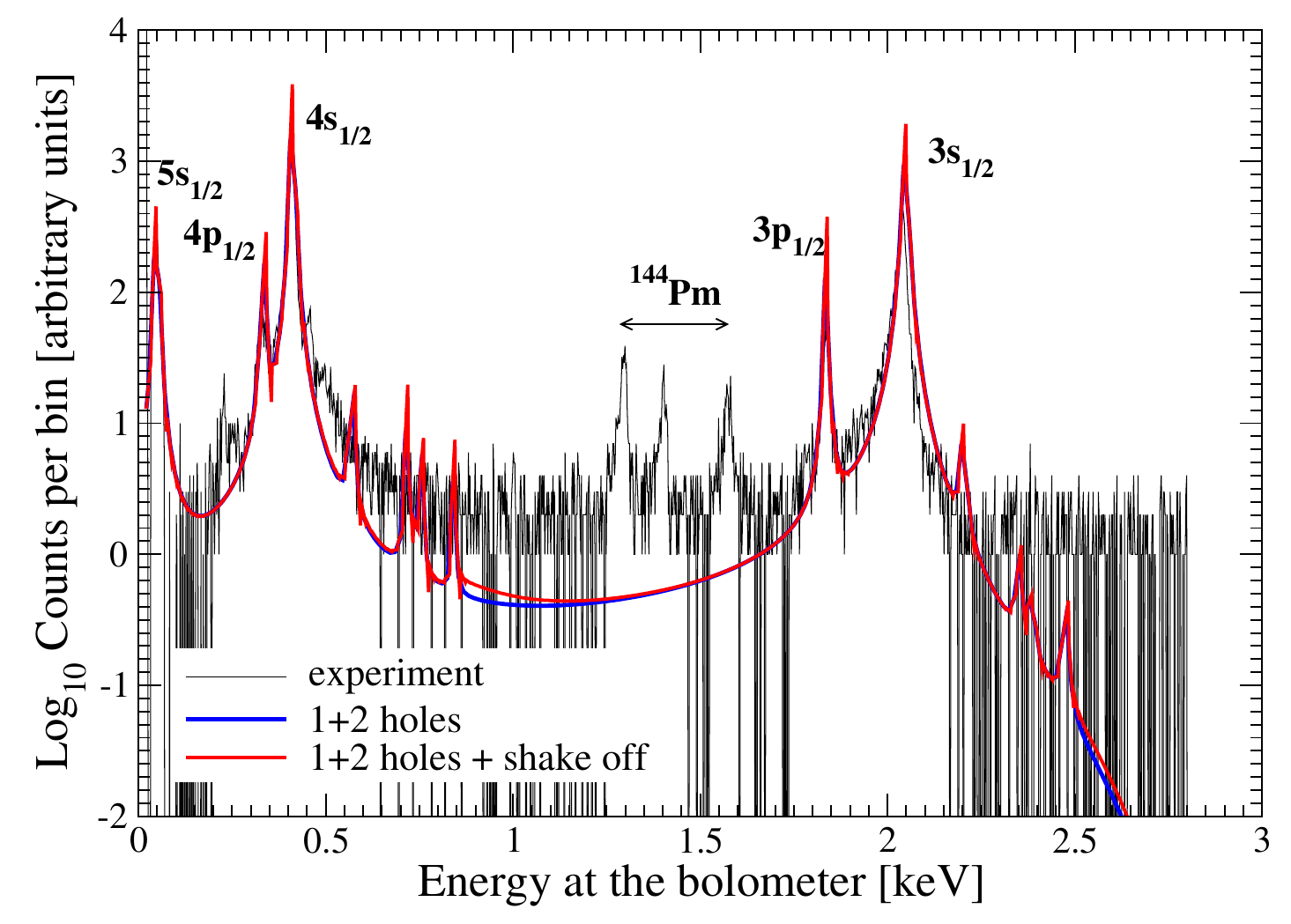}
    \caption{Calculated final-state spectrum for the electron-capture decay of $^{163}$Ho by Faessler {\em et al.} \cite{Faessler:2016hxd} (solid red, blue curves), compared to the experimental data of the ECHO collaboration \cite{echo:2014}.}
    \label{fig:faessler}
\end{figure}

\begin{figure}[hbtp]
    \centering
    \includegraphics[width=5in]{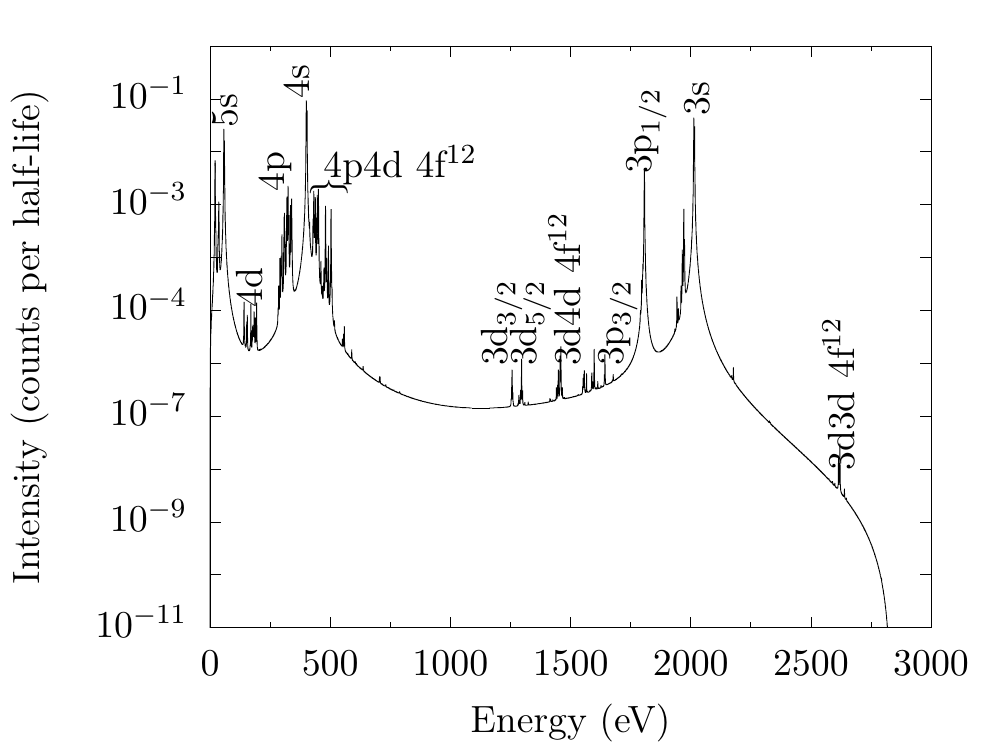}
    \includegraphics[width=4.5in]{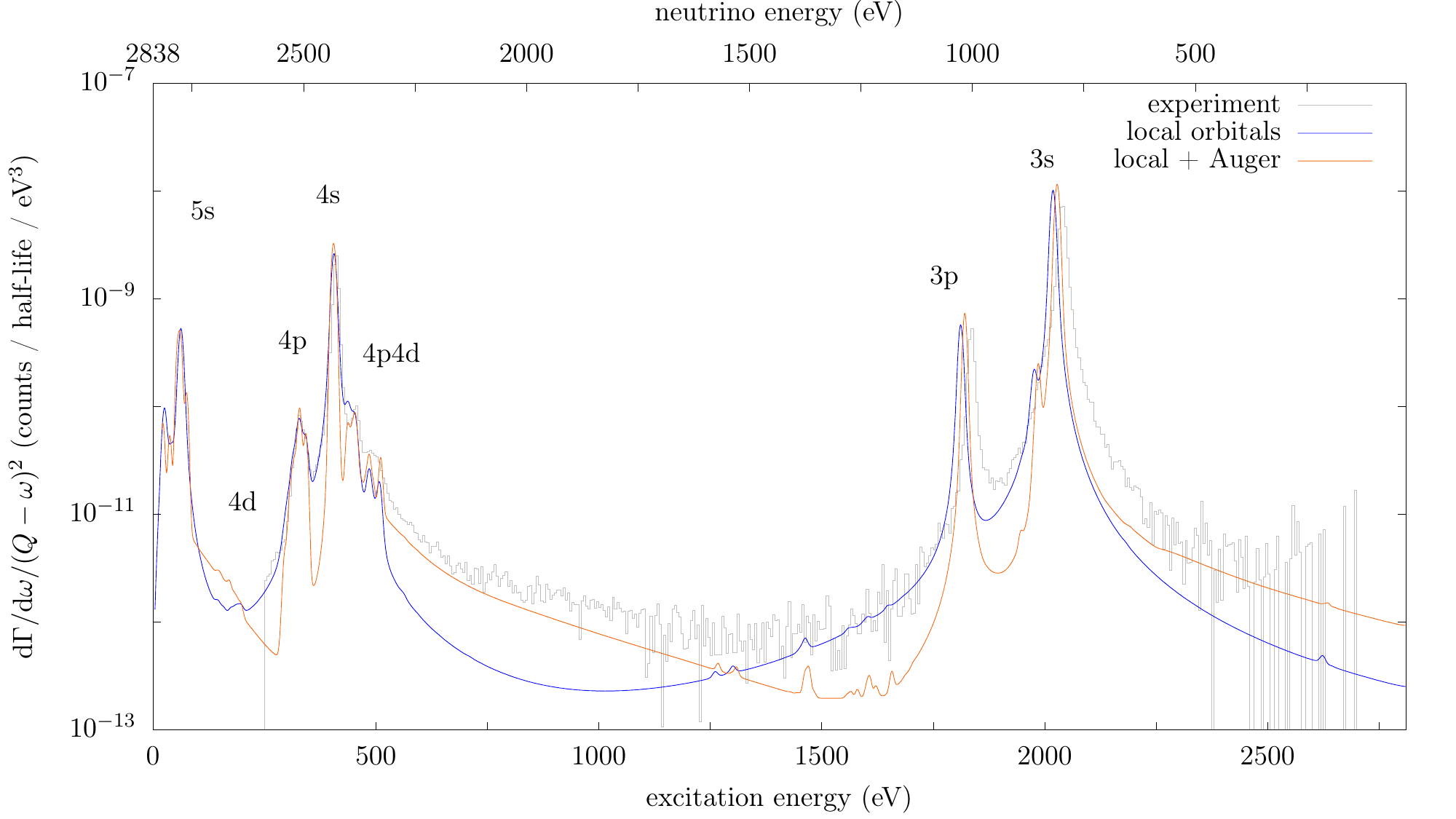}
    \caption{Top: Calculated final-state spectrum for the electron-capture decay of $^{163}$Ho by Brass {\em et al.} \cite{Brass:2017kov}.  Bottom: From Brass \etal \cite{Brass:2020spy}, the same calculation convolved with the instrumental resolution (solid curve lowest between 500 and 1000 eV excitation, blue online) and compared with the data of Velte \etal \cite{Velte:2019jvx} (solid histogram, black online).  The third curve (red online) is a modified theoretical description including Auger emission into the continuum. ($\omega \equiv E_c$ in the figure.)}
    \label{fig:haverkort}
\end{figure}

A further modification \cite{Robertson_PhysRevC.91.035504} results from the change in the nuclear charge in electron capture.  The atomic wavefunctions of Dy are not the same as those of Ho, a final-state effect that requires expanding the final state in terms of the complete set of Dy levels, both bound and continuum.  In this expansion are many multi-vacancy states with electrons in the continuum that appear as relatively weak satellite structures in the spectrum (Fig.~\ref{fig:faessler}).
The calculations have been approached in different ways by \adr\ and Lusignoli \cite{DeRujula:2016fdu},  the T\"ubingen group \cite{Faessler:2014xpa,Faessler:2015pka,Faessler:2015txa,Faessler:2016hxd}, and the Heidelberg group  \cite{Brass:2017kov} and give quite different spectra (Fig.~\ref{fig:haverkort}). 
The Brass {\em et al.} calculation \cite{Brass:2017kov} appears to give a better account of the experimental features that are emerging as the statistical accuracy of the data mounts.  In particular, the satellite structure above the 4s peak is well represented.  The bottom panel of Fig.~\ref{fig:haverkort} displays the experimental and theoretical results with the phase-space factor divided out so that the comparison near the endpoint is easier to visualize.   

At least to this two-vacancy order, both calculations find the spectrum at the endpoint to be free of satellite structures.  The previously accepted Q-value near 2500 eV raised concerns that the spectrum shape at the endpoint might be so modified by satellites that a reliable extraction of neutrino mass would be impossible.  Even though satellite peaks raise the intensity, their shapes are heavily dependent on theory.  The SHIPTRAP measurement \cite{Shiptrap_PhysRevLett.115.062501} of 2833(34) eV, spectroscopically confirmed by ECHo \cite{Ranitzsch:PhysRevLett.119.122501,Velte:2019jvx}, fortunately moves the endpoint to a region that seems to have a locally smooth energy spectrum.  However, the additional 300 eV of Q-value reduces the intensity at the endpoint substantially.  

The theoretical spectrum reproduces to an impressive degree the general features revealed by the increasingly precise experimental data, but it does not explain the observed spectrum completely.  The original concept of IBEC in $^{163}$Ho has been set to one side while these subtle quantum-mechanical aspects of the atomic physics are being worked out, and the inner bremsstrahlung components wait to be included.  They interfere coherently (with a phase that is not yet known) with the resonance tails, and can be expected to change the spectral  shape between resonances and near the endpoint.

  Another consideration is that the decaying atom is located in a potential well in the host lattice and is subject to zero-point energy that broadens the neutrino lines.  This is a line-broadening effect that is likely small but needs to be accounted for in a neutrino mass analysis.   

\subsection{Theoretical corrections to the beta spectrum shape}

A number of small effects are known to modify the basic spectrum shape given in Eq.~(\ref{eq:simplebetaspectrum}).  In a 1991 paper  \cite{wilkinson:1991zz}, Wilkinson enumerates and calculates them: the Fermi function, screening,   exchange with atomic electrons, finite nuclear size effects in both the charge and weak-interaction distributions,  radiative corrections, and a collection of 4 recoil-order effects, namely weak magnetism, V-A interference, three-body (rather than two-body) phase space and the relative motion of the electron and nuclear charge.   A recent summary and calculation of these effects has been given by Kleesiek \etal \cite{Kleesiek:2018mel}. Interestingly, as the field has advanced and experiments have become more sensitive, these effects have become even less important rather than more so, because the spectral range for investigation has shrunk to mere tens of eV near the endpoint.  

The exception is the outer radiative correction, which modifies the spectral shape near the endpoint.  Kleesiek \etal \cite{Kleesiek:2018mel} use the correction from Repko and Wu \cite{repkoradiativePhysRevC.28.2433}. Were it to be neglected, the spectral distortion near the endpoint in the KATRIN experiment would resemble a neutrino mass of 46 meV.   

\begin{figure}[hbt]
    \centering
    \includegraphics[width=5in]{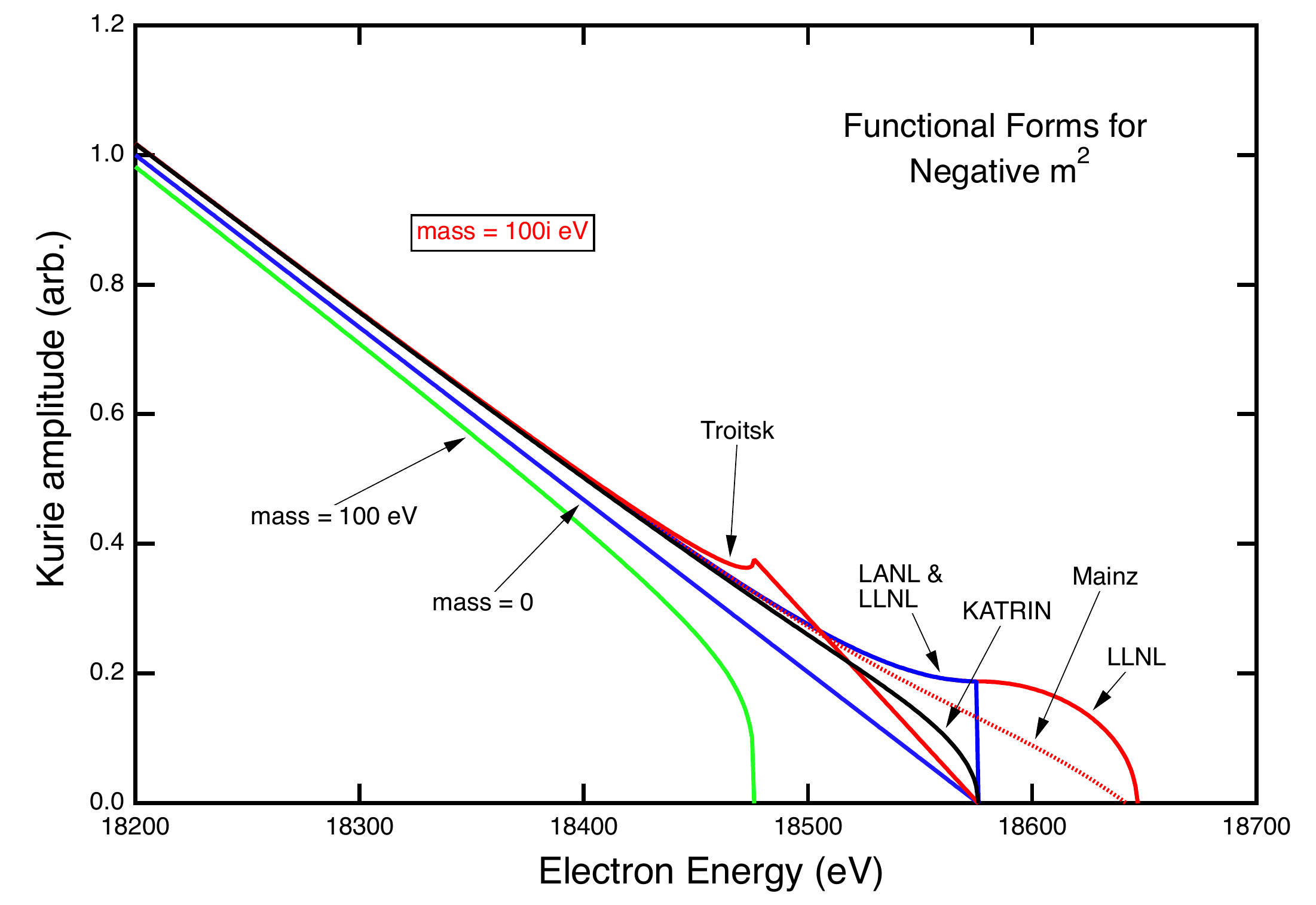}
    \caption{Illustration of the variety of functional forms chosen to continue Eq.~(\ref{eq:simplebetaspectrum}) into the negative-$m_\beta^2$ regime.  The functions are calculated for $m_\beta^2 = -10^4,\, 0,\, {\rm and}\, 10^4$ eV$^2$, and $E_0 = 18576$ eV. The Kurie amplitude is the square root of the phase-space density.}
    \label{fig:negm}
\end{figure}

We turn next to a spectral effect that arises not from subtle theoretical corrections, but from statistical considerations.  The spectrum in Eq.~(\ref{eq:simplebetaspectrum})
is valid only for positive or zero values of $m_\beta^2$.  Measurements will produce statistically too many counts in the endpoint region half the time, and too few half the time.  The analysis of data from an experiment calls for a way to handle smoothly the transition across this boundary.  For that purpose, a function must be defined that describes the spectrum for negative values of the fit parameter, $m_\beta^2$.  The function is a matter of choice for each type of experiment, but the criterion it needs to meet is that the fit parameter (for example, $\chi^2$) should display the parabolic behavior that is expected of a normally distributed quantity and that the form of the parabola matches the positive-$m_\beta^2$ regime.  Experimental groups (except KATRIN) have chosen a variety of different functions that meet this criterion in the circumstances of their apparatus and data sets, as illustrated in Fig.~\ref{fig:negm}.
The functions used by the different groups are as given in Table~\ref{tab:negmass}.  In each case,
\begin{eqnarray}
\epsilon &=& E_0 - E \\
 k^2 &=& - m_\beta^2
\end{eqnarray}
and for $m^2 < 0$ one replaces the $ \epsilon \sqrt{\epsilon^2 - m_\beta^2}\ \Theta (\epsilon-m_\beta)  $ part of the phase-space expression in Eq.~(\ref{eq:simplebetaspectrum}) by the expression in the table.
\begin{table}[htb]
    \centering
    \caption{Empirical functions used by various experimental groups to parameterize the tritium spectrum in the negative-mass-squared regime.}
    \begin{tabular}{lcllr}
    \hline \hline
    Group & Function & \phantom{aa}& Remarks & Ref.  \\
    \hline
    LANL &  $(\epsilon^2 + \frac{k^2}{2})\  \Theta(\epsilon) $  & & & \cite{robertson:1991aa} \\
    LLNL &$|\epsilon^2 +
\frac{k^2\epsilon}{2|\epsilon|}|\  \Theta(\epsilon + k)$ & & & \\
     Mainz & $\left[\epsilon +
\mu\exp(-1-\frac{\epsilon}{\mu})\right]\sqrt{\epsilon^2 +k^2}\ \Theta(\epsilon+\mu)$ &  & $\mu = 0.66k $  & \cite{Kraus:2004zw} \\
Troitsk & $[2\epsilon^2 -\epsilon\sqrt{\epsilon^2 - k^2}]\ \Theta (\epsilon) $ & & Smoothed with  & \cite{Aseev:2011dq}\\
& & & bidimensional splines. & \\
  KATRIN & $\epsilon\sqrt{\epsilon^2 +k^2}\ \Theta(\epsilon)$ & & & \cite{Aker:2019uuj} \\   
  \hline\hline
     \end{tabular}
    \label{tab:negmass}
\end{table}

The functions do not play a major role in the reporting of upper limits from experiments, which are derived from fits to positive values of $m_\beta^2$.  The reported central value and its uncertainty, however, clearly depend on the functional parameterization when the central value happens to fall in the negative-$m_\beta^2$ regime. The central values and their uncertainties are valuable for combining the results of different experiments and for assessing the statistical likelihood of a given upper limit.  The KATRIN analysis \cite{Aker:2019uuj} uses the same function as the original phase space, simply allowing $m_\beta^2$ to go negative, which gives an asymmetric $\chi^2$ function. However, the central value found is the same as when a more symmetrizing function from the Mainz group  \cite{Weinheimer:1993pd} is used, and the phase-space form was a simplification.  

For the general analysis of tritium beta decay data, a clear overview of the different methodologies, classical, unified, and Bayesian, may be found in Ref.~\cite{Kleesiek:2018mel}.

\noindent

 \section{Experimental sensitivity}
 \label{sec:sensitivity}
 \setcounter{equation}{0}
 
  It is useful to ask, what is the smallest mass
detectable in beta decay, and what experimental approach is most likely to be
fruitful?  The main aspects that enter into the answer are readily identified.  They are the statistical accuracy, the energy resolution, the background, and the systematic uncertainties.    

Three classes of instrument  can be distinguished: 
\begin{enumerate}
    \item Differential filters: A point in the spectrum is counted,
    \item Differential spectrometers:  The spectrum is counted as a whole with each event sorted by energy, and 
    \item Integral filters: The intensity above a point in the spectrum is counted. 
\end{enumerate}
Examples of these instruments are, 1) magnetic `spectrometers' (really filters) of the Tret'yakov type, 2) microcalorimeters and CRES (cyclotron radiation emission spectrometers, see Sec.~\ref{sec:future}), and 3) MAC-E filters. 

In the next section we develop a simple method for estimating the physics reach of a direct neutrino mass experiment, taking into account the distinct attributes of the instrument.

\subsection{Estimation method}
\label{sec:estimation}

The statistical sensitivity to neutrino mass is fundamentally determined by the number of events in an `analysis window' as given in Eq.~(\ref{eq:signalrate}), reproduced here:
\begin{eqnarray}
N_s 
&\simeq&rt (\Delta E)^3\left[1-\frac{3}{2}\frac{\sum_{i=1,3}|U_{ei}|^2m_i^2}{(\Delta E)^2}\right]. \nonumber
\end{eqnarray}
More realistically, if there is an additional differential background rate $b$ that is energy-independent, the total number of events in an analysis window of width $\Delta E$ is
\begin{eqnarray}
N_{\rm tot} &=& rt (\Delta E)^3\left[1-\frac{3}{2}\frac{m_\beta^2}{(\Delta E)^2}\right] +bt\Delta E
\end{eqnarray}
This background description is appropriate for differential instruments.  For integral filters, the integral background rate $b_{\rm int}$ does not scale appreciably with the size of the analysis window.  We deal first with the differential description.  

The statistical uncertainty $\sigma_{m_\beta^2} $ is thus related to the variance in the total number of events:
\begin{eqnarray}
\frac{\partial N_{\rm tot}}{\partial m_\beta^2} &=& -\frac{3rt\Delta E}{2} \\
\sigma_{m_\beta^2} &=& \frac{2}{3rt\Delta E}\sqrt{N_{\rm tot}} \label{eq:sigraw} \\
 &\simeq & \frac{2}{3rt}\sqrt{rt\Delta E + \frac{bt}{\Delta E}},  \label{eq:sig}
\end{eqnarray}
It is assumed that the neutrino mass is small compared to the width $\Delta E$.  There is an optimum choice of $\Delta E$ that minimizes the uncertainty,
\begin{eqnarray}
\Delta E_{\rm opt} &=& \sqrt{\frac{b}{r}}.
\end{eqnarray}
For this choice,
\begin{eqnarray}
 \sigma_{m_\beta^2} &\simeq&  \frac{2^{3/2}b^{1/4}}{3t^{1/2}r^{3/4}}  \\
 &\equiv & \sigma_{\rm opt}.
\end{eqnarray}
The minimum is broad -- setting the analysis window width  incorrectly by a factor $m$ results in an increase in the statistical uncertainty by a factor
\begin{eqnarray}
\frac{\sigma}{\sigma_{\rm opt}} &=& \sqrt{\frac{1}{2}\left(m+\frac{1}{m}\right)}.
\end{eqnarray}
For example, if $\Delta E$ is 2 times wider than the optimum, there is a 10\% increase in the statistical uncertainty.  As a practical matter, the ratio $b/r$ may be very small when rates are high or backgrounds low.  The optimum analysis window $\Delta E$ may then be determined by other factors, for example  instrumental broadening with a standard deviation $\sigma_{\rm instr}$, because the neutrino-mass effect on the spectrum is now smeared over this larger interval.  In turn, improving the instrumental resolution beyond a certain point is not useful if one encounters a limit set by the broadening caused by the final-state distribution (FSD).  In the decay of molecular T$_2$ to T$^3$He$^+$, the molecular final-state distribution of the ground-state rotational and vibrational manifold has a standard deviation $\sigma_{\rm FSD}\simeq 0.4$ eV~\cite{PhysRevLett.84.242,Bodine:2015sma}.    The temperature also plays a role through translational Doppler broadening $\sigma_{\rm trans}$  \cite{Bodine:2015sma}. The quadrature of the individual contributions forms a quantitative basis for fixing $\Delta E$:
\begin{eqnarray}
\Delta E &=& \sqrt{ \frac{b}{r} + C^2(\sigma_{\rm FSD}^2+ \sigma_{\rm trans}^2 +\sigma_{\rm instr}^2 + ... )}  \label{eq:eqten}
\end{eqnarray}
where $C=\sqrt{8\ln{2}}=2.35$.

The rms width $\sigma_i$ of each resolution component  is determined experimentally by some means with an associated uncertainty $u(\sigma_i)$  that propagates into the square of the neutrino mass, leading to additional, non-statistical, contributions to $\sigma_{m_\beta^2}$ of the form
\begin{eqnarray}
\sigma_{m_\beta^2}&=&\sqrt{\sum_i\left[\frac{\partial(m_\beta^2)}{\partial(\sigma_i)}u(\sigma_i)\right]^2}, 
\label{eq:thirteen}
\end{eqnarray}
 There is a simple relationship~\cite{robertson:1988aa} between an error in the width  of a resolution contribution and the corresponding error introduced into the neutrino mass:
\begin{eqnarray}
\sigma_{m_\beta^2} &\approx & -2  u(\sigma_{i}^2) \\
&=& -4u(\sigma_i)\sigma_i
\label{eqtwenty}
\end{eqnarray}
Equation (\ref{eq:thirteen}) then becomes,
\begin{eqnarray}
\sigma_{m_\beta^2}&=&4\sqrt{\sum_i\sigma_i^4\left(\frac{u(\sigma_i)}{\sigma_i}\right)^2},
\end{eqnarray}
to be combined with Eq.~(\ref{eq:sig}) to yield the total uncertainty.  This approach to estimating the sensitivity of an experiment is approximate but can be quite accurate.  Care is needed with the choice of resolution parameters.  For example, the variance of the full FSD in molecular tritium decay is very large (695 eV$^2$) but experiments with high sensitivity can focus on the last $\sim25$ eV of the spectrum where only the ground-state rotational-vibrational manifold is present. The variance of that part of the FSD is much smaller, only $\sim 0.16$ eV$^2$.  A similar caveat applies to contributions from inelastic scattering, which have a large total variance but do not set in until about 13 eV below the endpoint.  We show later the application of this ansatz to estimating the sensitivity of Project 8 and $^{163}$Ho experiments.

The main difference for integral filter instruments is that background events accumulate at an integral rate $b_{\rm int}$ independent of $\Delta E$:
\begin{eqnarray}
N_{\rm tot} &=& rt (\Delta E)^3\left[1-\frac{3}{2}\frac{m_\beta^2}{(\Delta E)^2}\right] +b_{\rm int}t.
\end{eqnarray}
In this case the optimum window is
\begin{eqnarray}
\Delta E_{\rm opt} &=& \left(\frac{2b_{\rm int}}{r}\right)^{1/3}, \label{eq:maceoptimumwindow}
\end{eqnarray}
and the statistical uncertainty becomes
\begin{eqnarray}
 \sigma_{m_\beta^2} &\simeq&  \left(\frac{16}{27}\right)^{1/6}\frac{b_{\rm int}^{1/6}}{t^{1/2}r^{2/3}}.  \label{eq:macesensitivity}
\end{eqnarray}

For the purposes of experiment planning in establishments equipped only with cocktail napkins, Eq.~(\ref{eq:sig}) provides the necessary estimations for differential spectrometers.  The 90\% CL limit that could be set on a small neutrino mass in a background-free, systematics-free experiment is
\begin{eqnarray}
m_\beta&\lesssim&\left(\frac{\Delta E}{rt}\right)^{1/4}.
\end{eqnarray}
For example, with $\Delta E = 1$ eV and a source producing 1 detectable count per day in the last eV, a limit of about 0.23 eV could be set on the mass in a year.

The rule is the same for integral filters,  except that an adjustment to the running time $t$ is required to account for the additional time needed for measuring the background, the spectrum intensity, and the endpoint energy.  This adjustment has been done by Monte Carlo for KATRIN and produces the measurement-time distribution (MTD) that assigns an optimal run time to each energy point. 

With the publication of the first neutrino mass limit from KATRIN \cite{Aker:2019uuj,aker2021analysis}, it becomes possible to compare the prediction of Eq.~(\ref{eq:macesensitivity}) to the performance of the actual experiment.  Since the MAC-E filter is a point-by-point instrument, what emerges is a `time-expansion ratio', the increase in running time needed in order to provide the additional information beyond the number of events in the analysis window (i.e. the background, spectrum intensity, and endpoint energy).  Table 
~\ref{tab:katrintime} gathers the necessary information.
\begin{table}[htb]
    \centering
    \caption{The time expansion ratio for KATRIN as derived from its first neutrino mass measurement \cite{Aker:2019uuj}.}
    \begin{tabular}{lrlr}
    \hline \hline
    Parameter & Value & Unit & Ref.  \\
    \hline
    Fraction of nuclear T events in last eV \phantom{aaa} & $2.93\times 10^{-13}$ & & Eq.~(\ref{eq:mother}) \\
    T$_2$ branch to electronic ground state & 0.57 &  & \cite{saenz00} \\
    Maximum pitch angle & 50.4 & deg & \cite{Aker:2019uuj} \\
    Detector area & 63.6 & cm$^2$ & \cite{Amsbaugh2015} \\
    Detector pixels used & 117/148 &  & \cite{Aker:2019uuj} \\
    Detector spectral region of interest & 0.95 & & \cite{Aker:2019uuj} \\
    Source column density $\rho d$ & $1.11\times 10^{17}$ & T$_2$ cm$^{-2}$ & \cite{Aker:2019uuj}  \\
    Unscattered fraction & 0.80 &  & \cite{Aker:2019uuj}  \\
    T atoms / total & 0.976 &  &  \cite{Aker:2019uuj}  \\
    Magnetic field at source & 2.52 & T & \cite{Aker:2019uuj} \\
    Magnetic field at detector & 2.52 & T &  \\
    Visible source activity & $2.46 \times 10^{10}$ & Bq & \\
    Accepted source activity & $2.68 \times 10^{9}$ & Bq & \\
    Accepted endpoint activity $r$ & $4.5\times 10^{-4}$ & Bq eV$^{-3}$ &  \\
    Background $b_{\rm int}$ & 0.293 & cps & \cite{Aker:2019uuj} \\
    Optimum window $\Delta E$ & 10.9 & eV & Eq.~(\ref{eq:maceoptimumwindow}) \\
    Statistical uncertainty in $m_\beta^2$ & 0.97 & eV$^2$ & \cite{Aker:2019uuj} \\
    Effective live time $t$ & $1.71\times 10^4$ &s & Eq.~(\ref{eq:macesensitivity}) \\
    Actual live time & 521.7 & h & \cite{Aker:2019uuj} \\
    Time expansion ratio & 110 &  &  \\
    \hline\hline
     \end{tabular}
    \label{tab:katrintime}
\end{table}
The optimum window size of 11 eV agrees well with the spectral emphasis chosen by KATRIN for data taking \cite{Aker:2019uuj}.  The time expansion ratio of 110 at the bottom of Table~\ref{tab:katrintime} is a quantitative measure of how much more time is needed to carry out an experiment with a point-by-point MAC-E filter than would be needed simply to accumulate the statistics in the optimum analysis window.  As Kleesiek \etal note \cite{Kleesiek:2018mel}, the choice of measurement time distribution for filter instruments cannot be just a numerical optimization exercise because it is necessary to obtain a spectrum rather than four numbers at strategically chosen energies.  There is  information about the quality of the data to be found in the spectrum that cannot be obtained in any other way.  The time expansion ratio is a motivation to seek spectrometric methods that allow simultaneous acquisition of data across (part of) the spectrum.  Of course, spectrometric measurements have their own time-expansion ratios too, because the parameters obtained from regions of the spectrum outside the analysis window contribute statistically, but simulations indicate that they are typically much closer to unity.  

The analysis of an actual experiment's data would be done without the approximations used in this section for illustration.

\subsection{Backgrounds}

The decay rate in the last eV of the tritium spectrum is roughly $2\times 10^{-13}$ of the total rate, and it scales as the cube of the energy interval.  Backgrounds become increasingly important as ever smaller neutrino masses are explored.  When physical detectors such as semiconductor detectors are used to register events from a spectrometer, natural radioactivity and cosmic-ray backgrounds are often the dominant source.  Every advance in sensitivity and scale brings a new and unanticipated background, and we described above the Rydberg atoms and trapped electrons from radon that have been discovered in the KATRIN background.

The method of Cyclotron Radiation Emission Spectroscopy (CRES), described in Sec.~\ref{sec:freqtechniques}, is expected to have low background because there is no physical detector with which electrons interact, and at no point are the electrons brought nearly to rest.  The registration and energy measurement of decays is carried out via microwave signals. Sources that can contribute include  cosmic-ray and radioactivity interactions with the tritium gas, false events produced by noise in the receiver chain, and noise from the abundant lower energy electrons that are inevitably present along with the few higher-energy ones near the endpoint.

A special but important case arises with a putative experiment on atomic tritium, such as that planned for the CRES experiment Project 8 \cite{Esfahani:2017dmu}.  The inevitable presence of molecular T$_2$ represents a background to the atomic spectrum near its endpoint because the Q-value of the molecular decay is 10 eV larger than for the atomic decay.  To estimate the influence of this background, the total rate in the analysis window in the absence of neutrino mass becomes
\begin{eqnarray}
N_{\rm tot}&\simeq&r_at(\Delta E)^3+bt\Delta E+r_mt[(E_m+\Delta E)^3-E_m^3],
\end{eqnarray}
where $E_m\simeq8$ eV, the effective endpoint energy difference when rovibrational excitations in the molecule are taken into account (see Fig.~\ref{fig:atmol}), and $r_a$ and $r_m$ are the rates in the last eV of the atomic and molecular spectra, respectively.  As was shown in Equation~\ref{eq:sigraw},
\begin{eqnarray}
\sigma_{m_\beta^2} &=& \frac{2}{3r_at}\sqrt{(r_a+r_m)t\Delta E+3r_mtE_m+\frac{(b+3E_m^2r_m)t}{\Delta E}}. 
\label{eq:molbackground}
\end{eqnarray}
The optimum analysis window for the background component similarly becomes,
\begin{eqnarray}
\Delta E_{\rm opt}&=&\sqrt{\frac{b+3E_m^2r_m}{r_a+r_m}}.
\label{eq:molbackground2}
\end{eqnarray}
If $b$ is small and $\Delta E_{\rm opt}$ is limited to 1 eV by other effects, the implied ratio $r_m/r_a < 0.005$.  In an experiment one would attempt to reduce the molecular contribution well below that value to deprive it of  significance. 

Attention to background is equally important in the calorimetric experiments.  The production of $^{163}$Ho in a reactor inevitably leads to co-production of other isotopes, and rigorous chemical purification followed by electromagnetic mass separation has been needed to produce essentially background-free spectra \cite{echo:2017,Velte:2019jvx}.  Control of contamination by $^{40}$K is particularly crucial because it produces a spectrum of Auger and X-ray lines around the Ho endpoint at 2.8 keV.  The airborne Rn daughter products must be controlled as in any material detector system.

\section{Future}
\label{sec:future}
\setcounter{equation}{0}

In the KATRIN experiment now in operation one finds the accumulated knowledge of decades of direct mass measurement experiments and the theoretical work on the final-state distribution of the tritium molecule. KATRIN was conceived at the time when neutrino oscillations were discovered and neutrino mass was shown to be non-zero, with a definite lower limit to the average mass.  It was designed for an order of magnitude improvement in mass sensitivity, which corresponds to four orders in simple statistical terms. The sheer scale of KATRIN dwarfs all previous experiments. KATRIN may indeed find the mass within its search range but, if not, the mass range between 0.2 and 0.01 eV remains inaccessible to laboratory experiment.  In either event, is there any approach that would allow either confirmation of a result from KATRIN, or exploration of the last remaining window?

As has been noted above, MAC-E filters are point-by-point instruments with a time expansion ratio of 100 or so.  In addition, the need to extract the electrons from the source places a limit on the source thickness, and hence also its activity.  These statistical restrictions may be susceptible to alleviation.  

The final-state spectrum has been a limitation and a source of systematic error since 1970, and motivated the search for methods in which its role is diminished or eliminated, while at the same time theoretical advances have greatly reduced the systematic uncertainties.  The low-temperature calorimetric methods used with $^{187}$Re and $^{163}$Ho arose from the quest to evade the final-state uncertainty because all the decay energy not carried away by neutrinos is delivered as heat to the lattice, collapsing the final-state distribution  into the desired delta function.  But even with that technique there are caveats, as has been described, and new kinds of final-state effects become manifest.  

In 2009 a novel idea, cyclotron radiation emission spectroscopy, was proposed by Monreal and Formaggio \cite{monreal:2009aa}.  As a spectroscopic method it can have a small, order-unity time expansion ratio, and there is no need to extract the electrons, lifting the source-size limitation.  In addition, it may make possible an experiment with atomic tritium, for which the final-state spectrum is free of uncertainty.  

We look into these new methods in more detail in this section. Promising though they are, the technical challenges are formidable.

\subsection{Calorimetric techniques}
\label{sec:calorimetrictech}

One of the difficulties inherent in the electromagnetic techniques discussed thus far, including those borne of magnetic adiabatic collimation, is that the electron needs to be transported away from the source before its energy can be measured.  In doing so, techniques need to be devised to minimize (or calibrate away) any energy losses collected in the transport process.  Calorimetric techniques potentially circumvent this fundamental limitation, since they collect {\em all} the energy released in the weak decay process (aside from that carried away by the neutrino).  As such, they offer a system with inherently different systematic uncertainties compared to those in electromagnetic methods.  Experimental calorimeters used thus far also make use of primogenitors that are distinct from tritium (mainly \isotope{Ho}{163} and \isotope{Re}{187}), providing further orthogonal checks on the validity of any potential positive mass signal.

\begin{figure}[htb]
  \begin{center}
  \begin{tabular}{c c}
  \includegraphics[width=3in]{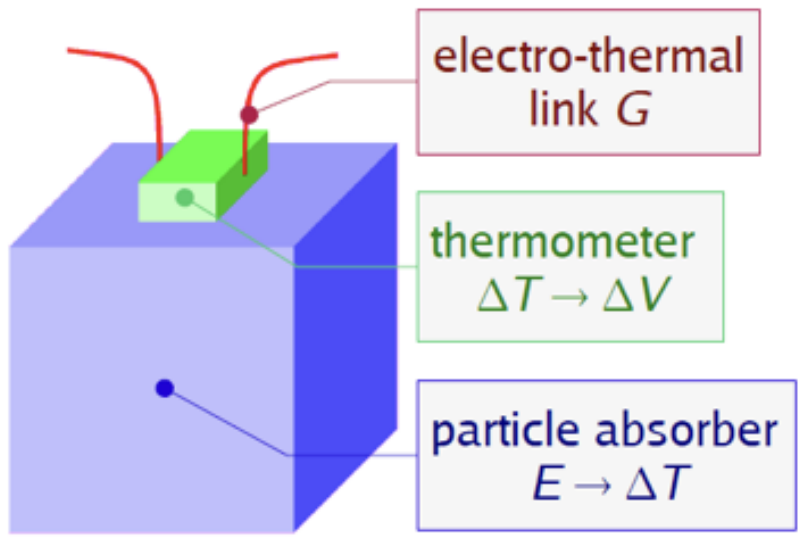} &  
  \includegraphics[width=2in]{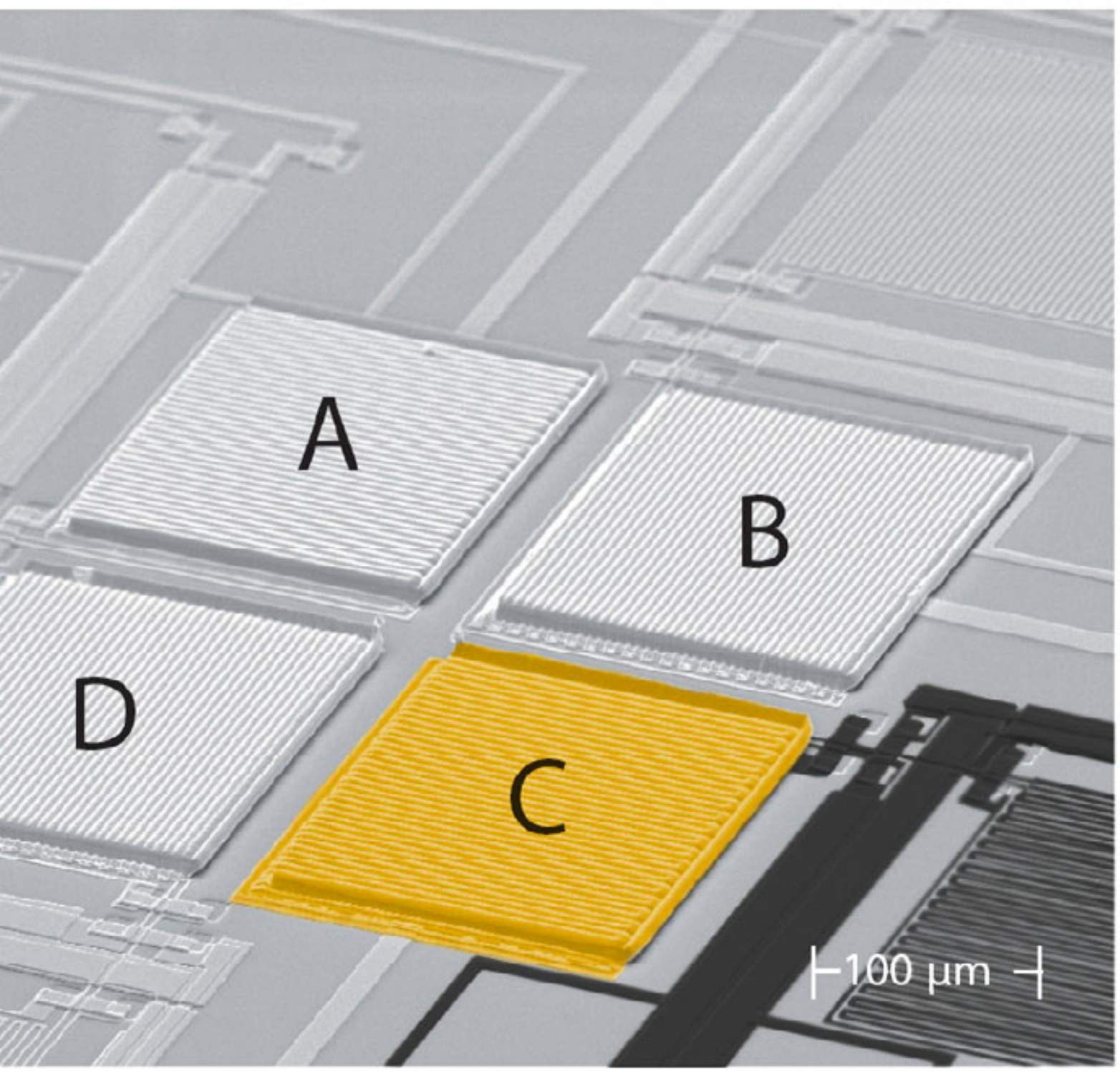} \\
  \end{tabular}
  \caption{Left:  Schematic of an idealized cryogenic microcalorimeter.  Energy deposited from a radioactive decay $(\Delta E)$ is transmuted into thermal energy (e.g. thermal phonons).  The thermal pulse $(\Delta T)$ is in turn converted to an electrical signal $(\Delta V)$ by coupling to an external system .  The timing and response of the microcalorimeter is determined via its electrothermal link $(G)$, and the heat capacity $C$ of the absorber.  Right: SEM picture of the central part of the fully microfabricated magnetic microcalorimeters with the four pixels (A, B, C and D) prepared to be irradiated. Figure reproduced from~\cite{gastaldo:2013}.}
  \label{fig:bolometer}
  \end{center}
\end{figure}

The most common form of calorimeter used to date for direct neutrino mass measurements has been the cryogenic microcalorimeter, so we will concentrate on devices of this type.  A schematic of an idealized version of such a detector is shown in Fig.~\ref{fig:bolometer}, although not all calorimeters have this specific configuration.  
Energy deposited from a radioactive decay $(\Delta E)$ is transmuted into thermal energy (e.g. thermal phonons).  The conversion from kinetic to thermal energy depends critically on the heat capacity, $C(T)$, of the target absorber and the surrounding system:
\begin{equation}
    \Delta T = \Delta E / C(T),
\end{equation}
where $\Delta T$ is the change in temperature induced by a deposition of energy $\Delta E$.  The specific heat capacity of the system depends strongly on temperature, with superconductors and dielectrics having a cubic dependence ($C(T) \propto T_b^3$), while normal metals have a linear dependence ($C(T) \propto T_b$).  As such, these detectors work best when the base temperature $T_b$ is extremely low, usually below 100 mK.   

The change in temperature must eventually be converted into a detectable (electrical) signal.  The conversion from a change in temperature $\delta T$ to a change in voltage $\Delta V$ depends on the specific technology deployed.  Typical methods include using the superconducting transition in thin materials, such as used by transition-edge sensors (TES, {\em e.g.} HOLMES), changes in the magnetization using metallic magnetic calorimeters, (MMC's, {\em e.g.} ECHo), or detecting changes in resistance as registered by neutron transmutation doped detectors (NTD, {\em e.g.} MANU).  Which technology is used often defines the kind of experiment being conducted, as each type of system often has trade-offs for its performance and sensitivity.

A critical metric for the performance of such microcalorimeters is the energy resolution of the detected energy deposition.  The energy resolution depends greatly on the specific configuration and readout method employed for the detector.  As an example, the theoretical lower limit for a cryogenic microcalorimeter read out by a transition edge sensor is given approximately by the formula:
\begin{equation}
    \sigma_E = \sqrt{\frac{4 k_b T_b^2 C_{\rm tot}}{\alpha_T}\sqrt{\frac{\beta_T + 1}{2}}}
\end{equation}
where $k_b$ is the Boltzmann constant, $C_{tot}$ is the total heat capacity of the microcalorimeter,  $\beta_T$ is the exponent of the temperature dependence of the thermal conductivity between the calorimeter and the thermal bath (typically $\beta_T \simeq 4$), and $\alpha_T$ is a dimensionless quantity that represents the sensitivity of the TES to the change in temperature ($\alpha_T = \frac{T}{R} \frac{dR}{dT}$, where $R$ is the resistance of the TES.  Values for $\alpha_T$ around 50-100 are not uncommon).  One thing to note is that the theoretical thermal noise limit is determined by the {\em total} heat capacity of the system.  Therefore, finer energy resolution scales with both temperature and the mass of each absorber.  As a result, such systems are operated in the millikelvin regime and  are constructed as {\em microcalorimeters}, so as to reduce the total heat capacity of the system.

Another critical parameter that dictates the performance of microcalorimeter systems is the timing response.  This is dictated both by the capacity of the system and by the electrothermal link conductance $G_{th}$ responsible for removing the excess heat from the absorber.
\begin{equation}
    \tau = \frac{C}{G_{th}}
\end{equation}
\noindent The rise and decay of the heat pulse depend critically on the thermal coupling between the absorber and the detector, as well as the coupling between the absorber and the heat bath.  The onset and decay of the signal pulses determine the maximum possible rate that a given absorber can tolerate before multiple pulses appear within a given time window (pile-up).  Pile-up leads to a mis-identification of the energy deposition.  Given that the density of events is far greater below the endpoint of the spectrum, pile-up manifests itself as a background, which scales approximately as $A \times f_{\rm pileup} \times (2 Q)$, where $A$ is the source activity and $f_{\rm pileup}$ is the fraction of pileup events within a given absorber.  A given calorimetry-based experiment needs to balance the inherent activity, the energy resolution, and the timing response of each absorber.  All these factors have moved the field toward {\em multiplexed} systems wherein thousands to hundreds of thousands of detectors need to be instrumented in order to achieve the desired neutrino mass sensitivity.

\begin{figure}[htb]
  \begin{center}
  \includegraphics[width=4.5in]{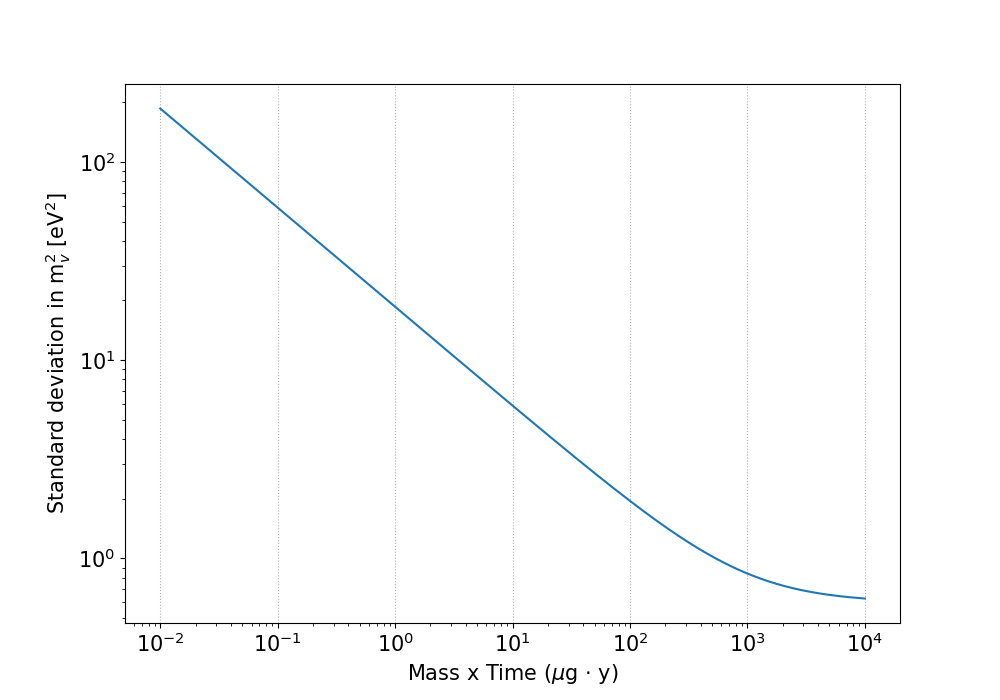}
  \caption{Uncertainty obtainable as a function of the \isotope{Ho}{163} mass under observation ($\mu$g$\cdot$y).  We assume a detector energy resolution of $1.00(15)$ eV and a background rate from a pileup fraction of $2.5 \times 10^{-4}$.  We have also assumed that approximately one part in $10^{12}$ decays occurs in the last eV of the holmium spectrum.}
  \label{fig:Ho163_sensitivity}
  \end{center}
\end{figure}

One can use the estimation method outlined in Sec.~\ref{sec:estimation} to provide guidance on the potential reach of the calorimetric effort.  Such is done in Figure~\ref{fig:Ho163_sensitivity}.  For reference, the current first phase of the HOLMES and ECHo experiments aim at a total detector mass of roughly $20 \mu$g.

\subsection{Frequency techniques} \label{sec:freqtechniques}

A new approach was suggested in 2009 by Monreal and Formaggio \cite{monreal:2009aa}.  Electrons from a gaseous beta emitter like tritium, when in a magnetic field, produce cyclotron radiation that can be detected in a sensitive receiver.  Because of a relativistic effect, the frequency of the radiation depends on the kinetic energy and so makes possible a new spectroscopy, Cyclotron Radiation Emission Spectroscopy (CRES). In a uniform magnetic field $B$, the frequency $f$ and  power $P$ radiated by an electron of kinetic energy $E$ are given by 
\begin{eqnarray}\label{eqn:f_cyclotron}
2 \pi f &=& \frac{2 \pi f_0}{\gamma} = \frac{eB}{m_e + E/c^2}, \\ 
\ P &=& \frac{2 \pi e^2 f_0^2}{3 \epsilon_0 c} \frac{\beta^2 \sin^2\theta}{1-\beta^2}.
\end{eqnarray}
The pitch angle $\theta$ is the angle between the momentum vector and field direction. The zero-energy electron cyclotron frequency $f_0$ is a fundamental constant \cite{Zyla:2020zbs},
\begin{eqnarray}
f_0&=& 27.992489872(8) {\rm \ GHz\ T}^{-1}.
\end{eqnarray}
The maximum power radiated by an 18-keV electron in a 1~T field is about 1 fW.  

\begin{figure}[htb]
  \begin{center}
  \begin{tabular}{c c}
  \includegraphics[width=3.2in]{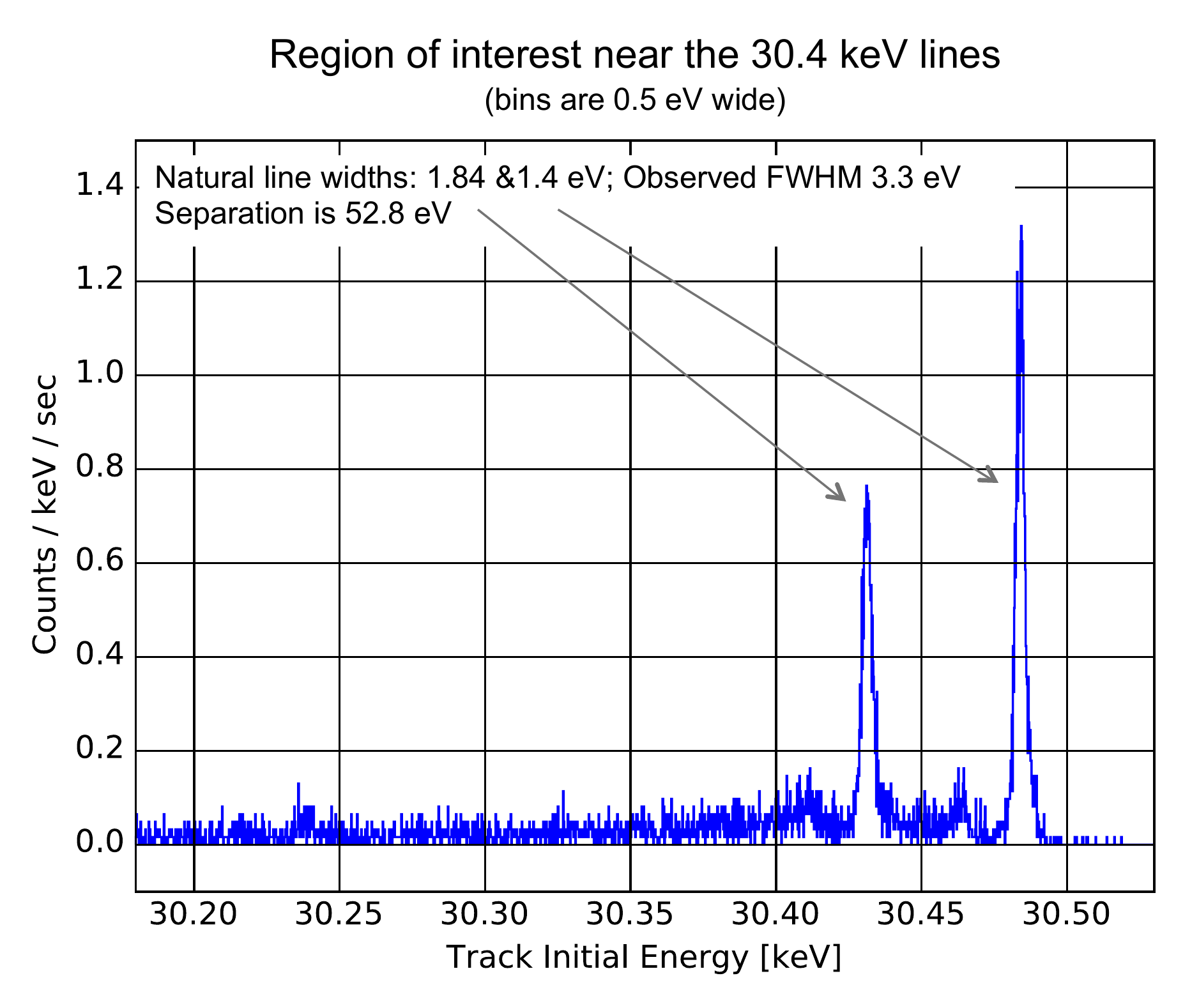} &
  \includegraphics[width=3.2in]{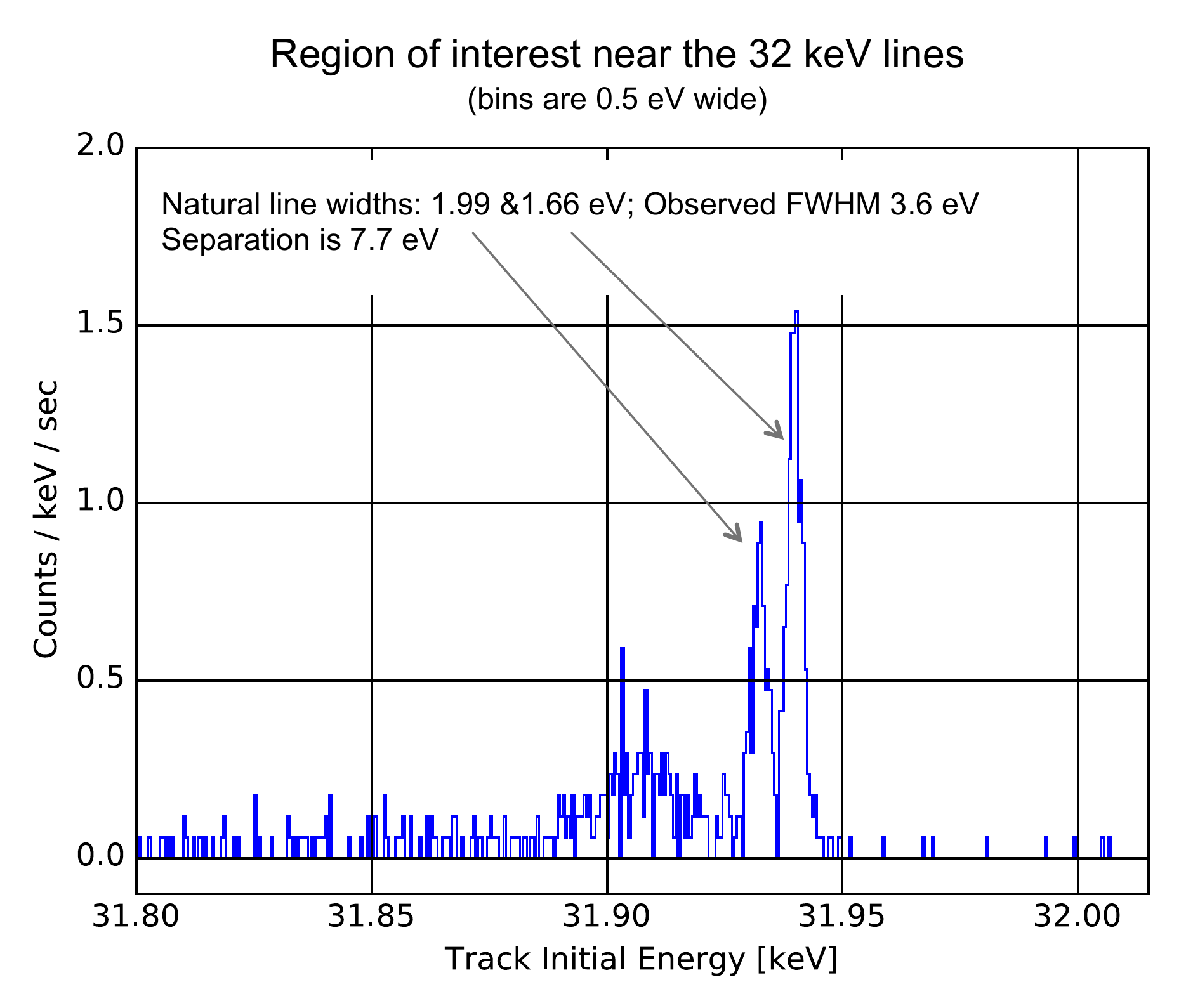} \\
  \end{tabular}
  \caption{Internal-conversion electron lines in the decay of $^{83m}$Kr measured by the Project 8 collaboration with the CRES method \cite{Esfahani:2017dmu}. Left: The L2 and L3 lines from the 32-keV isomeric decay transition. Right: The M2 and M3 lines from the same transition.  The events not in the sharp peaks arise mainly from shakeup and shakeoff processes in the decay \cite{Robertson:2020boa}, and partly from scattering in the residual gas.}
  \label{fig:krlines}
  \end{center}
\end{figure}

The first experimental demonstration of CRES was made in 2014 by the Project 8 Collaboration \cite{Asner:2014cwa} with the isotope $^{83m}$Kr, the decay of which produces sharp internal conversion lines.  The experimental cell consisted of a section of WR-42 rectangular waveguide having a cross section of 10.7 $\times$ 5.0 mm.  Because the electrons  travel a great distance in the several microseconds needed to make an accurate measurement of the frequency, a magnetic trap formed by a coil around the waveguide was used to trap electrons having pitch angles $\theta$ near $\pi/2$.  Signals produced by electrons were transmitted by the waveguide to a low-noise cryogenic amplifier, superheterodyne receiver, and digitizer.  
\begin{figure}[htb]
  \begin{center}
  \begin{tabular}{c c}
  \includegraphics[width=4.0in]{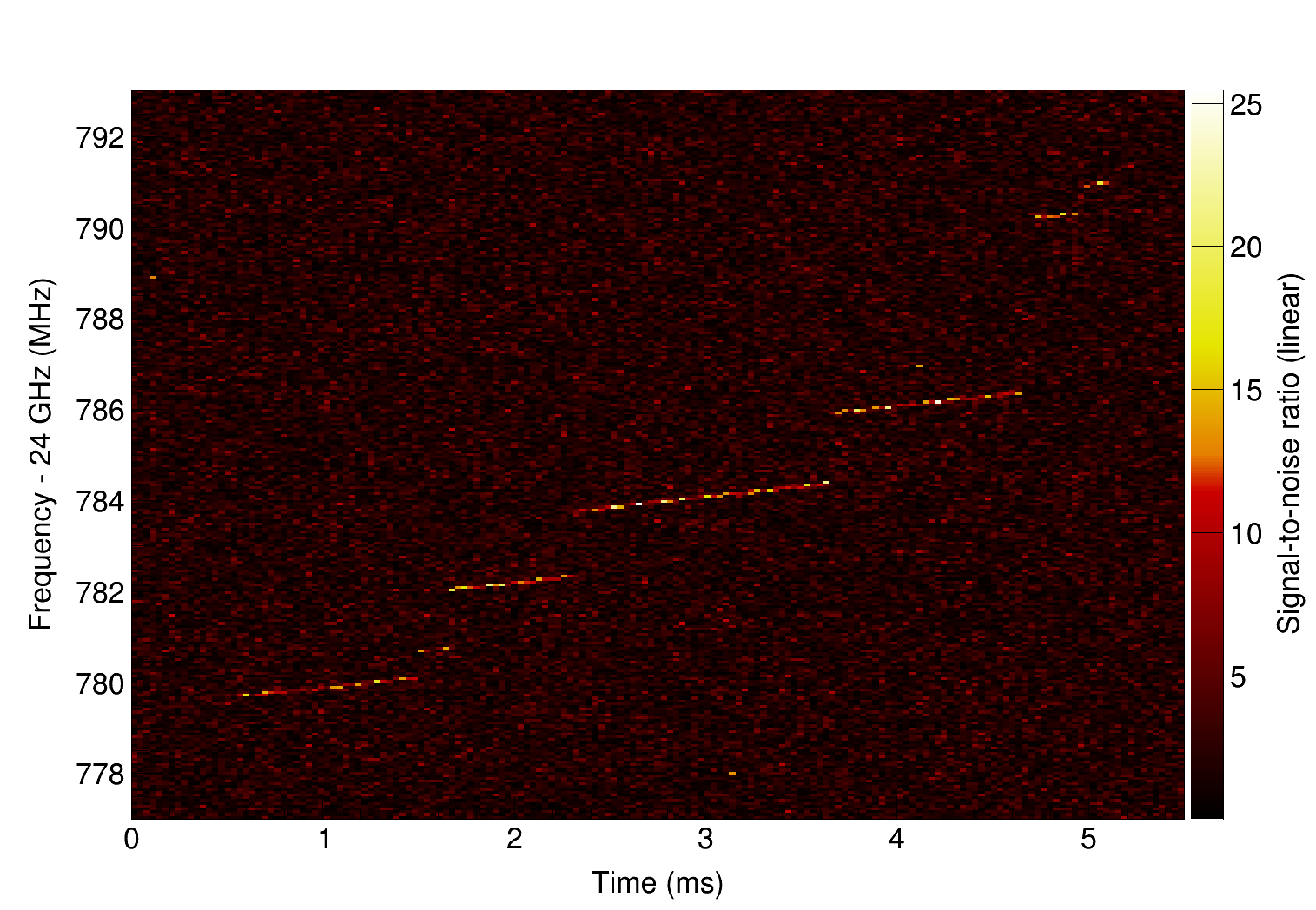} &
  \includegraphics[width=2.0in]{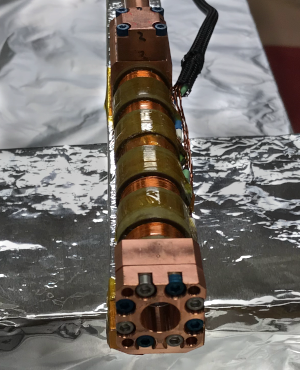} \\
  \end{tabular}
  \caption{Left: First event observed by the CRES method (Project 8 Collaboration \cite{Asner:2014cwa}). The spectrogram shows RF power in 25-kHz frequency bins and 40-$\mu$s time bins.  An electron is created by $^{83m}$Kr decay near the lower left corner and forms a track that slopes upward due to radiation loss.  The discontinuities from track to track are caused by the electron scattering inelastically from the background gas, which is mainly hydrogen.  The most probable jump size corresponds to about 14 eV. Eventually the electron scatters out of the trap and is lost. Right: Waveguide insert used for the CRES technique by the Project 8 collaboration. The inner diameter of the waveguide is 1 cm.}
  \label{fig:eventzero}
  \end{center}
\end{figure}
The first event recorded is shown in the iconic plot reproduced in Fig.~\ref{fig:eventzero}.  For each such decay event, the initial electron energy is derived from the frequency at the onset of power in the first track.  The potential for good energy resolution is visually apparent from the narrowness, approximately one frequency bin, of the tracks.  From the relationship between energy and frequency, 
\begin{eqnarray}
\frac{dE}{E}
&=&-\frac{\gamma}{\gamma-1}\frac{df}{f}. 
\label{eq:freqenergyresolution}
\end{eqnarray}
A 30-keV electron has $\gamma=1.059$, which means that a frequency resolution of 1 ppm corresponds to 18 ppm in energy, or roughly 0.5 eV. Other contributions arise from system noise because there is uncertainty as to which bin is the first, and from differences in the average magnetic field experienced by electrons with different pitch angles in the trap.  Examples of the good resolution obtainable with the CRES method are shown in Fig.~\ref{fig:krlines} from \cite{Esfahani:2017dmu}. The instrumental resolution is about 3 eV FWHM for these data. 

The need for a trap leads to complications in the energy spectrum.  In the experiment under discussion, the electron moves back and forth in the trap along a magnetic field line that is aligned with the axis of the waveguide.  The received signal is frequency modulated by the axial motion because of the Doppler effect. Amplitude or frequency modulation of a steady carrier introduces sidebands that are spaced from the carrier by multiples of the modulating frequency.   When frequency modulation causes shifts in the frequency that are greater than the frequency of modulation itself, sidebands proliferate at multiples of the modulation frequency.  The power is spread excessively and it becomes difficult to detect and identify the many weak sidebands, while the carrier power also becomes small and can disappear altogether.  The critical parameter is the modulation index, which is defined as
\begin{eqnarray}
h&=& \frac{\Delta f}{f_a}
\end{eqnarray}  
where $\Delta f$ is the peak one-sided frequency shift and $f_a$ the modulating frequency, the axial frequency in this application.   When $h \le 1$ the power is mostly in the carrier and the two lowest-order sidebands, and sideband proliferation sets in above that.  When $h = 2.405 $ (the first zero in a related Bessel function) the carrier vanishes completely.  The phenomenology of this effect is described in \cite{Esfahani:2019aa}, and it strongly restricts the usable axial amplitude and pitch angle in the trap.  This effect coupled with the small dimensions of the waveguide result in an effective volume of the gas in the source, the equivalent volume that contributes detectable events, that is only a fraction of a cubic millimeter. 

Application of CRES to the neutrino mass problem requires much larger effective volumes of tritium for statistical accuracy.  Two approaches to circumventing these limitations are being explored by the Project 8 collaboration. One is to collect the cyclotron radiation emitted perpendicular to the motion of the electron's guiding center in the trap \cite{Esfahani:2017dmu}.  The resulting reduction in solid angle places severe demands on the noise level of the amplifiers and the temperature of the system.  The second approach is the use of a cavity to collect the RF energy emitted.  Certain TE modes are Doppler-free but a large overmoded cavity is complex to use and analyze. 

The ultimate reach of experiments based on molecular tritium is limited by the final-state distribution, which smears the beta spectrum at the endpoint.  The CRES method raises once again the long-sought goal of an atomic tritium experiment.  Since only microwave photons and not the beta electrons themselves need to be directly detected, a magnetic trap for atomic tritium is usable.  The fact that molecular tritium levels have  magnetic moments at least thousands of times smaller than free atomic tritium means that molecules are essentially not trapped.  This property helps in providing the low ratio of molecules to atoms that is necessary for a background-free measurement at the atomic endpoint.  Atomic tritium contained within Tesla-scale magnetic walls is necessarily very cold, at a sub-Kelvin temperature, reducing thermal Doppler broadening.

\begin{figure}[htb]
  \begin{center}
  \includegraphics[width=6in]{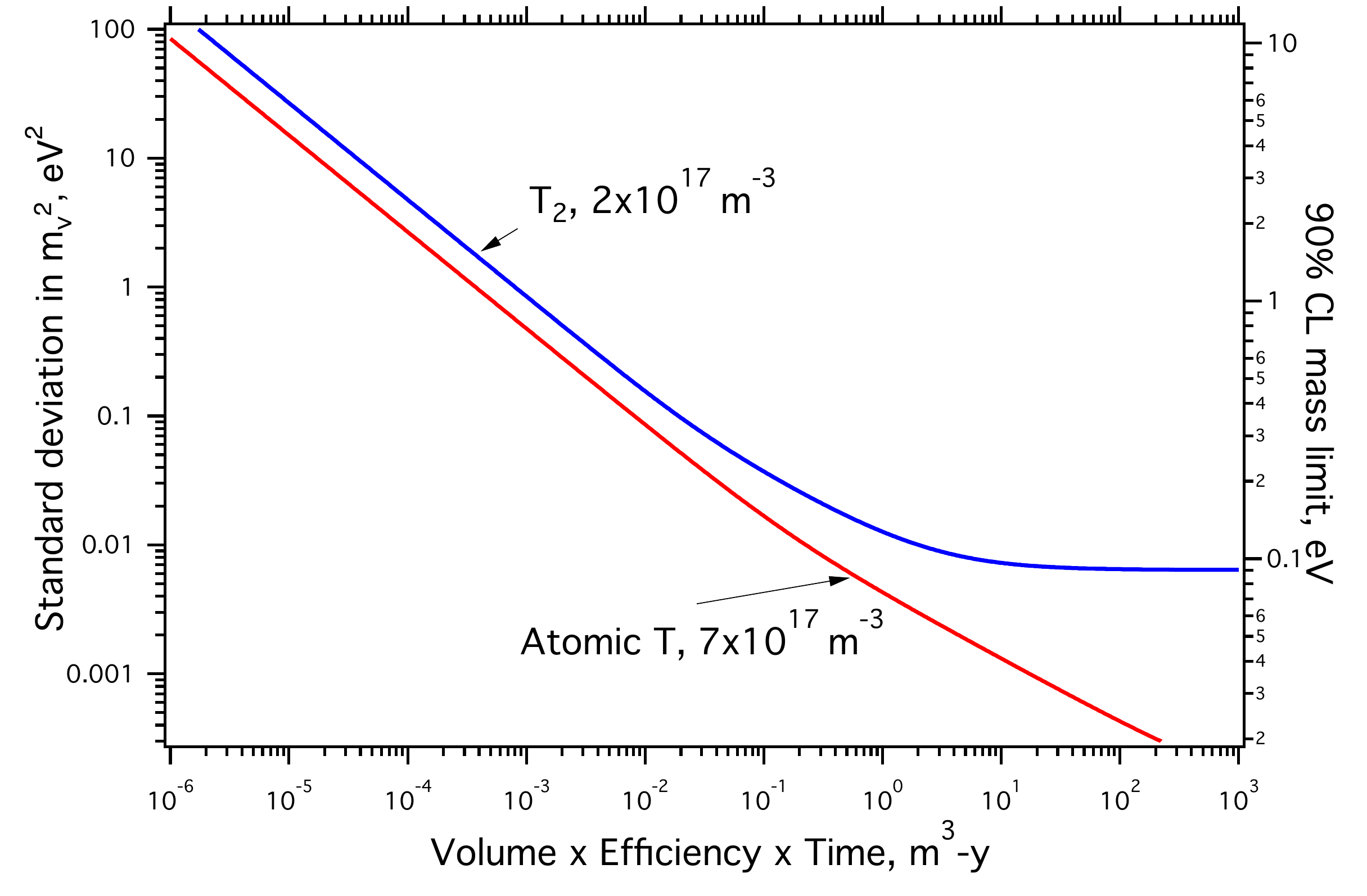}
  \caption{Uncertainty obtainable as a function of volume under observation for various choices of number density per cm$^3$.  Systematic uncertainties due to imperfect knowledge of contributions to the resolution are included.  The densities correspond to mean times between scatters of $200 \ \mu$s, as used in current Project 8 data. The frequency chosen is 26.5 GHz, the energy resolution is 3 ppm rms, the source temperature for molecular T$_2$ is 30K while for atomic T it is 1K, and the background is $10^{-6}$ per second per eV.  }
  \label{fig:sensitivity}
  \end{center}
\end{figure}

Figure~\ref{fig:sensitivity}   
shows calculated neutrino mass sensitivities for molecular and atomic tritium  with number densities chosen to reproduce the mean track durations presently used in Project 8, as a function of the product of volume, efficiency, and live time.  The estimation method of Sec.~\ref{sec:estimation} is used for these calculations.
The cross-section cited by Aseev {\em et al.}~\cite{aseev_energy_2000}, $3.4\times 10^{-18}$ cm$^2$, has been used for electron scattering by molecules, and for atoms we have used $9\times 10^{-19}$ cm$^2$ based on the work of Shah {\em et al.}~\cite{Shah:1987zz}.  For concreteness, we assume that the distributions  $\sigma_i$ in Eq.~(\ref{eq:eqten}) are each known to 1\%, i.e. $u(\sigma_i)/\sigma_i =0.01$.
For calculating the `sensitivity' shown here, the expected value for ${m_\beta^2}$ is taken to be 0, and, statistically, positive and negative values for this quantity are equally probable.  The 90\% CL is a one-sided interval derived by setting the 1.28-sigma upper threshold on ${m_\beta^2}$, which is assumed to be Gaussian distributed.  The square root of this number is displayed on the right-hand axis.

The physics reach of a Project 8 experiment depicted in Fig.~\ref{fig:sensitivity} is attractive, but should be regarded as an optimistic estimate of what could be done with this type of measurement.  The systematic uncertainties  on resolution-like parameters are assumed to be very small, 1\%,  and many effects are omitted.

As can be seen, an experiment with gaseous molecular T$_2$ reaches a limit in sensitivity of order 100 meV because of the width of the FSD combined with Doppler broadening associated with the minimum feasible operating temperature near 30 K.  It would be necessary to know the FSD to an accuracy of 0.1\% to reach the 40-meV level, and the running time would be 10 times longer than with an atomic experiment of the same size and efficiency.  For these reasons, the Project 8 collaboration is exploring the development of an atomic T source in a magnetic configuration that traps both spin-polarized atoms and the betas.  The density required is in an achievable range, of order $10^{18}$ m$^{-3}$. The mean energy of the atoms that can be stored depends on the magnetic wall height, and is about a factor of 10 to 20 below the energy equivalent to the height in order to slow evaporation and obtain lifetimes longer than tens of seconds.  A magnetic wall of height 1 T can retain 60-mK atoms for periods of a minute or so.

In addition to the resolution contributions from the gaseous source itself, specifically the FSD width and the translational Doppler broadening, there are instrumental contributions.
The instrumental resolution has two readily identifiable components, field inhomogeneity and noise.  Axial and radial variations of the trapping and background fields mean the average cyclotron frequency depends on the electron's  position within the trap and on its axial amplitude.  Moreover, the presence of a radial gradient in the magnetic field causes the electron's guiding center to circulate slowly around the axis, passing through regions that may have slightly different average field.  The drift velocity in the presence of a magnetic field gradient is given by \cite{Otten:2008zz}
\begin{eqnarray}
{\bf u_\perp} &=& \left[ \frac{(2E_\parallel+E_\perp)}{e}\frac{\bf \nabla_\perp B}{B^3}\right]\times {\bf B},
\end{eqnarray}
where $E_\parallel$ and $E_\perp$ are the parallel and perpendicular energies.  In the quasi-uniform field  that typifies a weak trap, a 20-keV electron moves azimuthally at 2 m/s if the field is 1 T and has a gradient of $10^{-4}$ T/m.  In stronger traps, the gradient is large and depends on the axial amplitude, leading to more complex behavior.  The sign of this ``grad-B'' motion can even change with radius \cite{Furse_thesis}.

Noise is fundamental  in the performance of a CRES experiment.  The signal must be detectable above receiver and thermal noise, which implies a bandwidth limitation.  That in turn implies a minimum observation time for reliable detection of a track, which, finally, sets a limit on the gas density $n$.  The mean free path  $\lambda=1/\sigma_0 n$, where $\sigma_0$ is the cross section.  For 18.6-keV electrons incident on molecular tritium, the inelastic cross section is \cite{aseev_energy_2000}, $3.4\times 10^{-18}$ cm$^2$, while for atoms it is $9\times 10^{-19}$ cm$^2$ \cite{Shah:1987zz} (the ratio of cross-sections confers almost a factor of 2 advantage on atomic experiments for a given activity density).  It was found in Project 8 \cite{Asner:2014cwa,Esfahani:2017dmu} that the track duration that optimized count rate and detection efficiency was about 200 $\mu$s. Noise directly plays a role in the energy resolution by introducing uncertainty into the frequency measurement and the start-time measurement.  Both quantities enter into the energy determination because radiation loss means the frequency is changing with time.  Noise also sets the background level for a given detection threshold because of random alignments of spectrogram pixels having above-average noise power.

\subsection{Atomic tritium}

Atomic tritium is an attractive candidate for freedom from final-state effects.  The ground state in the He$^+$ daughter is separated by 40 eV from higher excited states and receives 70\% of the decay intensity.  There are no rotational and vibrational excitations, and (in gaseous form) no lattice effects.  We mentioned above the low cross section for electron inelastic scattering on atomic tritium compared to molecular. The Los Alamos experiment \cite{Lanl82,robertson:1991aa} was designed to operate with atomic tritium, but the technical challenges proved insurmountable and molecular T$_2$ data were used.  As it turns out, both the Los Alamos and Livermore \cite{stoeffl:1995aa} experiments were systematically affected by the limitations of the contemporary theory of the molecular final state spectrum, as was subsequently found \cite{Bodine:2015sma}.  

Atomic hydrogen can be produced from molecular hydrogen either in an RF or DC discharge or by thermal `cracking' on a hot tungsten surface. That the latter strategy is efficient may be surprising because at 2600K $k_BT = 0.25$ eV while the binding energy of the H$_2$ molecule is 4.5 eV, but once dissociated the atoms cannot easily recombine because it is necessarily a 3-body process.  A comprehensive review and compendium of the dissociation of hydrogen and formation of beams  is given by Lucas \cite{Lucas:2014zz}.  Atomic beams of tritium have been produced only three times, by Nelson and Nafe in 1949 \cite{NelsonNafePhysRev.75.1194},  by Prodell and Kusch in 1957 \cite{ProdellKuschPhysRev.106.87}, and by Mathur \etal \cite{Mathur:1967zz} in 1967.  A DC discharge was used by Nelson and Nafe and RF discharges by the other two groups, and in each case the tritium was lost from circulation in a matter of hours.  Dilution with hydrogen (protium) also occurred quickly.

The Los Alamos research into atomic tritium used the RF discharge method, a 50-MHz electrodeless discharge in a borosilicate or silica glass tube. Tritium was never used, and it was always found that the dissociation fraction with protium and deuterium decreased over time periods of hours to days, and the tube became discolored.  Similar experiences were reported in research on Bose-Einstein condensation in hydrogen  \cite{WalravenSilveraRSI53}, and it was commonly attributed to contaminants in the gas or vacuum system.  Recently we reexamined the Los Alamos logbooks and find possible evidence for a different mechanism.  The RF discharge produces fast electrons, hard UV light, and reactive atoms, and is capable of attacking the glass envelope.  The glass surface was probably reduced to silicon monoxide, which is brown in color and was incorrectly ascribed to contaminants.  At the same time, the hydrogen was converted to water that was subsequently found on vacuum-system cold traps after operation of the discharge, but not otherwise.  Slevin and Stirling, however, report operation of a seemingly similar RF discharge tube over thousands of hours \cite{SlevinStirlingRSI52}, so it is evident that not all the relevant factors have been identified.  Inventory conservation and equipment longevity are important in a tritium experiment.  We venture that the RF discharge method will be unsatisfactory for an atomic tritium experiment where efficient recycling is essential, and that thermal cracking will be preferable.  It is furthermore prudent to avoid ionizing the gas at any point because it leads to ion pumping of the tritium (some thermal crackers use electron bombardment heating, which can ionize the background gas). Ferrous metals are known to take up and release hydrogen in high-vacuum systems, and it is of interest to explore the use of other materials such as copper and aluminum.

Another challenge of atomic tritium is that the Q-value is 8 eV less than for the molecule \cite{Bodine:2015sma} (see Fig.~\ref{fig:atmol}). The tritium spectrum rises parabolically below the endpoint, and, as described above and shown in Eq.~(\ref{eq:molbackground}), a very small molecular contamination at the 0.01\% level will represent an important background to the atomic spectrum in future sensitive experiments.  Typically the dissociation fraction from a dissociator is not better than 90\%, insufficient for an atomic tritium measurement if there is no further purification.  Adding to the difficulty, the Los Alamos spectrometer resolution, about 25 eV FWHM, was insufficient to resolve the atomic and molecular components, so the group planned to use frequency doubled and tripled Nd:YAG laser light for an absorption measurement to determine and monitor the molecular fraction.  However, that technology was in its infancy at the time.  These factors played a major role in the retreat to molecular tritium at both Los Alamos and Livermore.   Much higher resolution is now the norm in tritium beta decay, such that the molecular and atomic components would be readily resolvable in the spectrum itself.  This will not mitigate the need for high atomic purity but it does at least allow a continuous internal calibration of it.  

As experiments push below the 1-eV level, ever larger activities of tritium must be under observation to keep measurement times within reason.  The KATRIN experiment has approximately 1 Ci under observation in the source tube.  Will it be possible to develop a similar activity in atomic tritium?  The answer hinges on the production and containment of atomic tritium.  We consider containment first.

\begin{figure}[htb]
    \centering
    \includegraphics[width=4in]{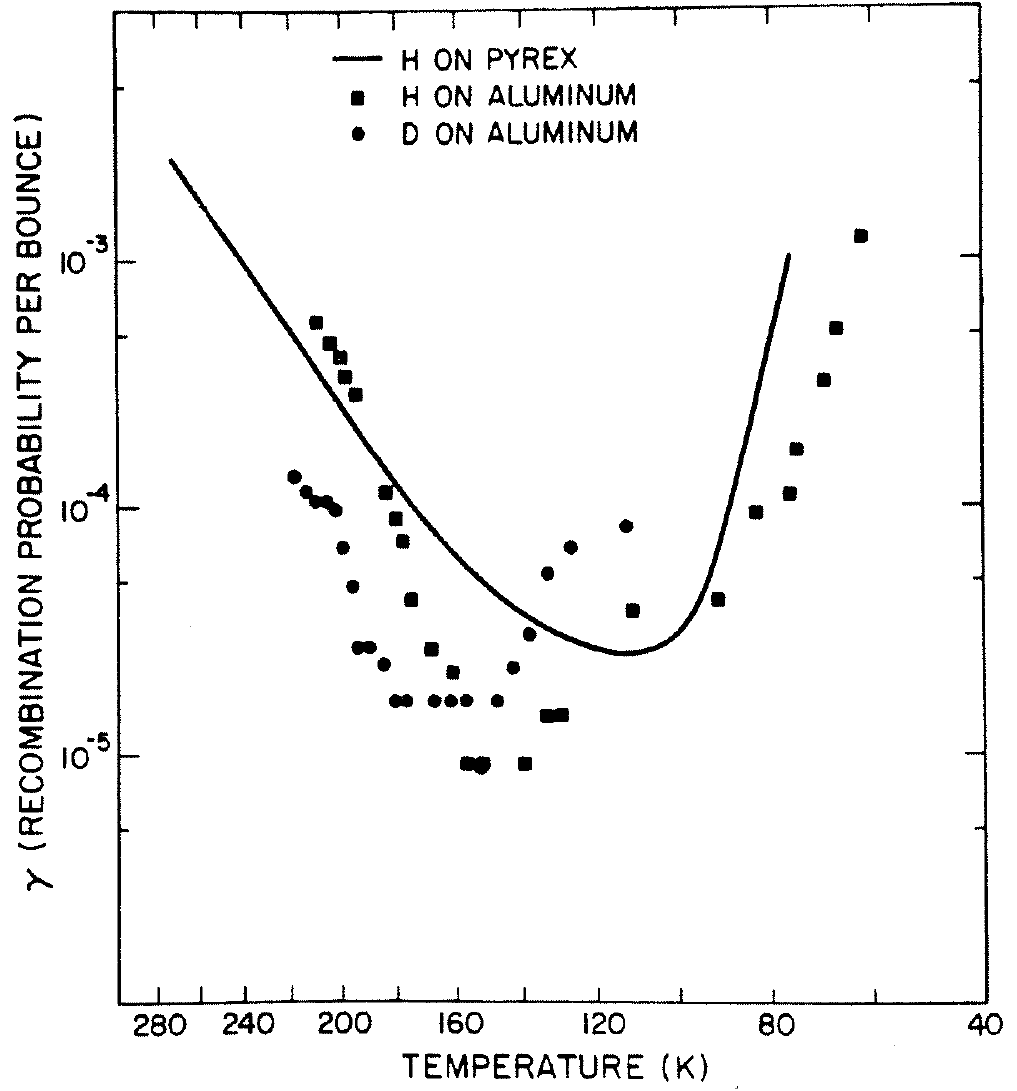}
    \caption{Recombination probability per bounce for hydrogen and deuterium atoms on aluminum \cite{Lanl82}. The solid curve is for borosilicate glass \cite{Wood&Wise1962}.}
    \label{fig:recombination}
\end{figure}

Physical bottle containment has long been used in hydrogen masers. Glass surfaces, sometimes coated with hydrocarbon or fluorocarbon films, are quite effective in inhibiting both recombination and spin flip.  For a tritium experiment, however, insulating surfaces are a danger because of the accumulation of unknown surface charges that affect the measured spectrum. This motivated the Los Alamos group to carry out an experimental search for a metallic surface that did not encourage recombination.  It was found that aluminum (a standard 6069 alloy) performed as well as glass (Fig.~\ref{fig:recombination}).
The source tube in the Los Alamos experiment was therefore made of aluminum, 5 cm in diameter and internally polished to reduce declivities where recombination would be enhanced.  Today, this approach to an atomic trap is disqualified by the molecular background, which cannot be reduced to the $\lesssim 10^{-4}$ level that a sensitive atomic experiment requires.  On the other hand, aluminum, glass, and possibly other materials are well suited to the accommodation step where dissociated atoms are first cooled before entering the trap.  Fluorocarbons, however, are not suitable with tritium because they are degraded by radiolysis \cite{Fox2017_FusionScience}, which leads to the formation of TF, a corrosive gas.

\begin{figure}[htb]
    \centering
    \includegraphics[width=4in]{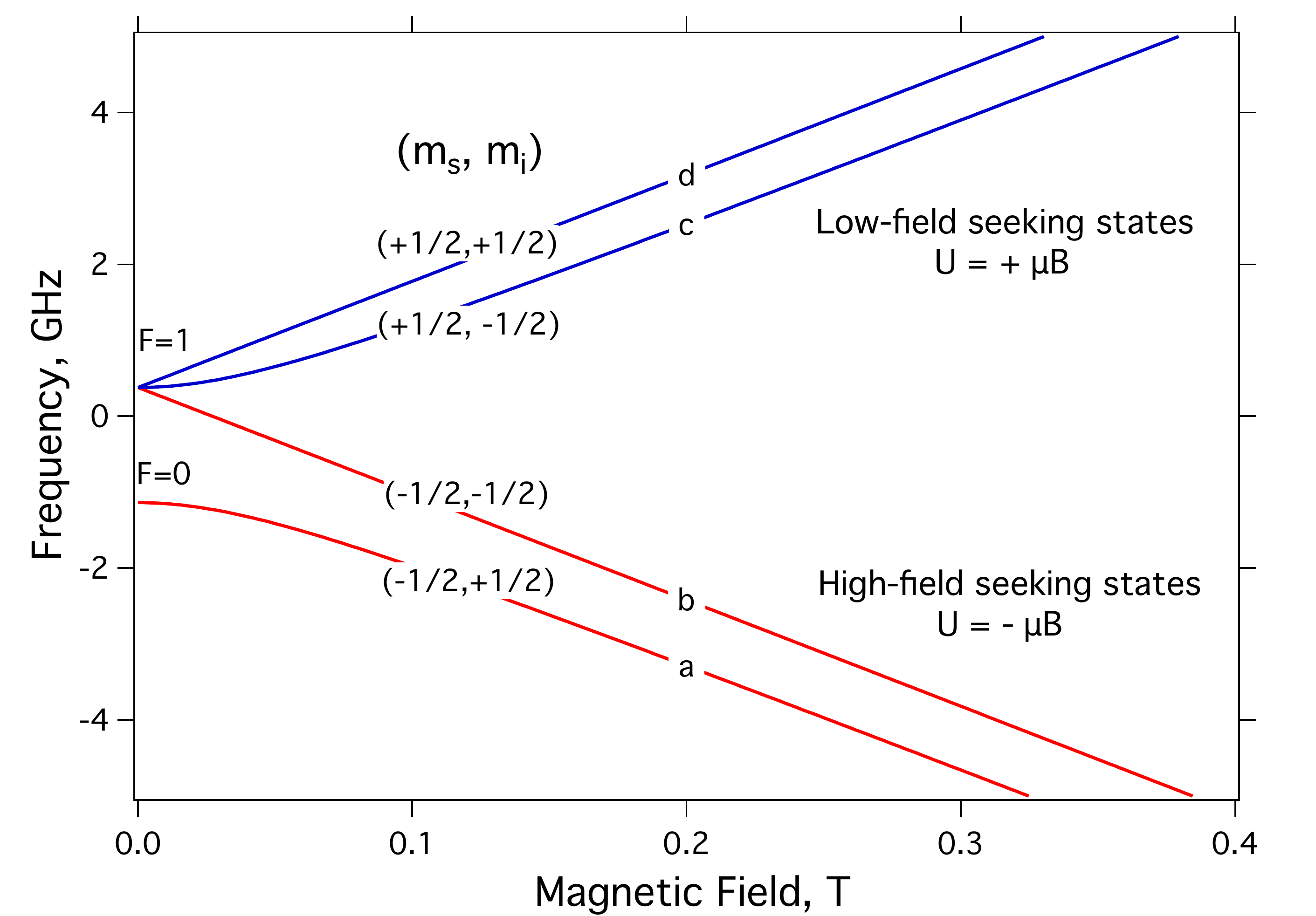}
    \caption{Breit-Rabi diagram for atomic tritium, with parameters from \cite{SilveraWalraven1986}. The effective magnetic moment is proportional to the slope of the hyperfine levels. At low fields the total angular momentum and its projection, $(F,m_F)$, are good quantum numbers, whereas at high field the electronic and nuclear spins orient independently and their projections $(m_s,m_i)$ become good quantum numbers.}
    \label{fig:breitrabi}
\end{figure}

Atomic hydrogen has a magnetic moment $\mu$ of approximately 1 Bohr magneton, which makes possible a purely magnetic trap without physical walls.  A major advantage is that the magnetic moment of molecular hydrogen is 3 orders of magnitude smaller, and molecules will be able to escape the trap with relative ease, improving the atomic purity. The magnetic energy is 
\begin{eqnarray}
U&=&\boldsymbol{\mu}\cdot\boldsymbol{B}.
\end{eqnarray}
The magnetic energy is displayed in a Breit-Rabi diagram, Fig.~\ref{fig:breitrabi}. 
If the magnetic moment is  in the ``low-field-seeking'' orientation with respect to the magnetic field, an atom in a local field minimum is attracted to the minimum. Maxwell's equations do not permit a local field maximum, but high-field-seeking atoms can be trapped in a saddle-point region if physical walls can also be used to prevent their escape to higher fields. The reason for using the evidently more complicated high-field trap is related to dipolar spin-flip interactions \cite{Lagendijk_PhysRevB.33.626}, which are energetically forbidden in this configuration.  They dominate the loss rate from field-minimum traps.    

It was discovered in the course of experiments to make a Bose-Einstein condensate of hydrogen \cite{BEC_PhysRevLett.81.3811} that atomic hydrogen could be efficiently contained in a high-field bottle trap in which the walls were coated with a film of superfluid He \cite{SilveraWalraven1986}.  Unfortunately the method does not work with atomic tritium, because the adsorption energy of atomic tritium on superfluid He is too high \cite{SilveraWalraven1986}.  From theory, the adsorption energy for hydrogen (protium) is 0.85 K and for tritium 3.2 K, a prediction that is experimentally supported at the $\sim 15$\% accuracy level for protium.  The surface recombination rate depends exponentially on this quantity.   

 Two kinds of field-minimum magnetic trap have been devised, the Ioffe-Pritchard trap \cite{Pritchard_PhysRevLett.51.1336} and the Halbach array \cite{walstrom200982}.  In both types the magnetic `walls' of the trap are regions of relatively high magnetic field orthogonal to the central field.  If an atom approaches these regions not too quickly, its magnetic moment $\mu$ orients with the local direction of the field $B$.  Atoms in low-field-seeking states are repelled by the magnetic walls back to the central region, while atoms in high-field-seeking states are ejected from the trap.
 
 The simplest form of the Ioffe-Pritchard trap is a quadrupole magnet with pinch coils at either end, and it has been successfully used to trap hydrogen atoms \cite{Hess_PhysRevB.34.3476} and free neutrons \cite{O'Shaughnessy:2009zw}. Increasing the multipolarity of the trap magnet gives a larger central volume of relatively uniform field (provided by a separate solenoid), and the ALPHA collaboration use an octupole trap in their antihydrogen experiments \cite{Ahmadi:2020ael,Amole:2014vna}. In principle, a large superconducting Ioffe-Pritchard trap of high multipolarity combined with a solenoid can provide both the magnetic walls for trapping atomic tritium and the uniform central magnetic field for carrying out CRES, although such a large trap has never been built.   
 
 Halbach arrays are periodic assemblies of permanent-magnet blocks arranged to produce a field near the surface that is $>1$  T and which falls off with a characteristic distance given by the block period \cite{walstrom200982}.  One of the most noteworthy applications of this approach is the UCNtau experiment to measure the lifetime of the free neutron in a gravitomagnetic Halbach trap \cite{Pattie:2017vsj}.  The loss rate for processes other than beta decay translates to a lifetime of order months for trapped ultracold neutrons.  The permeability of rare-earth magnetic materials commonly  used to make the blocks is quite low (1.05 for NdFeB) which permits the superimposition of an external field for CRES measurement in the central region of the trap.  
 
 The basic ingredients of an atomic experiment can all be identified: dissociator, accommodator, velocity and state selector, and magnetic trap.  The detection of CRES signals from the larger volume has not yet been demonstrated, and is the next step in development of this new technology.

\section{Conclusion}
\label{sec:conclusion}
\setcounter{equation}{0}

In his formulation of the theory of beta decay in 1933-4, Fermi remarked \cite{fermi:1934} ``...we conclude that the rest mass of the neutrino is either zero, or in any case, very small in comparison to the mass of the electron.''  In the intervening years experimentalists have relentlessly pressed onward to find the mass, their efforts recorded in a perfect Moore's Law of technical progress (Fig.~\ref{fig:mooreplot}).  We know now that there are not just one but three different kinds of neutrino.  The revolutionary discovery of neutrino oscillations at the end of the last century showed that neutrinos indeed have mass, and that the particles with the well-defined flavors -- electron, mu, or tau --  are actually linear superpositions of particles with well-defined mass. Oscillations reveal the differences between the squares of the masses and set a minimum value because no mass can be less than zero, but they do not yield the masses themselves.  Nevertheless, this information enormously simplifies the (still difficult) task of the experimentalist because only the `easy' mass that is coupled mainly to electrons needs to be measured to determine them all, assuming the mass ordering is also  known.  And, perhaps even more significant, neutrino mass is finally confined within a window having both upper and lower bounds, a window that is steadily shrinking with each new experimental idea. 

At each stage in this odyssey, the next step has seemed insurmountable, but always new ideas and insights have opened a path.  The development of magnetic spectrometers, gaseous tritium sources, the MAC-E filter, microcalorimetry, and cyclotron radiation emission spectroscopy are examples. 

The KATRIN experiment is in operation, after 17 years of construction.  The ingenuity and care in its design are bearing fruit with a factor of 2 improved limit in only a month's data taking.  It will in due course reach its limit of sensitivity, either finding the neutrino mass or setting a limit in the vicinity of 0.2 eV.  Otten and Weinheimer \cite{Otten:2008zz} note KATRIN's singular nature, ``This could lead to a somewhat uneasy situation, in particular, if a finite but small mass signal happens to appear. How can the requirement to independently check a new result be fulfilled?'' The appearance on the scene of a novel technology, cyclotron radiation emission spectroscopy (CRES) \cite{monreal:2009aa}, may offer an answer to this important question.  It seems likely that a CRES atomic tritium experiment at {\em some} scale could be mounted, but whether it can approach or exceed KATRIN's reach is at present unknown.  The technical challenges are great.

Neutrino mass is the only fermion mass for which the minimal Standard Model makes a firm prediction: zero.  That the prediction was incorrect is the first {\em contradiction} of the Standard Model, rather than simply something omitted.  Neutrino mass seems to arise from a mechanism different from the Standard Model's Higgs mechanism and finding out what that is will illuminate the way to a more comprehensive theory.  At the same time, we are convinced that the universe is filled with neutrinos from the big bang, and their mass has affected the formation of the largest structures.  The cosmological model with a cosmological constant and cold dark matter, like the Standard Model, is an extraordinarily predictive theory, and yet it is assembled from ingredients with which we have no earthly familiarity.  In both theories there are signs of tension.  Laboratory measurement of the mass of the neutrino is one of the keys needed to unlock the mysteries.  

\section{Acknowledgments}
\label{sec:acknowledgments}
We wish to express our thanks to Martin Fertl, Stephan Friedrich, Loredana Gastaldo, David Kaiser, Kyle Leach, Angelo Nucciotti,  Walter Pettus, Alan Poon, Val\'{e}rian Sibille, Martin Slez\'{a}k,  Thomas Th{\"u}mmler, and Brent VanDevender for valuable discussions and materials for this report.  The work of AdG is supported in part by the DOE Office of Science award \#DE-SC0010143, the work of JAF is supported in part by DOE Office of Science award \#DE-SC0011091, and the work of RGHR is supported in part by DOE Office of Science award \#DE-FG02-97ER41020.

\bibliographystyle{apsrev-title}
\bibliography{massreview}

\end{document}